\documentstyle[12pt,nat_julie,psfig,layout2e,amssymb,longtable]{book}
\makeindex

\renewcommand{\labelenumi}{\roman{enumi}.)}

\renewcommand{\thefootnote}{\fnsymbol{footnote}}
\newcommand{\beg}{\begin{equation}}
\newcommand{\ee}{\end{equation}}
\newcommand{\bea}{\begin{eqnarray}}
\newcommand{\eea}{\end{eqnarray}}
\newcommand{\bfi}{\begin{figure}}
\newcommand{\efi}{\end{figure}}

\newcommand{\s}{{\rm\thinspace s}}
\newcommand{\bmin}{minimum energy magnetic field}
\newcommand{\sig}{$\sigma_{RM_z}$}
\newcommand{\si}{$\sigma_{RM}$}
\newcommand{\ao}{$a_\circ^\beta\rho_\circ$}
\newcommand{\p}{p$_{lobe}$}
\newcommand{\DLa}{D$_{lobe_1}$}
\newcommand{\DLb}{D$_{lobe_2}$}
\newcommand{\Po}{P$_{151}$}
\newcommand{\DL}{D$_{lobe}$}
\newcommand{\Ds}{D$_{source}$}
\newcommand{\so}{s$_\circ$}
\newcommand{\li}{$_{151\, {\rm{MHz}}}$}
\newcommand{\rad}{rad m$^{-2}$}
\newcommand{\DMz}{DM$_z$}
\newcommand{\qo}{Q$_\circ$}
\newcommand{\ah}{$\alpha_{high}$}
\newcommand{\ahm}{\alpha_{high}}
\author{Julie-Ann Goodlet}
\title{Environments of
powerful radio galaxies through the cosmic ages}

\begin{document}
\pagenumbering{roman}
\maketitle
\date{}
\makeabstract
\tableofcontents
\listoffigures
\listoftables
\newpage
\setcounter{page}{0}
\pagenumbering{arabic}
\thispagestyle{empty}

\psfull

\chapter*{Preface} 

All work presented here has been carried out by the
author, in collaboration with others, at Southampton University
between October 2001 and August 2004. The results presented in
chapters 2 and 3 have been published by \citet*{me} in MNRAS.

The data used in chapters 2 and 3 uses archival observations provided
by the National Radio Astronomy Observatory, details of which can be
found in Tables \ref{vlaobs} to \ref{vlaobsc}. AD429 uses observations
taken by J. Dennett-Thorpe and AD444 uses observations taken by P. Best.
All observations used in this body of work were re-reduced and calibrated
by the author.

The author uses models of the Faraday screen presented by \citet{burn}
and \citet{tribble}. The author also uses a model for the evolution of
FRII sources presented by \citet{ka97} and \citet*{kat97}.

JAG 23/08/04
\thispagestyle{empty}
\newpage
\chapter*{Acknowledgements}

I would like to thank PPARC for their financial support in the form of
a PPARC studentship.

Above all I would like to thank my supervisor Christian Kaiser for his
support and advice throughout my three years. I would also like to
extend my thanks to Philip Best at the University of Edinburgh. His
advice on the use of AIPS was essential. I also would like to thank
the NRAO VLA facility for providing me with all the archival data I
needed.

I would also like to thank my fiance, David for all his love and
laughter. I would never have got to the end with out you. I would also
like to thank my family especially my Nan, my Mum and my Dad for their
support during the dark days.
\vspace{2cm}

This thesis is dedicated to my grandfather, Kenneth Tennant Goodlet,
who sadly passed away this year before I could finish and who would
have been very proud.

\noindent I miss you..
\thispagestyle{empty}

\newpage
\thispagestyle{empty}
\vspace*{8.5cm}
\begin{quote}
``In the beginning, the Universe was created. This  has made a lot of people very angry, and is generally considered to have been a bad move.''
\end{quote}
\vspace*{10.5cm}

\begin{flushright}
{\scriptsize - Douglas Adams}
\end{flushright}
\newpage 

\newpage
\setcounter{page}{1}
\thispagestyle{empty}
\chapter{Introduction}

The formation of all massive galaxies and also understanding their
evolution is a vital component in understanding how the Universe
evolved. Some of the most extreme, distant and powerful sources in the
Universe have been observed in the radio band. AGN and radio galaxies
were first solidly detected in radio in the 1950's using large
aperture radio telescopes such as the Cambridge University and Sydney
University radio telescopes \citep{book_agn}. The publication of the
3$^{\rm rd}$ Cambridge survey (3C) in 1962 \citep{b62} represented the
first comprehensive list of radio sources in the northern
hemisphere. Unfortunately it was not the first complete survey using
the Cambridge telescope. Embarrassingly the 1$^{\rm st}$ and 2$^{\rm
nd}$ Cambridge surveys were found to be highly confusion limited and
had to be abandoned \citep{ms57}. The Cambridge telescope was later
upgraded and the well known 3C survey was produced. The revised 3C
sample (3CR) is flux-limited, only sources brighter than 12 Jy at 178
MHz are included and hence only contains sources out to
$z\sim2$. Current technology has vastly improved on this limit, the
new Texas - Oxford Survey (TOOT) \citep{hr03} has a flux limit 100
times fainter than the 3CR sample and reaches as far back as $z\sim4$.

Radio sources are visible in all wavebands ranging from X-ray, through
the optical and into the radio. The Cosmic Microwave Background
radiation scatters off the radio synchrotron emitting electrons into
the X-ray band and 3C 273 was the first radio source to be found in
X-rays \citep{blm70}. The Compton scattered emission from radio jets
emits in the X-ray but AGN also emit X-rays directly. It was not until
the late 1970's that the {\it Ariel V} survey found a sub-class of
radio galaxies that were strong X-ray emitters \citep{elvis}. Radio
sources often have optical counterparts that have been observed with
the large optical telescopes such as Keck and the Hubble Space
Telescope. By observing these sources in the optical, observers can
determine the host galaxy properties. For many sources this is the
only way to determine where the nucleus of the radio source should lie
as it may be too faint to be present at radio wavelengths. Recent
studies by \citet{sid03} using SCUBA in the sub-mm have found that AGN
may be present in many if not {\it all} large elliptical
galaxies. Radio galaxies have been found to contain some of the most massive and distant
black holes observed \citep{wmj03,alb99}.

Observing in the radio offers an unique opportunity that is absent
from any other waveband. Radio observations are not limited by obscuration
of dust such as in the UV and soft X-ray. Radio observations also do not
depend on the time of day and only the most extreme of weather
conditions can disrupt observations. Telescopes can be built near
cities, e.g. Jodrell Bank, and do not need to be built on high
mountain tops. Radio astronomy is often connected to the large
impressive arrays of dish antennas that span many kilometres such as
the Very Large Array (VLA) or the globe with Very Long Baseline
Interferometry (VLBI). However, radio astronomy can and is being done
on much smaller and simpler scales. LOFAR (low frequency array) for
example is currently being built and will use 10,000 wire antennas,
spread out across the Netherlands and Germany.

It is perhaps the high redshift Universe that is best suited to radio
observations and by studying how the environments of populations of
radio galaxies evolve we can begin to understand how the Universe
itself evolved and also understand how the properties of the host
galaxy are affected by changes in the source environment.

\section{Classifying radio galaxies}
\index{FRI / FRII classification} 
\citet{fr74} were the first to suggest that radio galaxies could be
effectively split into two distinct classifications, the so called
FRI's and FRII's. FRI's are bright near their centers and fade out
towards the edge whereas FRII's are brightest at their edges and
fainter towards the center.  Around $ P_{\rm 178 MHz}\sim 10^{25}$
Hz$^{-1}$ there is a break between the two classifications, all FRI's,
in general, lie below this break and all FRII's above. \citet{lo94}
found that the break in radio luminosity extends into the
optical and that FRIIs reside in the more optically-luminous galaxies.

The distinction between the two classifications becomes even more 
complicated when their morphologies are considered: FRI's generally 
have both jet and counter-jet structures visible and exhibit 
complicated non-uniform structure \citep[e.g. 3C 48][]{dragon}. FRII's 
on the other-hand have very rarely been observed to have both jet and 
counter-jet structures. FRII jets are well confined and their lobe 
structure can be seen to extend up to several Mpc (e.g. 3C 236 and 
3C 326). The luminosity break and the associated change in the source 
morphology has been suggested to be due to magnetohydrodynamical 
properties of the jet flow which determine the stability of the jet 
\citep{raw,ka97}. 
 
\begin{figure}[!h] 
\centerline{\psfig{file=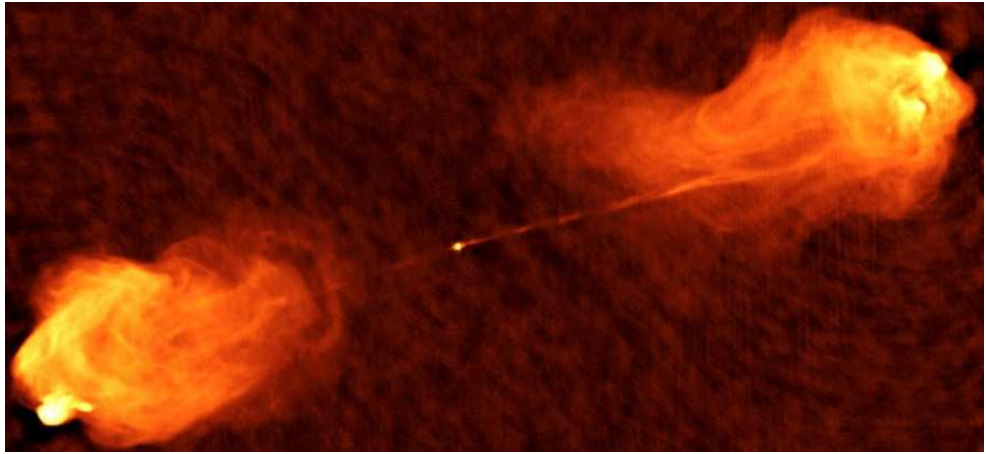,width=13cm}} 
\caption[Cygnus A] 
{High resolution image of Cygnus A by \citet{carilli}. This image was 
generated with data from telescopes of the National Radio Astronomy 
Observatory, a National Science Foundation Facility, managed by 
Associated Universities. Inc.}\label{CygA} 
\end{figure} 
\index{Cygnus A} 
The most famous example of a FRII is 3C 405, otherwise known as Cygnus 
A (see \citet{carilli} for a good overview on this source). Figure
\ref{CygA} shows a high resolution image of Cygnus A, a ``Classical
Double'' source.  The low redshift \citep[$z\sim 0.06$,][]{olm97} of
this object means that its radio structure was been well studied. At
the top-right and bottom-left of Figure \ref{CygA} there are two large
diffuse structures known as the radio lobes. Within each lobe is a
bright locale, known as the radio hotspot. Cygnus A is atypical in the
fact that it shows a strong core (near the center) with both jet and
counter-jet structures propagating outward. Both lobes show extensive
filamentary structure, see \citet{feb89} for a discussion on the
nature of these filaments. Cygnus A is often said to be over-luminous
\citep{carilli} compared to most sources at this redshift.
The relationship between redshift and radio-luminosity is discussed
later in this chapter.
In this thesis I concentrate on FRII objects only.

\section{Environment or Orientation?}
Many authors have used polarisation and flux measurements of powerful
radio sources to argue that any observed asymmetries in these
measurements are caused by orientation and/or environmental
differences. Observing in different wavebands gives an insight into
different regions of the source. In the optical, emission line
properties of a source can give an indication of the richness of the
source environment. X-ray observations sample the hot gas around a
source and can determine how this gas influences the radio source. In
the following sections I review the evidence for the interaction of
powerful radio sources with their environment and explain how
orientation effects also have a strong impact on what is observed at
radio and other wavebands. I will then outline my own approach for
determining the relationship between the observed source properties and
the source environment.


\subsection{The Laing-Garrington effect}\label{rad}
\index{Laing-Garrington effect}
\citet{glc88} and \citet{l88} defined a sample of FRII sources with
well defined jets. Both studies found that depolarisation, which is
defined as the ratio of percentage polarisation at two separate
frequencies, was almost always larger on the counter-jet side. This
was deemed to be a simple orientation effect. The emission from the
counter-jet, which is pointing away from an observer, has to travel
along a longer path through the magnetised plasma in the vicinity of
the source, causing it to appear more depolarised due to an
inhomogeneous Faraday rotation (see chapters 4 and 5). This is known as
the Laing-Garrington effect. \citet{gc91} proved that depolarisation
by gas in the lobes themselves was unlikely to cause the observed
asymmetry. This is now the common assumption: the depolarising medium
in radio galaxies is external to the lobes and any internal component
will not significantly contribute to the overall depolarisation.

\subsection{Liu-Pooley effect}
\index{Liu-Pooley effect} \citet{lp91b,lp91} found that the least
depolarised lobe in a source also has the flatter spectral
index. General arguments for the asymmetry of the spectral index in
the lobes can be attributed to Doppler beaming of the
hotspots. However, \citet{lp91b,lp91} found that the entire lobe shows
a flatter spectral index, and thus the observed spectral asymmetry
must be due to some intrinsic difference of the source
environments. Later studies by \citet{ics01} and \citet{dtb99} found
that the Liu-Pooley effect was preferentially stronger for smaller
sources and was independent of whether the sample comprised of radio
galaxies, quasars or a mixture of both, see chapter 4 for a more detailed discussion.

\subsection{Evidence for a non uniform medium around radio galaxies}
\index{Faraday screen} 

Many authors have tried to determine if the so called Faraday screen
around powerful radio sources is uniform or clumpy. The Faraday screen
can cause the intrinsic polarisation vector of a source to rotate. It
also causes a drop in the observed percentage polarisation if the
structure of the magnetic field, within the Faraday screen, varies on
scales smaller than the size of the beam of the telescope used in the
observations. Both effects are highly wavelength dependent. The angle
over which the polarisation vector rotates is directly related to the
line--of--sight strength of the magnetic field and the column density
of electrons surrounding a source. \citet{carilli} found that Cygnus A
showed a magnetic field structure that was uniform over $\sim$ 10 kpc.

\index{depolarisation}
\citet{prm89} found that the lobe length directly related
to the strength of the depolarisation. Almost all of their high
redshift, high radio-luminosity sources showed a preference for the
smaller lobe to be more depolarised. This was explained as an
environmental difference between the two lobes of an individual source
and \citet{prm89} concluded that the Faraday medium around their radio
galaxies was distinctly clumpy. If one lobe was embedded in a denser
environment, it would not be able to expand as freely as the other
lobe and the denser environment would cause a larger degree of
depolarisation. This trend with size has also been found in many other
studies
\citep[e.g.][]{strom,sj88,blr97}. \citet{strom} also concluded that
the depolarisation-size trend could be explained by the presence of a
large gaseous halo around a source. Small sources would be deeply embedded in
this Faraday screen and would be expected to show more depolarisation
than a source that was only partially embedded.  This is consistent
with observations that there is a distinct lack of linear polarisation
at long wavelengths ($\lambda>49$cm), which can be explained as a
result of a large halo of gas around the source, generally associated
with the host galaxy \citep{sj88}. This halo then creates a large
Faraday rotation, vastly reducing the degree of polarisation observed.

The extent to which the Faraday screen affects the observed radio
properties of a source was first investigated by \citet{burn}. More recent
and perhaps realistic interpretations have been suggested by
\citet{tribble} and
\citet{ev03}.

\subsection{X-ray}
\index{X-ray}
A radio emitting source is generally assumed to be surrounded by a
halo of hot, diffuse X-ray emitting material. \citet{gc91} found that
the X-ray emission from elliptical galaxies was comprised of a
symmetric halo, independent of whether the galaxy was in a group or a
cluster. By combining X-ray observations with radio observations it is
possible to show that in at least some sources, the diffuse halo of
material is associated with the Faraday medium causing significant
depolarisation \citep{gc91,carilli}. 


Observations of radio sources using ROSAT and CHANDRA have shown that
FRII sources have a strong effect on the material surrounding them. A
good example of this interaction is Cygnus A where the radio lobes are
seen to forcibly expand into the X-ray emitting material
\citep{carilli}. \citet{ywm01} also found this effect in their CHANDRA
observations of M87. X-ray observations can be used to directly
determine the gas density in the source environment. In a recent study
by \citet{hw00}, it was found that low redshift FRII's do not reside
in dense environments which is consistent with the results using
galaxy correlation methods.

\subsection{Galaxy correlation methods}
\index{galaxy correlation}
Galaxy-correlation methods are an indirect method of sampling the
environment around powerful radio galaxies. We expect that the density
of the gas in-between galaxies is higher in `rich' environments,
e.g. galaxy clusters, compared to `poorer' environments, e.g. isolated
galaxies.  The extent to which galaxies cluster around each other can
be analysed using cross-correlation methods
\citep{ls79,pp88,ymp89,yg87}. This method is independent of any other
observational method for sampling the source environment and only
depends on how accurate galaxies can be correlated with each other.

\citet{ls79} were among the first to try and determine the extent to
which radio sources in the 3CR and 4C samples (with $z<0.1$),
associated with the Zwicky galaxies \citep{zwicky}, are
clustered. \citet{ls79} found that extended powerful radio galaxies
were in richer regions of space than galaxies selected at
random. However, they also found that the sources with the strongest
emission line spectra were isolated and very massive. This indicated
that some sources could be massive enough to provide their own
atmospheres and did not need to be within clusters or groups of
galaxies. A later study by \citet{pp88} using a much larger sample of
galaxies out to $z<0.25$, did not find this tendency for strong
emission galaxies to be isolated, but did find that the most powerful
galaxies were in rich clusters.

The conflicting results lead to the question as to whether the
difference in the richness of the environment that surrounded a source
was due to the differences in the redshift or radio-luminosity of the
samples used. To try to overcome this degeneracy in redshift and
radio-luminosity \citet{ymp89} studied yet a larger volume of space
with $0.15 <z<0.82$. \citet{ymp89} found that it was probably the
differing radio-luminosity of the sources indicating differences in
the richness of the environment, but they still could not rule out the
possibility that there was some residual cosmic epoch effect.

\subsection{Alignment effect}\label{align}
\index{alignment effect}
Powerful radio sources are often found to be located in giant
elliptical galaxies and also to contain extended regions producing
strong emission lines
\citep[e.g][]{ibr02,blr001,blr002,tmr98,mbk91,rsl04,ibr03}. These
regions producing optical emission lines have often been found to
align closely to the radio emission. A striking case is the aligned
radio galaxy 3C 280 \citep{rsl04}, which displays a tight correlation
between its rest frame UV line emission and its radio emission. An
expansion of a radio source into the surrounding medium can trigger
star-formation \citep{slm96,dbv97}. 4C 41.17 at $z\sim3.8$ is perhaps
the strongest case for triggered star-formation. Optical measurements
of the stellar population show a striking correlation with the
position of the radio lobes. \citet{cmb87} also found that higher
redshift sources displayed a stronger alignment effect, with very
little alignment present in lower redshift sources \citep{ibr02}.
Interestingly \citet{ibr03} found that the strength of the alignment
effect was also dependent on the radio-luminosity of the sample, with
the strength of the correlation weakening in the fainter 6C sample
compared to the 3CR sample.

The observed alignment effect has also been found to depend on
size, \citep{blr96} and the wavelength at which the observations were
taken \citep{ibr03}. \citet{ibr03} found that the alignment effect was
non-existent in their infrared observations of the 6C and 3CR sources
at $z\sim1$. \citet{mbk91} found that there was more emission-line gas
present in the near side of a source compared to the far side which
argues for an orientation dependence of the emission-line gas.

\subsection{Spectral Ageing}\label{specage}
\index{spectral ageing} 

The spectral index of a source has been found, in general, to be
flatter in the vicinity of the hotspot and steeper nearer the core
regions \citep[e.g][]{al87}. This can be explained through energy
losses of the electrons due to synchrotron radiation. The electrons
towards the core are older than those closer to the hotspot and so
have undergone more spectral ageing. Spectral ageing thus offers a
unique way to determine the ages of radio sources. There are three
main assumptions in this method:
\begin{enumerate}
\item The main site of particle acceleration is the hotspot at the end of the jet, 
\item The magnetic field is constant through-out the source,
\item The distribution of electrons can be described as a power-law.
\end{enumerate} 
\citet{br01} argue against spectral ageing in radio sources,
suggesting that the observed trend could be caused by a gradient in
the magnetic field combined with a curved power-law
distribution. Spectral ageing also assumes that there is no mixing
between different populations of particles and that the radiative
lifetime of synchrotron electrons is longer than the spectral
age. \citet{br01} suggest that both assumptions may be unrealistic.

Although the spectral ageing method may be too simplistic it is
interesting to note that \citet{al87} find that the environmental
effects on a source only become important around the FRI/FRII
break. In other words, more luminous sources do not show spectral
signatures caused by environmental effects.

\subsection{Unified Model}\label{uni}
\begin{figure}[!h]
\centerline{\psfig{file=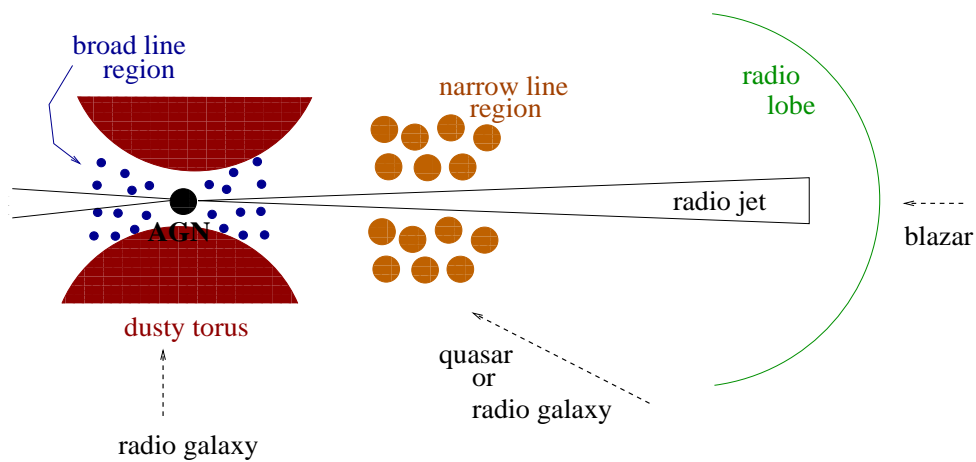,width=13cm}}
\caption[Unified Model]
{Simple cartoon picture of the unified model. The line-of-sight
orientation determines if a source is observed to be a radio galaxy,
quasar or blazar.}\label{or}
\end{figure}

\index{unified model} There are many sub-classes within the FRII
definition. Perhaps the most well known are the terms, radio galaxy,
quasar and blazar. Originally these sources were thought to be
separate and independent populations. It was proposed
\citep[e.g][]{uni} that the difference between these three populations
was simply the angle at which the source was viewed, see Figure
\ref{or}. At large angles to the line-of-sight an observer would
simply see a radio galaxy with narrow line emission properties. As the
angle decreases, the flux ratio of the lobes increases and the
emission becomes dominated by the lobe pointing towards the
observer. The narrow-line emission can no longer be seen and only
broad line regions are visible, which is known as a broad line
quasar. At very small angles to the line-of-sight the emission is
completely swamped by the highly polarised jet emission. The separate
FRII sub-classes can be simply stated a function of orientation angle
\citep{scheuer2,peacock87,uni}.

If the orientation angle was a factor in whether of not a source was
observed to be a quasar or a `normal' radio galaxy would expect all
quasars and blazars to show evidence of beaming. \citet{b89} did
indeed find that all the quasars in his sample showed some evidence of
beaming.


\subsection{Source or environment?}
Sections \ref{rad} to \ref{uni} indicate that it is very
hard to disentangle whether or not the gaseous environments of FRII
radio sources change systematically as a function of redshift and/or
radio luminosity. This is further complicated by orientation effects
\citep[e.g.][]{glc88}, Doppler boosting
\citep[e.g.][]{lp91a,lp91b}, size \citep{blr96} and wavelength
\citep{ibr03}.
\index{z-P degeneracy}

Studies are further complicated by the use of flux-limited samples and
the resulting degeneracy between redshift and radio-luminosity.

\section{Flux limited samples}
\begin{figure}[!h]
\centerline{\psfig{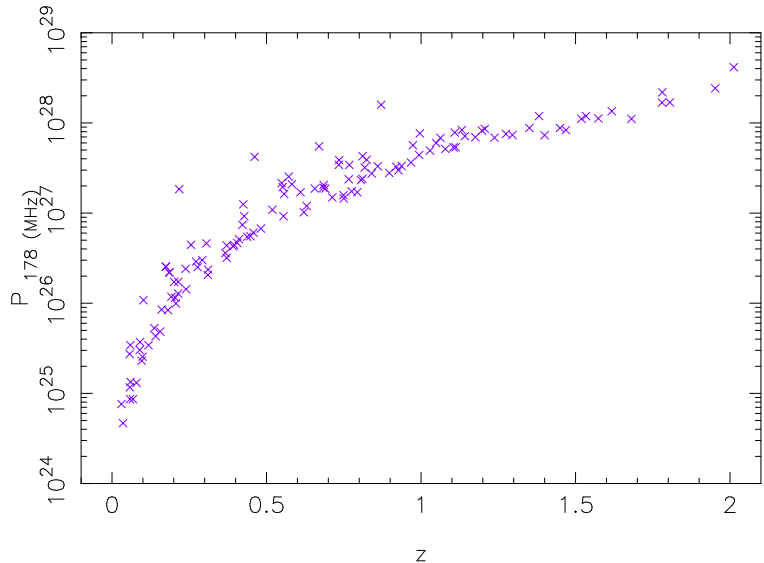}}
\caption[A radio luminosity-redshift plot for the 3CRR sample]
{A radio luminosity-redshift plot for the 3CRR sample.}\label{3CRR}
\end{figure}

\index{flux-limit}
A purely observational effect hampering the study of source
environments is the Malmquist bias inherent in any flux-limited
sample. Figure \ref{3CRR} demonstrates how severe this problem is in
the case of the 3CRR sample by \citet{lrl83}. All complete surveys in
the radio are flux-limited thus any analysis of the environments of
these sources with respect to cosmic epoch or radio-luminosity will be
greatly limited. To overcome the degeneracy between redshift and
radio-luminosity it is possible to use sources taken from several
samples, generally with different flux-limits, e.g. combining the
3CRR, 6CE and 7CRS samples
\citep{brw99}. 
\index{3CRR}

\index{z-P degeneracy}
Figure \ref{3CRR} shows that there is a distinct lack of high
radio-luminosity sources at lower redshifts. In fact, all low redshift
sources (albeit a few over-luminous sources like Cygnus A) are at much
lower luminosities than their high redshift cousins. As we observe at
higher redshifts we are observing larger volumes of space. At low
redshift the chance of finding a relatively powerful source is small
because the radio luminosity function is steep. As the sampled volume
of space increases, the probability of finding a rare high-luminosity
source naturally increases. One of the aims of radio astronomy is to
`fill in' the z-P plane and remove the strong bias that is present in
all surveys, even when using several complete samples, at different
flux limits. \citet{hr03} gives a concise review of the flux-limit
problem.

\section{Modelling FRII's and their environment}
\index{FRII modelling}
In the previous sections I have reviewed observational techniques to
test the environments of powerful radio sources. However, we can also
attempt to model these sources and try to gain insights into their
environments from a more theoretical point of view.

\begin{figure}[!h]
\centerline{\psfig{file=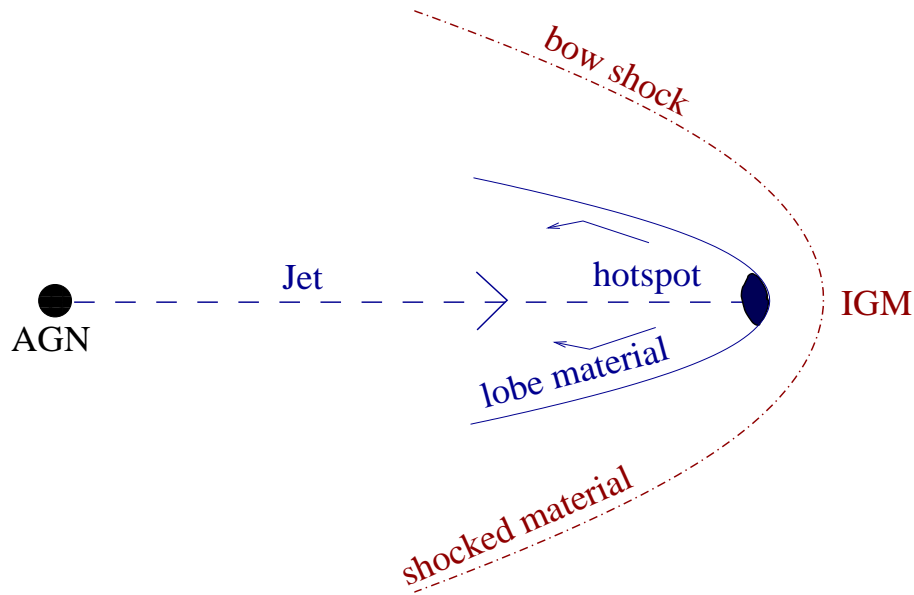,width=12cm}}
\caption[A simple impression of a FRII source ]
{A simple impression of a FRII source.}\label{pic}
\end{figure}

\citet{scheuer} and \citet{br74}  were the first to explain the 
double-lobed structure. In the simplest model, energy is transported
from the nucleus of the source to the radio lobes by a relativistic
beam, more commonly known as the radio-jet. \citet{scheuer} predicted
that only a small fraction of the energy in the jet would be radiated
away from the tip of the beam, known as the hotspot. The jet
terminates at the hotspot and creates a jet shock. The lobe is
generally assumed to be over-pressured and expands supersonically into
the external medium, creating a bow shock \citep{a02}. A simple picture of
the dynamics of a radio source is presented in Figure \ref{pic}. 

FRII's are generally assumed to expand with time along the jet axis
and the pressure exerted by the jet is balanced by the ram pressure of
the external gas \citep[e.g][]{scheuer,ka97,kat97,hw00}. The jet must
have enough ram pressure to overcome the external gas pressure,
otherwise the source will not have a well defined hotspot
\citep{pp88}. The rate of expansion of the lobe is highly dependent on
the strength of the jet and the density of the source environment
\citep{ka97}. If a source was strongly constrained perpendicular to
the jet axis, then there would be a tendency for sources to be long
and thin which has not been observed \citep{mlf85}.  \index{density of
source environment} \index{lobe pressure} \index{jet power}
\index{source age}

In this thesis I apply the model of \citet{ka97} and \citet*{kat97} to
the radio observations of FRII radio galaxies and quasars. Using this
model I obtain estimates for the gas density in the environment of
these sources. I then compare these model-dependent estimates with my
observational results on the depolarisation of the radio emission.

\section{This work}
It is often difficult to obtain information on the properties of the
black holes at the centres of active galaxies, their jets and their
large scale environments from direct observations. As I have
demonstrated, many studies to try and disentangle the
source/environment interaction suffer from the redshift-radio
luminosity degeneracy present in all flux-limited samples.
\index{z-P degeneracy}

To break the degeneracy between redshift and radio luminosity I chose
3 subsamples of sources from the 3CRR and 6C/7C catalogues. The
control sample consists of 6C and 7C sources at $z \sim 1$
with radio luminosities of around $10^{27}$ WHz$^{-1}$ at 151
MHz. Another sample at the same redshift consists of 3CRR sources with
radio luminosities around a magnitude higher at 151 MHz. The final
sample consists of 3CRR sources at $z \sim 0.4$, again
with radio luminosities of around $10^{27}$ WHz$^{-1}$ at 151 MHz.
The observations of these sources can then be used to study the
source properties and the medium around the source, thus discovering
which correlate with redshift and which correlate with radio
luminosity enabling the following questions to be answered:

\begin{enumerate}
\item Does a relationship exist between radio luminosity and the 
environment in which a given radio source lives?

\item Do the source environments evolve with redshift?

\end{enumerate}

To arrive at answers to these questions I investigate in detail the
depolarisation properties of my sample. I also apply the models of
source evolution mentioned above to my observational results. Finally,
I apply the later technique to a much larger sample.

As previously stated depolarisation and Faraday rotation are
wavelength dependent. At low frequencies the degree of polarisation
would be too low for an in-depth analysis of the environment and at
high frequencies some of my sources would have too low
signal-to-noise. Thus an intermediate frequency range of 1.4 GHz to
4.8 GHz was chosen.

The body of this work is dedicated to answering the question as to
whether it is the redshift or the radio-luminosity of powerful radio
sources that is the dominant factor in determining their observed
characteristics.

\section{Synopsis}
\index{synopsis}
In {\it chapter 2} I give a detailed description of the sample
selection and subsequent methods used in the data reduction. I also
present individual notes for each source used in the sample. All
sources which have pre-published maps between 1.4 GHz and 4.8 GHz are
detailed in Table \ref{maps}.

\noindent In {\it chapter 3} I present the polarisation and flux
properties for each source present in my sample and analyse the bulk
trends present between my sub-samples. I also present the results of
correcting the depolarisation measurements to a common redshift, $z=1$,
using the \citet{burn} law.

\noindent In {\it chapter 4} I present a more detailed analysis of the results
presented in chapter 3. I use the Spearman Rank, Partial Spearman Rank
and Principal Component Analysis tests to determine how the rotation
measure and polarisation parameters relate to the fundamental
parameters of redshift, radio-luminosity and radio size. I also
analyse the asymmetries present in the flux and polarisation results
to determine how the environments of these 26 sources differ and if
these differences can be related to the fundamental parameters.

\noindent In {\it chapter 5} I present the results of using the
\citet{tribble} models to correct the depolarisation measurements to a
common redshift, $z=1$, and the results are compared with those from
chapter 3 using the simpler \citet{burn} prescription. I then use the
\citet{tribble} models to determine if there is any evolution in the
Faraday screen with respect to redshift or radio-luminosity.

\noindent In {\it chapter 6} I use the \citet{ka97} and \citet{kat97}
models to estimate the density of the source environment and lobe
pressure for each source described in chapter 2. I compare the density
of the source environment and the lobe pressure, from the model, with
the observed depolarisation and variations in the rotation measure to
determine if the polarisation parameters are affected by the source
environment. I then apply the models to the FRII sources from the
combined 3CRR, 6CE and 7C III samples to estimate the density of the
source environment, lobe pressure, jet-power and lobe age for each
source. This allows a global approach to determine how the
environments of powerful radio sources evolve with cosmological epoch
and also with radio-luminosity.

\noindent {\it Chapter 7} details the conclusions drawn from my
findings and gives possible avenues for an extension to this study.

\noindent In appendices \ref{Amap} to \ref{Cmap} I present the radio
maps of all 26 sources from the small sample. Each source has an
associated polarisation intensity and polarisation angle map at 4.8
GHz and 1.4 GHz, a spectral index and depolarisation map between 1.4
GHz and 4.8 GHz and a rotation measure and magnetic field angle map
between 1.4 GHz and 4.8 GHz. 3C 457 is the exception as this source
does not have any rotation measure or magnetic field maps associated
with it. All chapters use a cosmology which assumes $H_o= 75 $
kms$^{-1}$Mpc$^{-1}$, and $\Omega_m = 0.35$ ($\Lambda = 0.65$). For
full details of the cosmology used see appendix \ref{Cosm}. Finally,
appendices \ref{ap:3CRR} to \ref{ap:7C} detail the archive information
used in the modelling of the 3CRR, 6CE and 7C III samples.
 
\newpage
\thispagestyle{empty}
\chapter[Observations]{Observations and Data Analysis}

Observations of the polarisation properties of extragalactic radio
sources can provide information on the relationships between the radio
source properties and their environments as well as the evolution of
both with redshift. Many previous studies of variations in
polarisation properties have suffered from a degeneracy between radio
luminosity and redshift due to a Malmquist bias, present in all flux-limited
samples. A good example of this effect is the depolarisation
correlations (i.e. the ratio of percentage polarised flux at two
frequencies) found independently by \citet{kcg72} and \citet{mt73}.

\index{depolarisation} \index{z-P degeneracy} \index{magnetic field}
\index{Faraday screen} 

\citet{kcg72} found that depolarisation of the radio lobes generally
increased with redshift whereas \citet{mt73} found depolarisation to
increase with radio luminosity. Due to the flux-limited samples (PKS
and 3C) used by both authors it is difficult to distinguish which is
the fundamental correlation, or whether some combination of the two
occurs. Both suggestions have ready explanations: (i) If radio sources
are confined by a dense medium then synchrotron losses due to
adiabatic expansion are reduced, the internal magnetic field is
stronger and a more luminous radio source results; if this confining
medium also acts as a Faraday medium, more luminous sources will tend
to be more depolarised. (ii) Sources at different cosmological epochs
may reside in different environments and/or their intrinsic properties
may change with redshift.

\citet{hl91} observed that galaxy densities around FRII radio sources
increased with redshift out to $z \sim 0.5$ and beyond, but
\citet{wls01} did not find this trend in a recent study.
\citet{wpk84} argued that the increase in rotation measure with
redshift is primarily attributable to an increasing contribution of
intervening matter. However, depolarisation asymmetries within a
source, e.g the Laing-Garrington effect, increase with redshift which
imply an origin local to the host galaxy \citep{gc91a,l80,lp80}.


\section{Observational program}
\index{observational program}
\subsection{Sample selection}
 \begin{figure}[!h]
\centerline{\psfig{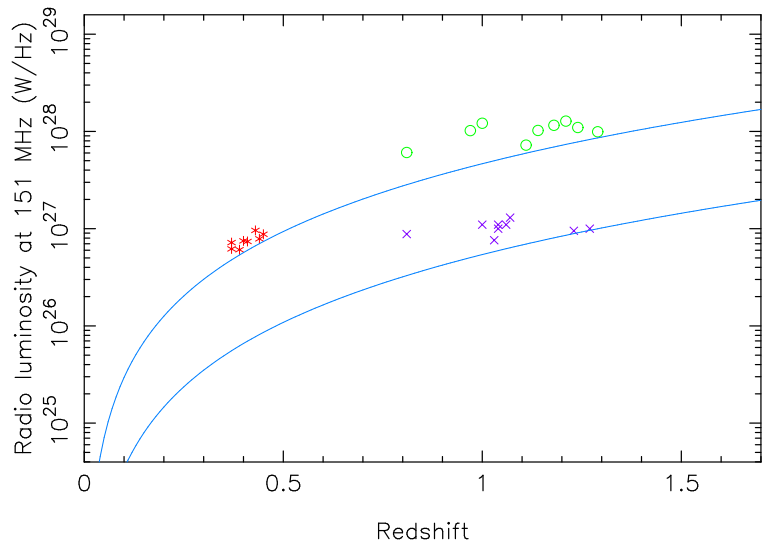}}
\caption[A radio luminosity-redshift plot for the 3 subsamples.]  {A
radio luminosity-redshift plot showing the 3 subsamples used in the
observations. Sample A is represented by `$\times$', sample B by
`$\circ$' and sample C by `*'. The lines mark the flux limits for the
3CRR and 6C samples. A spectral index of -0.75 was used to shift the
3CRR data, originally observed at 178 MHz to 151 MHz.}\label{pzplot}
\end{figure}
\begin{figure}[!h]
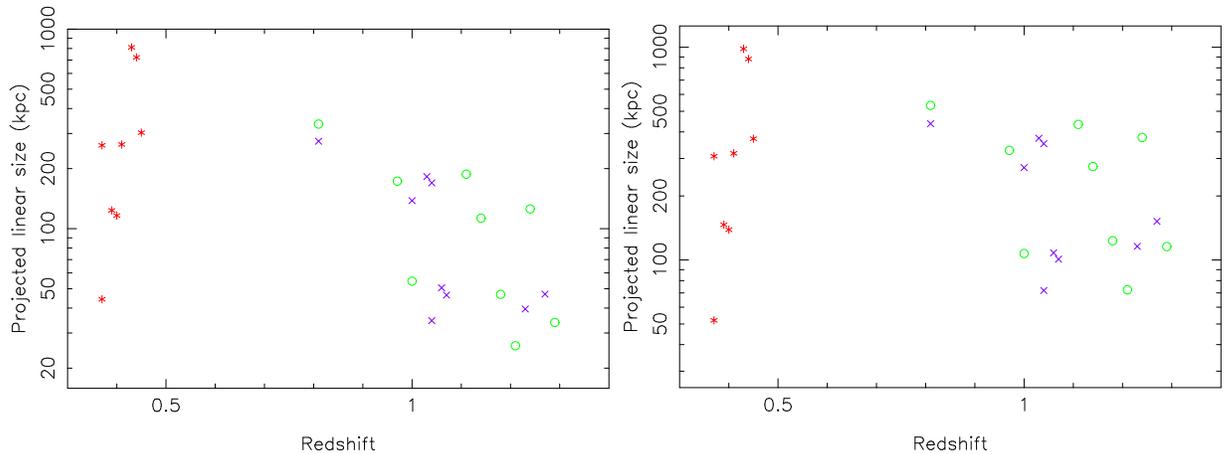

\centerline{\psfig{file=dis3_c.ps,width=8cm,angle=270}
\psfig{file=dis4_c.ps,width=8cm,angle=270}}
\caption[Linear size-redshift plots of the 3 subsamples.]
{Linear size-redshift plots of the 3 subsamples used in the
observations. Symbols as in Figure \ref{pzplot}. Figure (a) assumes
$H_o= 75$ kms$^{-1}$Mpc$^{-1}$, and $\Omega_{m} = 0.5$,
$\Omega_{\Lambda} = 0$. Figure (b) assumes $H_o= 75$
kms$^{-1}$Mpc$^{-1}$, and $\Omega_{m} = 0.35$, $\Omega_{\Lambda} =
0.65$.}\label{sizeplot}
\end{figure}

\index{3CRR}\index{6CE}\index{7C III}
Sample A was defined as a subsample chosen from the 6CE \citep{erl97}
subregion of the 6C survey \citep{hmw90}, and the 7C III subsample
\citep{lrh99}, drawn from the 7C and 8C surveys \citep{pwr98}. The
selected sources have redshifts $0.8<z<1.3$, and radio luminosities at 151
MHz between $6.5 \times 10^{26}\, {\rm WHz^{-1}}<P_{151\, {\rm MHz}}<1.35
\times 10^{27}\, {\rm WHz^{-1}}$. Sample B was defined as a subsample
from the revised 3CR survey by \citet{lrl83} containing sources within
the same redshift range, but with luminosities in the range $6.5 \times
10^{27}\, {\rm WHz^{-1}}<P_{151\, {\rm MHz}}<1.35 \times 10^{28}\, {\rm
WHz^{-1}}$. Sample C is also from the 3CRR catalogue; it has the same
radio luminosity distribution as the control sample, sample A, but with
$0.3<z<0.5$.  I only include sources that are more luminous than the
flux limits of the original samples at 151 MHz (Figure
\ref{pzplot}). In all samples only sources with angular sizes $
\theta\ge 10''$ (corresponding to $\sim$ 90 kpc at z=1) were
included, see Table \ref{setup}. This angular size limit is imposed by
the depolarisation measurements. I required a minimum of ten
independent telescope beams (1'' per beam) over the entire source to
ensure sufficient signal--to--noise. The distributions of linear sizes
of the radio lobes are reasonably matched across all the samples
(Figure
\ref{sizeplot}), but note the two `giant' sources in sample B.

Each sample initially contained 9 sources. The source 3C 109 was
subsequently excluded from sample C as the VLA data for that source
was of much poorer quality than that for the rest of the sample.  The
sources in the 3 subsamples are representative of sources with similar
redshifts, radio-luminosities and sizes. However, the samples are not
statistically complete because of observing time limitations. Sources
that were fairly well observed at 4.8 GHz at B array (and C array if
needed) were selected from the archives, this ensured that only a
minimum of new observations were needed.

\begin{table}[!h]
\centering
\begin{tabular}{ r@{ }l ccccc }\hline
\multicolumn{2}{c}{Source} & z&P$_{\rm 151\, MHz}$   & Angular size 
& $\sigma_{\rm 4.8\, GHz}$ & $\sigma_{\rm 1.4\, GHz}$\\
\multicolumn{2}{c}{ }& &(W/Hz)  &  (arcsec)   & ($\mu$ Jy) & ($\mu$ Jy)\vspace{1mm}\\\hline
6C& 0943+39&1.04&$1.0\times10^{27}$&10&20&55\\
6C& 1011+36&1.04&$1.1\times10^{27}$&49&15&65\\
6C& 1018+37&0.81&$8.8\times10^{26}$&64&24&50\\
6C& 1129+37&1.06&$1.1\times10^{27}$&15&21&60\\
6C& 1256+36&1.07&$1.3\times10^{27}$&14&18&70\\
6C& 1257+36&1.00&$1.1\times10^{27}$&38&33&100\\
{\it 7C}& {\it 1745+642}&1.23&$9.5\times10^{26}$&16&30&60\\
{\it 7C}& {\it 1801+690}&1.27&$1.0\times10^{27}$&21&24&75\\ \vspace{3mm}
{\it 7C}& {\it 1813+684}&1.03&$7.1\times10^{26}$&52&23&47\\
3C& 65&1.18&$1.0\times10^{28}$&17&22&61\\
{\it 3C}&{\it 68.1}&1.24&$1.1\times10^{28}$&52&41&100\\
3C& 252&1.11&$7.2\times10^{27}$&60&23&100\\
3C& 265&0.81&$6.5\times10^{27}$&78&32&70\\
3C& 268.1&0.97&$1.0\times10^{28}$&46&32&68\\
3C& 267&1.14&$1.0\times10^{28}$&38&25&61\\
3C& 280&1.00&$1.2\times10^{28}$&15&33&75\\
3C& 324&1.21&$1.3\times10^{28}$&10&26&70\\\vspace{3mm}
{\it 4C}& {\it 16.49}&1.29&$9.9\times10^{27}$&16&25&81\\
3C& 16&0.41&$7.4\times10^{26}$&63 & 36& 80\\
3C& 42&0.40&$7.5\times10^{26}$&28&29 &70\\
3C& 46&0.44&$7.9\times10^{27}$&168&20&53\\
3C& 299&0.37&$6.9\times10^{26}$&11&39&77\\
3C& 341&0.45&$8.8\times10^{26}$&70&26&88\\
{\it 3C}&{\it 351}&0.37&$7.6\times10^{26}$&65&31&100\\
3C& 457&0.43&$9.6\times10^{27}$&190&18&80\\
 4C& 14.27&0.39&$8.8\times10^{26}$&30&30&90\\ \hline
\end{tabular}
\caption[Source specifics]
{Details of the sources in sample A, B \& C. Sources in italics are
quasars. $\sigma _{\nu}$ gives the 1$\sigma$ noise level in the final
total flux maps at frequency $\nu$.} \label{setup}
\end{table}

The ratio of angular to physical size varies only by a factor 1.3
between $z=0.4$ and $z=1.4$. This ensured that all the sources were
observed at similar physical resolutions. All redshift values for 6C
sources were taken from \citet{erl97} except for 6C 1018+37 which was
taken from \citet{rel01}. Redshifts for the 7C sources were taken from
\citet{lrh99}. Redshifts for the 3CRR sources were taken from \citet{sdm97},
4C 16.49 was taken from \citet{bh01}, 4C 14.27 was taken from
\citet{hr92} and 3C 457 was taken from \citet{hb91}.

\subsection{Very Large Array observations}
\label{obs}
Observations of all 26 radio galaxies were made close to 1.4 GHz using
the A-array configuration and a 25--MHz bandwidth. This bandwidth was
used instead of 50 MHz to reduce the effect of bandwidth
depolarisation which occurs when there are large changes in the
rotation measure on the same scale as the beam size. The maximum
angular size that can be successfully imaged using A array at 1.4 GHz
is 38$''$. 12 sources were larger than this and were observed
additionally with B array. Observations were also made at 4.8 GHz
using a 50 MHz bandwidth. The maximum observable angular size in B
array at 4.8 GHz is 36$''$. The same 12 sources as before, were then
observed at 4.8 GHz with C array. This ensured that both 1.4 GHz and 4.8
GHz observations were equally matched in sensitivity and resolution.
Details of the observations are given in Tables \ref{vlaobs} to
\ref{vlaobsc}.

Sources in the 6C and 7C samples have a typical bridge surface
brightness of $\sim 70 \mu$ Jy beam$^{-1}$ in 4.8 GHz A-array
observations \citep{bel99}. In order to detect 10\% polarisation at
$3\sigma$ in B-array observations I required an rms noise level of $20
\mu$ Jy beam$^{-1}$, corresponding to 70 mins of integration time.  At
1.4 GHz, assuming $\alpha =-1.3$, bridges will be a factor of 4 more
luminous. At this frequency, the integration time is set by the
requirement to have an adequate amount of {\it uv\/}-coverage to map
the bridge structures. 20 minute observations were split into
4$\times$5 minute intervals. This observation splitting to improve
{\it uv\/}-coverage was also done for the 4.8 GHz data.

As Table \ref{vlaobs} demonstrates, for many of my sources the
integration times at 4.8 GHz are considerably less than the 70 min
requirement, due to telescope time constraints. At 1.4 GHz the
integration time on all the sources is above the minimum required for
good signal--to--noise. This is not the case at 4.8 GHz. Many of the
observed properties that depend on polarisation observations (e.g
depolarisation and rotation measure), are therefore poorly measured in
the fainter components at 4.8 GHz.  The values obtained are then only
representative of the small region detected and not the entire
component. Spectral index is independent of the polarisation
measurements and so it is relatively unaffected by the short
integration times.

\index{uv-coverage}
The 3CRR sources are more luminous but much of this is due to the
increase in the luminosity of their hotspots; their bridge structures
are only a few times brighter than those of the 6C/7C III sources. To
reach a $3\sigma$ detection of 7\% polarisation on the bridge
structures, a total integration time of 30 mins was required at 4.8 GHz
and 20 mins at 1.4 GHz, split into 3-4 minute intervals to improve
{\it uv\/}-coverage. The vast majority of the sources in sample B and
C had at least this minimum amount of time on source at each frequency
(see Tables
\ref{vlaobsb} and \ref{vlaobsc}).

\subsection{Observational Programs}
\index{integration time}
\begin{table}
\centering 
\begin{tabular}{ l ccccl }\hline
Source & Array  & Frequency & Bandwidth& Observing program& Int. \\
& configuration & (MHz)&(MHz)& (dd/mm/yy)  & (min)\\\hline 
6C 0943+39 & A &1465,1665&25  & 31/07/99  (AD429) & 16 \\
&  B & 4885,4535&50& 20/05/01 (AD444)&31\\
6C 1011+36 & A &1465,1665&25 & 31/07/99 (AD429) & 16\\
&  B & 1452,1652&25&  20/05/01 (AD444)& 17\\
&  B & 4885,4535&50 & 25/02/97 (AL397)&21\\
&  &4885,4535&50 & 20/05/01 (AD444)&30 \\
&  C & 4885,4535&50 & 12/06/00 (AD429)& 20 \\
6C 1018+37  & A & 1465,1665 & 25&31/07/99 (AD429)& 16\\
&  B & 1452,1652&25 & 20/05/01 (AD444)&17\\
&  B & 4885,4535&50 & 20/05/01 (AD444)&31\\
&  C & 4885,4535&50 & 12/06/00(AD429)& 20 \\
6C 1129+37  & A & 1465,1665&25  & 31/07/99 (AD429)& 16\\
&  B & 4885,4535&50  & 20/05/01 (AD444)& 17\\
&  & 4885,4535&50 &  25/02/97 (AL397)& 21\\
6C 1256+36  & A &1465,1665&25  & 31/07/99 (AD429)& 16\\
&  B & 4885,4535&50 & 27/02/93 (AR287)&15\\
6C 1257+36 & A &1465,1665&25  & 31/07/99 (AD429)& 16\\
&  B & 4885,4535&50  &20/05/01 (AD444)& 16\\
&  & 4885,4535&50 & 25/02/97 (AL397)& 22\\
7C 1745+642  & A &1465,1665&25  & 31/07/99 (AD429)& 16\\
&  B & 4885,4535&50  &20/05/01 (AD444)&11 \\
&  & 4885,4535&50 &  23/11/97 (AL401)& 31\\
7C 1801+690  & A &1465,1665&25  & 31/07/99 (AD429)&16 \\
&  B & 4885,4535&50  &26/03/96 (AB978)& 29\\
&  & 4885,4535&50 &  23/11/97 (AL401)& 17\\
7C 1813+684  & A & 1465,1665&25  & 31/07/99 (AD429)& 16\\
&  B & 1452,1652&25  & 20/05/01 (AD444)&16 \\

&  B & 4885,4535&50  & 20/05/01 (AD444)&39\\
&  & 4885,4535&50 & 23/11/97 (AL401)& 19\\
&  C &4885,4535&50  & 12/06/00 (AD429)& 20\\\hline
\end{tabular}
\caption[Details of the VLA observations for sample A]
{Details of the VLA observations for sample A with the integration
times included. See Tables \ref{vlaobsb} and \ref{vlaobsc} for samples
B and C respectively.} \label{vlaobs}
\end{table}
\begin{table}
\centering 
\begin{tabular}{ l ccccl }\hline
Source &Array  & Frequency &Bandwidth& Observing program&Int. \\
&  configuration & (MHz)&(MHz)& (dd/mm/yy)& (min) \\\hline 
3C 65& A &1465,1665&25 & 31/07/99 (AD429)& 16\\
&  B &4885,4535&50  & 20/05/01 (AD444) &20\\
3C 68.1  & A & 1417,1652&25  & 31/07/99 (AD429)& 16 \\
&  B &1417,1652&25  & 13/07/86 (AL113) &20\\
&  B &4885,4535&50  & 19/07/86 (AB369)& 300\\
&  C &4885,4535&50  & 12/06/00 (AD429) &20\\
3C 252  & A & 1465,1665&25 & 31/07/99 (AD429)& 16 \\
&  B &1465,1665&25  & 20/05/01 (AD444) & 27\\
&  B &4885,4535&50  & 19/07/86 (AB369)&97\\
&  C & 4885,4535&50  & 12/06/00 (AD429) &20\\
3C 265  & A & 1417,1652&25  & 31/07/99 (AD429)&16 \\
&  B & 1417,1652&25  & 13/07/86 (AL113) &30\\
&  B & 4873,4823&50  & 17/12/83 (AM224)& 238\\
&  C& 4873,4823&50 & 12/06/00 (AD429) &20\\
3C 267  & A & 1465,1665&25  & 31/07/99 (AD429)&16 \\
&  B & 4873,4823&50 & 17/12/83 (AM224)&56 \\
3C 268.1  & A & 1417,1652&25  & 31/07/99 (AD429)&16\\
 &  B & 1417,1652&25  & 13/07/86 (AL113) & 30\\
&  B & 4885,4835&50  & 15/08/88 (AR166)& 20\\
&  & 4885,4835&50 & 01/06/85 (AR123) &21\\
&  C & 4885,4835&50  & 06/11/86 (AL124)&102\\
3C 280  & A & 1465,1665&25 & 31/07/99 (AD429)&16 \\
&  B & 4873,4823&50 & 17/12/83 (AM224) &46\\
3C 324  & A & 1465,1665&25 & 31/07/99 (AD429)&16 \\
&  B & 4873,4823&50 & 17/12/83 (AM224)&51 \\
4C 16.49  & A &1465,1652&25 & 31/07/99 (AD429)&16 \\
& B & 4885,4535&50  & 04/03/97 (AB796)&30\\\hline
\end{tabular}
\caption{Details of the VLA observations for sample B} \label{vlaobsb}
\end{table}

\begin{table}
\centering
\begin{tabular}{ l ccccl }\hline
Source & Array  & Frequency & Bandwidth &Observing prog. & Int.\\
& configuration & (MHz)&(MHz)& (dd/mm/yy) & (min)\\\hline
3C 16  & A & 1452,1502&25 & 14/09/87 (AL146) & 59\\
&  B & 1452,1502&25 & 25/11/87 (AL146) & 39\\
&  B & 4885,4535&50  & 20/05/01 (AD444)&20\\
&  & 4885,4535&50 & 17/11/87 (AH271)&10 \\
&  C & 4885,4535&50  & 12/06/00 (AD429) &20\\
3C 42  & A & 1452,1502&25 & 14/09/87 (AL146)&40\\
&  B & 4885,4535&50  & 23/12/91 (AF213) &67\\
3C 46  & A & 1452,1502&25 & 31/07/99 (AD429)&16 \\
&  B & 1452,1502&25 & 25/11/87 (AL146) &35\\
&  B & 4885,4535&50  & 20/05/01 (AD444) &20\\
&  C & 4885,4535&50  & 12/06/00 (AD429) &20\\
3C 299  & A & 1452,1502&25 & 31/07/99 (AD429)&16 \\
&  B & 4835,4535&50  & 20/05/01 (AD444)&15\\
&  & 4885,4835&50 & 28/01/98 (AP331)&15 \\
3C 341  & A & 1452,1502&25 & 14/09/87 (AL146)&38 \\
&  B & 1452,1502&25 & 25/11/87 (AL146) &47\\
&  B & 4885,4535&50  & 20/05/01 (AD444)&11\\
&  & 4935,4535&50 & 26/10/92 (AA133) &25\\
&  C & 4885,4535&50  & 12/06/00 (AD429)&20 \\
3C 351 & A & 1452,1502&25 & 31/07/99 (AD429) &16\\
&  B & 1452,1502&25 & 25/11/87 (AL146)&56 \\
&  B & 4885,4835&50  & 19/07/86 (AB369)&12\\
&  & 4885,4535&50 & 20/5/01 (AD444)&16\\
&  C & 4885,4835&50  & 09/10/87 (AA64)&22 \\
3C 457 & A & 1452,1502&25 & 31/07/99 (AD429)&16 \\
&  B & 1452,1502&25& 25/11/87 (AL146)&30 \\
&  B &4885,4535&50  & 20/05/01 (AD444) &50\\
&  C & 4885,4535&50  & 12/06/00 (AD429)&20\\
4C 14.27  & A & 1452,1502&25 & 14/09/87 (AL146) &28\\
&  B & 4885,4535&50  & 20/05/01 (AD444)&17\\ \hline
\end{tabular}
\caption{Details of the VLA observations for sample C} \label{vlaobsc}
\end{table}

\index{AD429}
\paragraph{AD429}\label{AD429}
Most observations at 1.4 GHz in A array were obtained on 31/07/99 with
the observing program AD429. The data from this day is strongly
affected by a thunderstorm at the telescope site during most of the
observations. Even after removal of bad baselines and antennas the
noise level in this data remained at least twice the theoretical
value. However, careful calibration and {\small CLEAN}ing reduced this
effect to a minimum. Sample A was most affected by the thunderstorm
and the lack of observing time at all frequencies. However, I find
that the results obtained by \citet{bel99} for some of the sources in
sample A are in good agreement with my results. I am confident that my
data is reliable for fluxes above the 3$\sigma _{\rm noise}$ level.
The polarisation calibration of AD429 was compared to the B--array
data at 1.4 GHz (for the 12 sources that had B--array data). This
confirmed that the position angle (PA) of the polarisation vector in
both data sets agreed to within 15 degrees in all sources. This
additional check was used to ensure that the 1.4 GHz polarisation
angle calibration was accurate.  AD429 also contains observations of
sources in C array at 4.8 GHz. As these observations were taken on
12/06/00, these observations are obviously not affected by the
thunderstorm that affects the 1.4 GHz data.

\index{AD444}
\paragraph{AD444}
Tables \ref{vlaobs} to \ref{vlaobsc} show that AD444 was used to
observe some of the sources in B array at 4.8 GHz and also to
complement the AD429 1.4 GHz observations in B array. The observations
were all taken on the 20/05/01 and were not affected by any serious
problems.

\index{archival data}
\paragraph{Archival Data} 
For most sources some archival data has been used to supplement AD429
and AD444. Tables \ref{vlaobs} to \ref{vlaobsc} detail which archival
programs were used and the dates on which the sources were observed.
\section{Data Reduction}
\index{data reduction}
The data reduction of the VLA data was done using the National Radio
Astronomy Observatory's {\bf A}stronomical {\bf I}maging {\bf
P}rocessing {\bf S}ystem or {\small AIPS}. All data from the archives
was re-reduced to maintain consistency of data reduction methods when
analysing the results.
\subsection{Flux calibration.}
\index{flux calibration}
The first step in the data reduction process involves removing the
first 10s (usually) of each observation in the data set. This process
is used to remove any residual errors from antenna tracking problems
after moving to the target source. Once these initial scans are
removed then the data is `flagged' to remove any further errors in the
data. Data is usually flagged for many reasons, such as interference
from terrestrial sources (e.g mobile phones), errors with the antenna
and receivers and bad weather. If a thunderstorm is present during an
observing run, then this will cause large errors in the data due to
the highly ionised clouds. This was unfortunately the case for most of
the data set AD429 at 1.4 GHz and results in a higher noise level for
the sources, see section \ref{AD429}. All scans flagged are
entered into a `FG' table within {\small AIPS} and are no longer used
in the rest of the data reduction process.\\ 

\begin{figure}[!h]
\centerline{\psfig{file=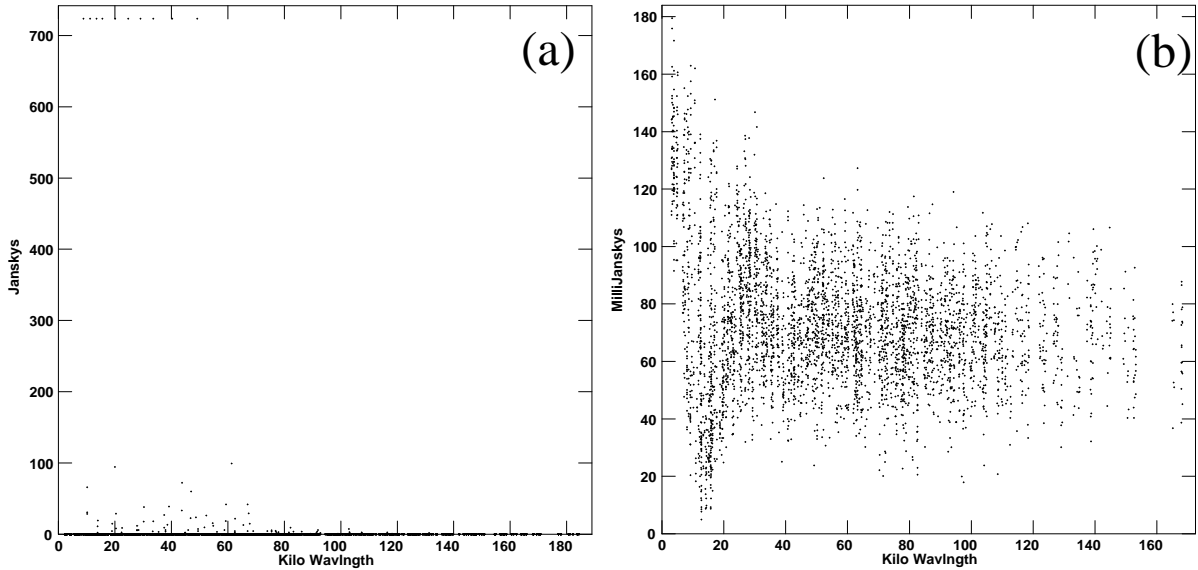,width=16.0cm}}
\caption[7C 1745+642: an example of flagging] {(a) A plot of the flux
against {\it u-v} for the source 7C 1745+642.  The radio flux of this
source lies in the mJy range indicating that the burst exceeding 700
Jy is from a lightning bolt near the array. The lightning bolt
affected 20 seconds of the data. (b) Removal of the affected 20
seconds results in a realistic UV spectrum. Only every 10th visibility
is plotted.}\label{thunder}
\end{figure}

\index{AD429} A good example of where flagging is essential can be
seen in the source 7C 1745+642 as observed as part of AD429. Figure
\ref{thunder}(a) shows the effect of the thunderstorm on 7C
1745+642. The usual flux of this source is $\sim 80$ mJy, but in
several timesteps the flux jumps to more than 700 Jy, completely
swamping the weak source underneath. By careful flagging of the bad
timesteps I was able to retrieve the flux from the source which can be
seen in Figure \ref{thunder}(b). Although this is an extreme case, it
does demonstrate the need for flagging. After all the corrupted data
has been removed the observations are ready to be calibrated.

\subsubsection{Calibrating the primary calibrator.}
\index{primary calibration}
The {\small AIPS} task {\small SETJY} uses a primary flux calibrator
such as 3C 286 or 3C 48 to determine the absolute flux density scale
on which all other calibrations are based. 3C 286 and 3C 48 are
observed every few years with the VLA in D--configuration. Their flux
density are recalculated accordingly using the modified Baars scale
\citep{baars},\index{Baars scale}
\begin{equation}\label{baarseq}
\centering
\log{S} = A +B\log{\nu} + C(\log{\nu})^2 + D(\log{\nu})^3
\end{equation}
where S is in Jansky, $\nu$ is in MHz and Table \ref{baarsval} gives
the values of the constants A, B, C and D for 3C 286 and 3C 48 which I
used here. {\small SETJY} uses the calculated flux densities of these
primary calibrators to set the absolute flux calibration.
\begin{table}[!h]
\centering
\begin{tabular}{c cccc}
Source&A&B&C&D\\ \hline
3C 48	&1.31572 & -0.74090 & -0.16708 & 0.01525 \\
3C 286 	&1.23734 & -0.43276 & -0.14223 & 0.00345 \\\hline
\end{tabular}
\caption{The constants for equation \ref{baarseq} with the 1999.2 values.}\label{baarsval}
\end{table}

The task {\small VLACALIB} was then used to determine the antenna
based calibration for the primary calibrator. Ideally the primary
calibrators should be point sources that do not have any time
variability. However, 3C 286 is partially resolved with most array
combinations of the VLA and most wave bands. It is resolved on two
scale lengths, it has a small secondary core located 2.5" from the
central core and the core itself is partially resolved at longer
baselines.  To over come this it is possible to restrict the {\it
uv}-range and (if necessary) the number of inner antennas used in the
calibration fit. 3C 48 is also resolved with some array configurations
and thus it also has some restrictions on its use. It also shows some
time variability on the time scale of years. However, both calibrators
are still the preferred primary calibrators and the restrictions do
not adversely affect the flux calibration. For full details see the
{\small AIPS} calibrator manual.
\subsubsection{Calibrating the secondary}
\index{secondary calibration}
The primary calibrator sets the absolute flux density of the
observations. Several secondary calibrators that lie as close to the
target sources as possible are observed as frequently as the target
sources. These secondary calibrators are essential in the phase
calibration of the sources. The secondary calibrators are calibrated
using the task {\small GETJY} which calculates the flux density of the
secondaries with respect to the primary, thus `bootstrapping' the
initial calibration on to the secondaries. {\small VLACLCAL} then
interpolates between individual observations of the secondary
calibrator, thus applying the calibration to the entire data set.
The last step in the flux calibration is to self-calibrate all the
calibrators, still using {\small VLACLCAL}. This ensures that the flux
calibration is complete. A final check on the flux density calibration
is done using the task {\small LISTR} with {\small OPTYPE=`MATRIX'}
and {\small DPARM = 3 1 0}. This displays all the rms noise values for
the calibrators. A high value of the rms noise indicates a bad
calibration. In this case the calibration process was repeated until a
satisfactory rms value was achieved.
\subsection{Polarisation calibration.}
\index{polarisation calibration} 

Polarisation calibration is more straight forward than the flux
density calibration. However, the range of polarisation angles (PA)
must be known for the calibrators as a successful calibration requires
a polarisation angle change of more than 90 degrees over the course of
individual observing programmes. {\small PCAL} reads in the {\it
uv}-data and determines the effective feed parameters for each antenna
and also for each IF (intermediate frequency)\footnote{The IF is the
frequency that comes from mixing the local oscillator (LO) frequency
with the radio frequency (RF). The RF is the incoming frequency of the
observations and the LO is a sinusoidal signal that converts the RF to
an IF.}. All corrections are then applied to the `AN' (antenna)
table. {\small PCAL} removes the parallactic angle from the phase of
the visibility data and applies the corrections to the instrumental
corrections that are placed in `AN'.

\subsection{Calibration of source files}
\index{source calibration} \index{Maximum Entropy Method} 

The flux calibration and the polarisation calibration is now complete
for the observations as a whole, but the sources must be individually
`cleaned'. There are several calibration methods that can be used, the
main two used are {\small CLEAN} and the Maximum Entropy Method
({\small MEM}). The {\small CLEAN} method by \citet{hogbom74}
represents a radio source as a collection of point sources in an empty
field. An iterative process is used to find the flux and position of
these point sources, removing 10\% from the flux of the point sources,
with the maximum flux, in each iteration. It then selects the most
probable image from a set of images. {\small MEM} is not a procedural
method, it selects the image with the largest entropy that fits the
data and lies within the noise level. For full details of both methods
see \citet{book_VLA}. I use {\small CLEAN} to calibrate all my
sources, but in order to check the quality of my maps I also use
{\small MEM}, implemented in the AIPS task VTESS, instead of the CLEAN
algorithm. The resulting maps are not significantly different from
those produced by the CLEAN algorithm. In fact, despite common opinion
VTESS is not necessarily superior to CLEAN in producing accurate maps
of extended low surface brightness regions \citep{rupen97}.

\subsubsection{Clean}
\index{CLEAN} 

Before the maps of individual sources can be cleaned they are first
split off from the entire data set using the task {\small SPLIT}. The
source is then loaded into {\small IMAGR} and the number of {\small
CLEAN} iterations set via {\small NITER}. {\small NITER} determines
the number of times {\small IMAGR} will subtract 10\% from the current
peak flux. By choosing too small a value of {\small NITER} the task
{\small IMAGR} will not reach the noise level. By choosing too large
an {\small NITER} the {\small CLEAN} algorithm will clean past the
noise level, i.e. it will assume that the noise is part of the source
structure and will try to incorporate it into a model of the
source. This `over--cleaning' can lead to errors in the final source
map. Thus it is a sensitive balance and a conservative value for
{\small NITER} is usually the best option for the first round of
cleaning. When {\small IMAGR} is finished it produces a set of clean
components ({\small CC}'s). The {\small CC}'s are essentially a
representation of a map of the source including the noise. Selecting a
set of {\small CC}'s excluding the noise provides a template map for
the observed source. This subset corresponds to a specific {\it
uv}--range of the source and this range must be used in {\small CALIB}
for self--calibration. The sub-set of the {\small CC}'s and their
corresponding {\it uv}--range are used in conjunction with {\small
SOLMODE}=`p' which means that the initial calibration is only
concerned with the phase of the source. The template map of the source
is used in self--calibration to reduce the residual noise, in other
words, the template is used to modify the {\it uv} data to more
closely fit the template. The new {\it uv} data set is then run
through {\small IMAGR} again and the whole process is repeated. In
general the self--calibration is used with {\small SOLMODE}=`p' for
two iterations and then {\small SOLMODE}=`a\&p' (i.e. both amplitude
and phase) for all remaining iterations. After the final self-cleaning
step the {\it uv} data set is ran through {\small IMAGR} with {\small
NITER} set to a much higher value to remove the last remnants of any
remaining noise from the stokes `I' image, i.e. the total intensity
image. {\small IMAGR} is then run on a reduced number of {\small
NITER} with stokes `$Q$' and then `$U$'\footnote{$Q=(RL+LR)/2$ and
$U={\rm i}(RL-LR)/2$, where $R$ is the right circular polarisation and
$L$ is the left circular polarisation \citep{formalont}.}  to produce
the raw maps that can be combined to produce polarisation maps
described below.

\section{Map production}\label{rmprobs}
\index{map production} 

Total intensity maps were made from the Stokes I parameters at each
frequency.  Polarisation maps were also made at all frequencies by
combining the Stokes $Q$ and $U$ polarisation parameters. A map was
then produced that contained the polarised flux, $P =(Q^2+U^2)^{1/2}$,
and the electric field position angle, $PA=0.5\tan^{-1}(\frac{U}{Q})$,
at a given frequency.  The {\small AIPS} task {\small POLCO} was used
to correct for Ricean bias, which arises when the Stokes $Q$ and $U$ maps
are combined without removing noise-dominated pixels. By careful
setting of the {\small PCUT} parameter this bias is removed.  All maps
only contain pixels where the polarised flux and the total intensity
flux are above $5\sigma_{\rm noise}$ at 4.8 GHz and $3\sigma_{\rm
noise}$ at 1.4 GHz. The lower threshold at 1.4 GHz was necessary
because the 1.4 GHz data had a higher noise level, so blanking flux
below $5\sigma$ resulted in large regions of polarised flux being
lost.

At all frequencies the individual maps were made such that the beam
size, the cell size of the image and the coordinates of the
observations were exactly the same.  If any of these properties of the
map differed between frequencies, then the resultant multi-frequency
map would contain false structures that would be directly related to
the mis-alignment of the maps. To make sure that the coordinates (and
cell size) were always within acceptable tolerances the {\small AIPS}
task {\small HGEOM} was used to realign maps at one frequency to maps
at another frequency. In sources where an identifiable core exists at
both frequencies, the core positions were used as a check on the
alignment from {\small HGEOM}.  In general {\small HGEOM} is adequate
in aligning the multi-frequency data.  Sources with a distinct core at
all frequencies were aligned within 0.03'', where no core existed the
hotspots were aligned within 0.045''. In 4 sources this was not
sufficient. 3C 68.1 had to be shifted 0.05'' east and 0.07'' north, 3C
265 had to be shifted 0.04'' west and 0.02'' north, 3C 299 had to be
shifted 0.1'' east and 0.03'' north and finally 3C 16 had to be
shifted 0.1'' east and 0.1'' north. All shifts were applied to the 4.8
GHz observations.

\index{spectral index}
\index{depolarisation}
Spectral index maps were made between 4.8 GHz and 1.4 GHz, where the
spectral index, $\alpha$, is defined by $S_{\nu} \propto
\nu^{\alpha}$.  Depolarisation maps were made by dividing the map of
the percentage polarised flux at 4.8 GHz by the corresponding map at
1.4 GHz.  Rotation measure maps were made using the polarisation angle
maps at three frequencies,\index{rotation measure}
\begin{equation}\label{rmeq}
PA(\lambda) = PA_{\circ} + RM \lambda^2,
\end{equation}
where $PA$ is the observed polarisation position angle of a source,
$PA_{\circ}$ is the initial polarisation angle before any Faraday
rotation, RM is the rotation measure and $\lambda$ is the wavelength
of the observations. Clearly, polarisation angle measurements are
ambiguous by $\pm$n$\pi$ and this can introduce ambiguities in the
rotation measure maps. A change of $\pi$ between 1.4 GHz and 4.8 GHz
introduced by the used fitting algorithm will cause a change of
$\approx$ 80 rad m$^{-2}$ in rotation measure.  To determine if any
strong rotation measure change is real within a rotation measure map a
plot of the polarisation angle (measured on both sides of the observed
jump) against $\lambda^2$, including $n\pi$ ambiguities can be
produced. The best fit from {\small AIPS} is then overlayed. Any true
feature will not show any ambiguities in $n\pi$. This has been done
for two sources: 4C 16.49 and 6C 1256+36. The resulting fits are
presented in Figures
\ref{6lambda} to \ref{4lambda} and their corresponding $\chi^2$ values
for the fits are discussed in the relevant notes on these sources
below. Another test that a feature is real is that a true jump in
rotation measure causes depolarisation near the jump, but the magnetic
field map shows no corresponding jump in the region.

\index{AD429}
\index{n$\pi$ ambiguities}
The rotation measure maps of 7C 1813+684, 3C 65 and 3C 268.1 contained
obvious jumps in position angle which I was not able to remove. Plots
analogous to Figures \ref{6lambda} to \ref{4lambda} indicated that
there were regions that obviously contained errors caused by n$\pi$
ambiguities. As previously noted the A array AD429 data was
problematic and this was found to be the cause of the jumps. To
overcome this problem I artificially shifted the position angles at
1.4 GHz data down by 10 to 15 degrees before the $PA$ maps were
produced. This shift is within the position angle error and resolved
any ambiguities.

Table \ref{vlaobsc} shows that all sources in sample C were observed
with IFs separated by only 50 MHz or less at around 1.4 GHz. This
means that they were not well enough separated at 1.4 GHz to overcome
the $n\pi$ ambiguities.  To compensate for this lack of separation the
4.8 GHz observations were split into their two component frequencies,
4885 MHz and 4535 MHz. I then used 4 frequencies for the fit instead
of 3, but the sources in sample C are still only marginally sensitive
to $n\pi$ jumps.  The resulting rotation measure maps cover the same
frequency range as samples A and B, but use different frequencies for
the fit. This was not possible in the case of 3C 351 and 3C 299,
resulting in larger uncertainties in the rotation measurements for
these sources.

In the case of sample C any source that has a large range of rotation
measures ($> 80$ rad m$^{-2}$), the {\small AIPS} task {\small RM}
will force the rotation measure into a range $\pm 40$ rad m$^{-2}$
around the mean rotation measure. This is due to the lack of frequency
separation at 1.4 GHz and it can cause jumps. In the case of 3C 457
these jumps were severe and I was unable to resolve them. The rotation
measure and magnetic field maps for this source are not included in
the analysis. The rotation measure varies smoothly over all other
sources in this sample. The error affects the absolute value of the
rotation measure for each source and therefore it does not affect the
difference in the rotation measure between the two lobes of a given
source, dRM or the rms variation in the rotation measure.

Depolarisation and rotation measure properties are discussed in more
detail in chapter \ref{sampleresults}.

\subsection{Notes on individual sources}
\index{source notes}
In appendices \ref{Amap} to \ref{Cmap} (Figures \ref{0943.1} to
\ref{4c14.1}) I present the maps of the radio properties discussed above
for all the sources from the 3 samples. Each figure shows the
depolarisation map (where the polarisation was detected at all
frequencies), the spectral index map, the rotation measure map and the
magnetic field direction map (when a rotation measure is
detected). Table \ref{maps} contains a listing of previously published
maps for all sources, for completeness I have included all my
reduced maps in the appendices but these maps are not published in
Goodlet et~al.(2004).  In all cases only regions from the top end of
the grey--scales saturate, as I have always kept the lowest values
well inside the grey--scales, to ensure that no information has been
lost.
\index{archival data}
\begin{table}
\centering
\begin{tabular}{ l lll }\\\hline
Source & Map&Frequency & Reference\\ & & (GHz)& \\\hline 6C 0943+39&
P&4.8 & \citet{bel99}\\ 
6C 1011+36&P&4.8 & \citet{bel99}\\ &TI&1.4 &
\citet{lla95}\\ 
6C 1129+37&P&4.8 & \citet{bel99}\\ &TI&1.4 &\citet{lla95}\\ 
6C 1256+36&P&4.8 & \citet{bel99}\\ &TI&1.4 &\citet{lla95}\\ 
6C 1257+36&P&4.8 & \citet{bel99}\\ &TI&1.4 &\citet{lla95}\\ 
3C 65& TI& 1.4, 4.8 & \citet{pwx95}\\ 
3C 68.1 & P & 4.8& \citet{bhl94}\\ &TI &1.4 & \citet{lms89}\\ 
3C 252& P &4.8 &\citet{fbb93}\\ 
3C 265& P & 4.8 & \citet{fbb93}\\ 
3C 267 & TI & 4.8 &\citet{blr97}\\ & & 1.4 & \citet{lms89}\\ 
3C 268.1 & P & 4.8 &\citet{l81}\\ &TI &1.4 & \citet{lms89}\\ 
3C 280 & P & 1.4, 4.8 &\citet{lp91}\\ & S, D & 1.4, 4.8 & \citet{lp91}\\ 
3C 324 & TI & 4.8 &\citet{bcg98}\\ &P & 1.4 & \citet{fbb93}\\ 
4C 16.49 & P & 4.8 &\citet{lbm93}\\ 
3C 16 & TI & 4.8 & \citet{gfg88}\\ &P & 1.4 &\citet{lp91a}\\ 
3C 42 & P & 4.8 & \citet{fbp97}\\ 
3C 46 & TI & 4.8 &\citet{gfg88}\\ & & 1.4 & \citet{gpp88}\\
3C 299 & P & 1.4, 4.8 & \citet{lp91}\\
 & S, D & 1.4, 4.8 & \citet{lp91b}\\
3C 341 & P & 1.4 & \citet{lp91a}\\
3C 351& TI&  4.8 & \citet{bhl94}\\
& P & 1.4 & \citet{lp91a}\\
3C 457& P & 1.4&\citet{lp91a}\\
4C 14.27 & P & 1.4 & \citet{lp91a}\\\hline
\end{tabular}
\caption[Details of previously published maps.]
{Details of previously published maps. TI = total intensity, P =
polarisation, S = spectral index \& D = depolarisation} \label{maps}
\end{table}
\subsubsection{Sample A:} 
\begin{figure}[!h]
\centerline{\psfig{file=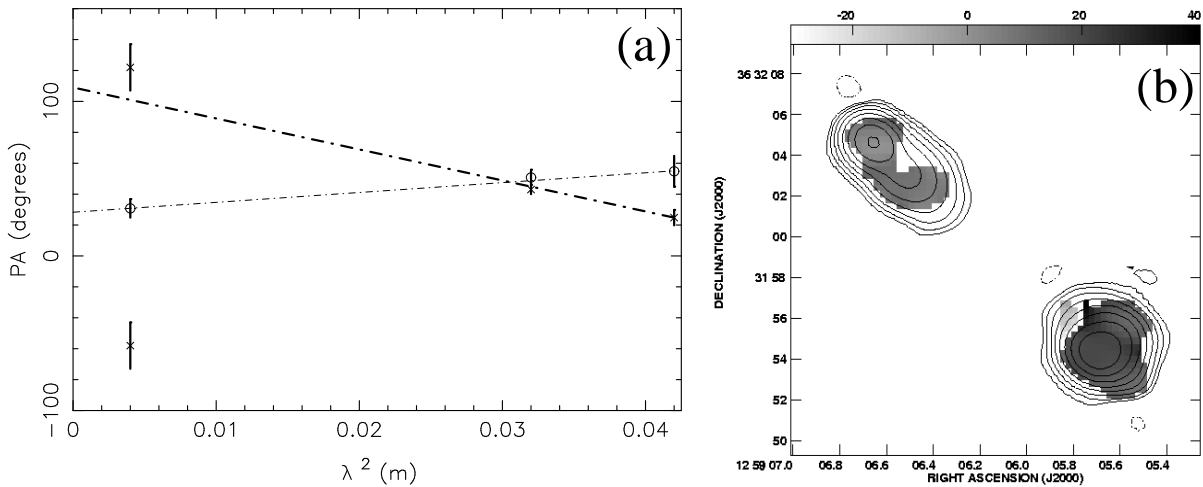,width=16.0cm}}
\caption[Rotation measure fits for 6C 1256+36] {(a) A plot of the
polarisation angle against $\lambda^2$, allowing for n$\pi$
variations, for the southern lobe of 6C 1256+36 with the dashed lines
showing the best fit models. All reasonable n$\pi$ solutions for the
4.7 GHz data are considered and plotted.  Both sides of the jump are
plotted with `$\circ$' indicating one side of the jump and `$\times$'
the other. (b) The rotation measure map for 6C 1256+36 (rad m$^{-2}$)
between 4710 MHz, 1665 MHz and 1465 MHz. All contours are at $5
\sigma$ at 4710 MHz (0.25 mJy beam$^{-1}$)$\times$(-1, 1, 2,
4...,1024) with a beam size of 2.5'' $\times$ 1.4''.}\label{6lambda}
\end{figure}

\paragraph*{{\it 6C 0943+39:}}(Figure A.1) 
No core is detected in my observations. \citet{bel99} detected a core
at 8.2 GHz and minimally at 4.8 GHz. My non-detection is probably due
to the different resolution of the two data sets. The value of the
rotation measure in the Eastern lobe must be considered with caution
as it is based on only a few pixels.
\paragraph*{{\it 6C 1011+36:}}(Figure A.2) 
This is a classic double-lobed structure, showing a strong core at
both 4.7 GHz and 1.4 GHz with an inverted spectrum.
\paragraph*{{\it 6C 1018+37:}}(Figure A.3) 
The maps were made with the smaller arrays only at each frequency. In
the 1.4 GHz A-array data set the lower lobe was partially resolved out,
but this was compounded by the high noise. So no feasible combination
of the A and B array was possible.  To maintain consistency the
B-array 4.7 GHz data was also excluded.
\paragraph*{{\it 6C 1129+37:}}(Figure A.4) 
The SE lobe contains two distinct hotspots, \citet{bel99} found 3
hotspots. This discrepancy is probably due to the different
resolutions of the two observations. The source shows distinct regions
of very strong depolarisation, however these regions are slightly
smaller than the beam size.
\paragraph*{{\it 6C 1256+36:}}(Figure A.5) 
The rotation measure map shows distinct changes in the values of the
rotation measure. Figure \ref{6lambda} shows that although the jump in
RM, in the southern lobe, does not correspond to a jump in
depolarisation, it is not due to any error in the fitting program. The
corresponding reduced $\chi^2$ values for the fits are given in Table
\ref{values}.

\begin{table}[!h]
\centering 
\begin{tabular}{cc r@{.}lr@{.}lr@{.}l c}\hline
Source &Lobe&\multicolumn{2}{c}{$n=-1$}&\multicolumn{2}{c}{$n=0$} &
\multicolumn{2}{c}{$ n=1$} &Symbol\\\hline
6C 1256+36&S&382&0    &51&0     &1&9&$\times$\\
&&981&6    &1&2   &976&5&$\circ$\\\hline
\end{tabular}
\caption{Reduced $\chi^2$ values for the rotation measure fits for
6C 1256+36.}\label{values}
\end{table}
\paragraph*{{\it 6C 1257+37:}}(Figure A.6)
A core was detected at 4.7 GHz but was absent from the 1.4 GHz
data. The high noise level and short observation time meant that the S
lobe had very little polarised flux above the noise level. Reliable
values for the rotation measure were found in only a few pixels around
the hotspots.
\paragraph*{{\it 7C 1745+642:}}(Figure A.7) 
This is a highly core dominated source, with the northern lobe
appearing faintly. There is an indication of a jet-like structure
leading down from the core into the southern, highly extended,
off-axis, lobe. The source is a weak core dominated quasar
\citep{bh01}.
\paragraph*{{\it 7C 1813+684:}}(Figure A.9) This is the faintest
of the sources in sample A and is also a quasar \citep{bh01}.  The
source shows a compact core that is present at all observing
frequencies, but it is too faint to detect any reliable polarisation
properties.
\subsubsection{Sample B:}
\paragraph*{{\it 3C 65:}}(Figure B.1)
The W lobe shows a strong depolarisation shadow that is smaller than
the beam size. \citet{b00} found the source to lie in a cluster which
might account for the presence of the depolarisation shadow and the
large depolarisation overall.
\paragraph*{{\it 3C 68.1:}}(Figure B.2) The source is a 
quasar \citep{bhl94}.  A core has been detected by \citet{bhl94} in
deeper observations.
\paragraph*{{\it 3C 252:}}(Figure B.3) 
The SE lobe shows a sharp drop in the polarisation
between the 4.7 GHz and 1.4 GHz observations.
\paragraph*{{\it 3C 265:}}(Figure B.4) The NW
lobe shows evidence of a compact, bright region with a highly ordered
magnetic field which at
higher resolutions \citet{fbb93} show is the primary hotspot.
\paragraph*{{\it 3C 267:}}(Figure B.5) 
The E lobe is highly extended, reaching to the core position, which
can be seen in the 1.4 GHz image. The large depolarisation region in
the W lobe coincides with a region with no observed rotation
measure. The core is strongly inverted with $\alpha =
0.48$.

\paragraph*{{\it 3C268.1:}}(Figure B.6) 
This is a classic double--lobed source but there is no core detected
at either 1.4 GHz or 4.8 GHz.

\paragraph*{{\it 3C 280:}}(Figure B.7)
The value of the rotation measure and the magnetic field direction in
the E lobe must be treated with caution as it is based on only a small
region of the entire lobe. The sharp changes in the rotation measure
map are not seen in the magnetic field map and the depolarisation map
shows a similar structure suggesting that is not due a fitting error.

\paragraph*{{\it 3C 324:}}(Figure B.8) 
The NE 
lobe shows evidence of a depolarisation shadow. \citet{b00} found the
source to lie in a cluster which may explain the faint shadow.

\begin{figure}[!h]
\centerline{\psfig{file=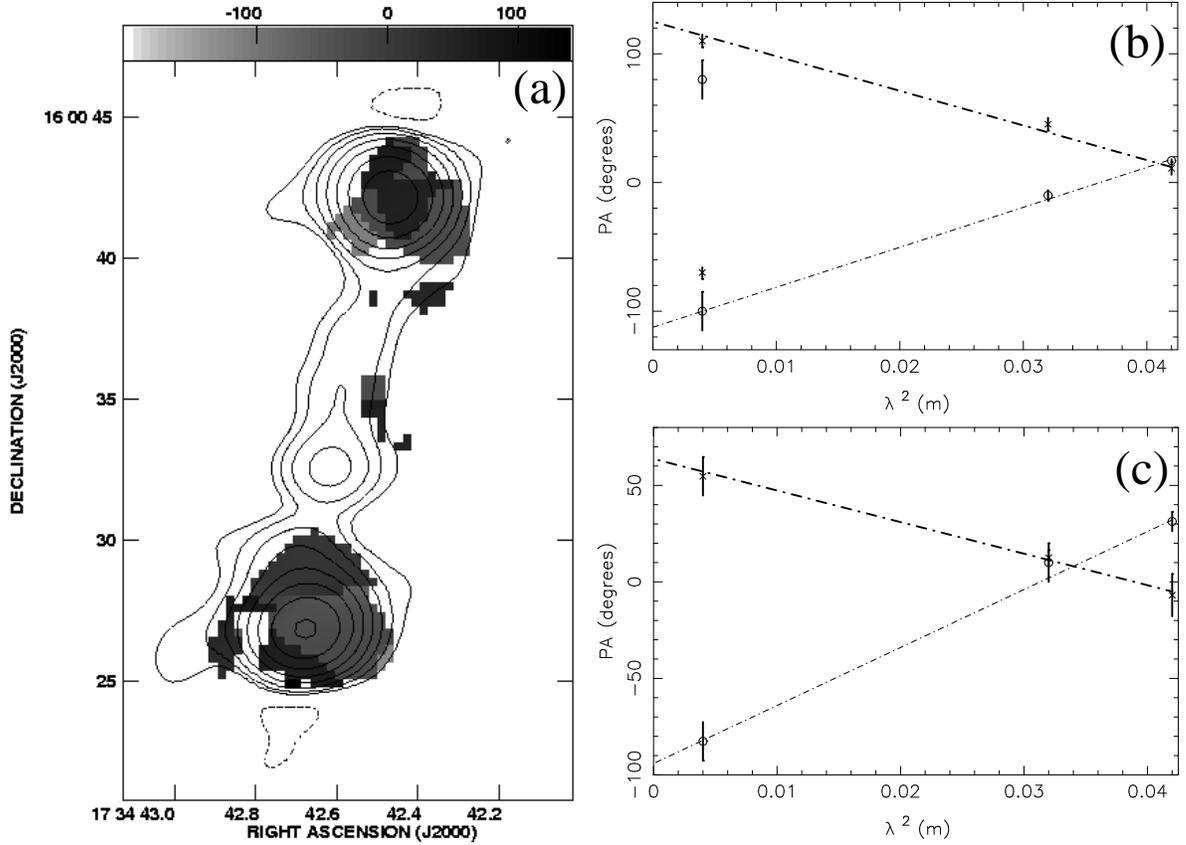,width=16.0cm}}
\caption[Rotation measure fits for 4C 16.49] {(a) Map of the rotation
measure of the radio source 4C 16.49 (rad m$^{-2}$) between 4710 MHz,
1665 MHz and 1465 MHz. All contours are at $5 \sigma$ at 4710 MHz (0.8
mJy beam$^{-1}$)$\times$(-1, 1, 2, 4...,1024) with a beam size of
2.2'' $\times$ 1.8''. (b) A plot of the polarisation angle against
$\lambda^2$, allowing for n$\pi$, for the northern lobe of 4C 16.49
with the dashed lines showing the best fit models. All reasonable
n$\pi$ solutions for the 4.7 GHz data are considered and plotted. Data
for two small regions, one on each side of the jump are plotted with
`$\circ$' indicating one side of the jump and `$\times$' the
other. The jump plotted lies SW of the central intensity contour. (c)
Same as (b) but for the southern lobe, the region plotted lie either
side of the jump south of the peak intensity contour.}\label{4lambda}
\end{figure}

\paragraph*{{\it 4C 16.49:}}(Figure B.9) 
The source is a quasar \citep{bh01}, that shows a strong radio core,
jet structure and possibly a small counter-jet. The source is highly
asymmetric with the southern lobe almost appearing to connect to the
core. It has a very steep spectral index, $\alpha < -1.0$ making it an
atypical source.  Figures \ref{4lambda}(b) and \ref{4lambda}(c)
demonstrates that the sharp changes in the rotation measure map,
see Figure \ref{4lambda}(a), are not due to any fitting errors. The
corresponding reduced $\chi^2$ values for the fits are given in Table
\ref{values2}.

\begin{table}[!h]
\centering 
\begin{tabular}{cc r@{.}lr@{.}lr@{.}l c}\hline
Source &Lobe&\multicolumn{2}{c}{$n=-1$}&\multicolumn{2}{c}{$n=0$} &
\multicolumn{2}{c}{$ n=1$} &Symbol\\\hline
4C 16.49&N (b)&1&6    &669&0  &2917&0&$\circ$\\
&&2772&0   &90&1&     1&7&$\times$\\
4C 16.49&S (c)&151&9    &0&9    &62&0  &$\circ$\\
&&45&4   &0&4     &34&8  &$\times$\\\hline
\end{tabular}
\caption[Reduced $\chi^2$ values for the rotation measure fits for 4C 16.49]
{Reduced $\chi^2$ values for the rotation measure fits for
Figure \ref{4lambda}.}\label{values2}

\end{table}
\subsubsection{Sample C:}
\paragraph*{{\it 3C 16:}}(Figure C.1) The source 
shows a strong SW lobe, with a relaxed NE lobe. The SW lobe shows a
strong depolarisation feature that is narrower than the beam size. The
strong rotation measure feature is evident in the depolarisation map,
but not the magnetic field map, indicating that it is not an error in
the fitting program.  No value for the rotation measure was obtained
for the NE lobe because the polarisation observed was too weak.
\paragraph*{{\it 3C 42:}}(Figure C.2)
The core was detected at 4.7 GHz, but was absent at the lower
frequencies. The source has been observed to lie in a small cluster by
\citet{vbq00}. \citet{fbp97} observed that the N hotspot was double,
but this is not evident in my observations which can be attributed to
the differences in the resolutions of the two observations.
\paragraph*{{\it 3C 46:}}(Figure C.3) 
The source has a prominent core at 4710 MHz, but it is
indistinguishable from the extended lobe at 1452 MHz.
\paragraph*{{\it 3C 341:}}(Figure C.4) The source is a classic double
with a resolved jet-like structure running into the SW lobe. The jet
is more prominent in the higher frequency observations than at the
lower frequencies.
\paragraph*{{\it 3C 351:}}(Figure C.5) The source is an extended and
distorted quasar \citep{bhl94}. Both lobes expand out to envelope the
core. The NE lobe is highly extended, off-axis and shows two very
distinct hotspots. The depolarisation increases towards the more
compact SW lobe supporting the idea that the environment around the SW
lobe is denser, stopping the expansion seen in the NE lobe. There is
evidence of a rotation measure ridge in the NE hotspots which
corresponds to a narrow ridge of depolarisation, but there is no
corresponding shift in the magnetic field map.
\paragraph*{{\it 3C 457:}}(Figure C.6) 
The SW lobe shows a prominent double hotspot. The small compact object
just south of the SW hotspots is most likely an unrelated background
object.  The inverted core was observed to be present at all
frequencies. This source has no rotation measure map or magnetic field
measure map as I was unable to remove all n$\pi$ ambiguities from this
source. This was due to the small separation of observing frequencies
around 1.4 GHz and 4.8 GHz, see section \ref{rmprobs}.
\paragraph*{{\it 3C 299:}}(Figure C.7) 
The source is the least luminous of all the sample members.  The source
shows a large change in rotation measure between the lobes but the
difference is probably due to the small number of pixels with rotation
measure information in the NE lobe.
\paragraph*{{\it 4C 14.27:}}(Figure C.8)
There is no core detected at any frequency even in a better quality
map by \citet{lp91a}.

\newpage%
\thispagestyle{empty}
\chapter[Trends]{Source properties and trends}\label{sampleresults}
\index{source properties}
The average spectral index, average depolarisation and average
rotation measure are given by $\alpha$, DM and RM respectively. Each
of these parameters are the average over each individual lobe of a
source. $\alpha$ is calculated using $S_\nu\propto\nu^{-\alpha}$
between 1.4 GHz and 4.8 GHz, where the $S_\nu$ are the Total Intensity
Flux values given in Tables \ref{A} to \ref{C}. The average DM is
calculated using the percentage polarised flux measurements
given in Tables \ref{A} to \ref{C}. The average RM is measured
directly from the rotation measure maps (see appendix \ref{Amap} to
\ref{Cmap}) using the {\small AIPS} task {\small IMEAN}.

The differential spectral index, differential depolarisation and
differential rotation measure over the lobe of an individual source
are given by d$\alpha$, dDM and dRM respectively. All differential
properties are taken to be the difference of the averages over
individual lobes of a source e.g. dDM = DM$_{lobe_1}$ -
DM$_{lobe_2}$. The rms variation of the rotation measure is defined by
$\sigma_{RM}$.

To test if any property is varying between samples I compare the
average of the property over a {\it sample} with the average taken
from the other two samples. This allows a simple statistical test to
be applied to the data. I also average properties over samples with
equal distributions in redshift (A+B) and equal distributions in
radio-luminosity (A+C). For example, if depolarisation increases with
redshift but not with radio-luminosity I would expect samples A and B
to show similar averages (within errors) but to be statistically
greater than sample C. In this case I would also expect to see the
average of sample A+C (low radio luminosity sources) to be
statistically similar to the average of sample B (high radio
luminosity sources). A more thorough statistical approach using
Spearman Rank, Partial Spearman Rank and Principal Component is
presented in chapter \ref{statsobs}.

\section{Polarised Flux and Spectral Index.}
\index{spectral index} \index{polarised flux}
\begin{table}
\centering
\begin{tabular}{ r@{ }l ll  r@{.}l r@{.}ll r@{.}l r@{.}l  l r@{.}l  }\hline
&&&\multicolumn{5}{c}{4800 MHz}&\multicolumn{5}{c}{1465 MHz}\\
\multicolumn{2}{c}{Source} & Comp. 
&&\multicolumn{2}{l}{Total} &\multicolumn{2}{l}{Polar-}&
&\multicolumn{2}{l}{Total} &\multicolumn{2}{l}{Polar-}
&&\multicolumn{2}{l}{\hspace{2mm} $\alpha$} \\
   & & &&\multicolumn{2}{l}{Flux}&\multicolumn{2}{l}{isation} 
&&\multicolumn{2}{l}{Flux}&\multicolumn{2}{l}{isation}
 &&\multicolumn{2}{l}{ }\\
    &     & &&\multicolumn{2}{l}{(mJy)} &\multicolumn{2}{c}{\%} 
&&\multicolumn{2}{l}{(mJy)} &\multicolumn{2}{c}{\%} 
&&\multicolumn{2}{l}{ } \\ \hline 
      6C&0943+37  &  W && 31&0 	& 11&8	 && 71&7 & 5&4 && -0&72 \\
   		& &E  &&  42&0 	& 2&7 	&& 136&4 & 0&7   && -1&01 \\
     6C&1011+36  &  N  && 44&3 	& 7&8 	&& 119&6 & 5&1&& -0&88\\
    		& & S  && 18&7 	& 12&2  	&& 50&9 & 11&5 && -0&89 \\
    	& &Core 	&&3&2 	&\multicolumn{2}{c}{--} 
	&\multicolumn{3}{r}{0.7 $\pm 0.5$} &\multicolumn{2}{c}{--} && 1&35\\
      6C&1018+37  & NE && 46&4 	& 10&0 	&& 125&8 & 8&1 & & -0&85
    \\ & &SW & 		& 28&7  	& 7&0 	&& 76&3 & 2&9  && -0&84\\
    && Core &\multicolumn{3}{r}{0.63 $\pm 0.2$}	&\multicolumn{2}{c}{--} 
	&\multicolumn{3}{r}{--}&\multicolumn{2}{c}{--}&\multicolumn{3}{c}{--}\\
      6C&1129+37 &  NW && 46&5	 & 7&3 &	& 129&1 & 2&9 && -0&85
    \\  && SE & &	73&2 	& 16&3 	&& 215&0  & 3&2 && -0&90 \\
      6C&1256+36  &  NE& & 57&8 	& 10&4 	&& 148&9 & 7&7 && -0&79
     \\  & & SW &  &	101&4 	& 8&9 	&& 288&5 & 8&8 && -0&87 \\
       6C&1257+36  &  NW && 43&5	& 17&0 	&& 102&2  & 13&4 && -0&71 
    \\  && SE 	& 	& 20&5 	& 9&8	&& 73&0  & 5&2 && -1&06\\
    && Core &\multicolumn{3}{r}{0.29 $\pm 0.1$}	&\multicolumn{2}{c}{--} 
	&\multicolumn{3}{r}{--}&\multicolumn{2}{c}{--}&\multicolumn{3}{c}{--}\\
      7C&1745+642 &  N && 23&5 	& 9&6 	&& 64&5 & 5&2&&-0&86
     \\&& Core & &	84&1 	& 4&9 	&& 69&4  &2&9  & \multicolumn{3}{r}{0.16 $\pm 0.1$} 
    \\ && S & 		&33&8 	& 8&6  	&& 98&5 & 9&5 && -0&91 \\
      7C&1801+690 &  N &\multicolumn{3}{r}{8.80 $\pm 3.0$}& 3&0 	&& 28&2 & 2&7 && -0&97
    \\& &Core & &	79&1 	& 1&8 	&& 78&4 & 4&0 && 0&007 \\
     & &S & 	&	28&7 	& 10&0	&& 75&8 & 7&2 && -0&81 \\
      7C&1813+684 &  NE &&15&0 	& 7&7 	&& 44&8 & 9&5&& -0&92
    \\ && SW &  &	30&2 	& 8&4 	&& 78&1 &7&4 && -0&79 \\
    && Core &	&	3&3 	&\multicolumn{2}{c}{--} 
	&&2&65 & \multicolumn{2}{c}{--}& \multicolumn{3}{c}{--} \\ \hline
\end{tabular}
\caption[Properties of the sample A radio source components]
{Properties of the sample A radio source components. Errors are 5\% or
less unless stated otherwise. The spectral indices are the mean values
for each component, calculated between approximately 4800 MHz and 1465
MHz.}\label{A}
\end{table}
\begin{table}
\centering
\begin{tabular}{ r@{ }l l l  r@{.}l l r@{.}l l r@{.}l r@{.}l  r@{.}l  }\hline
&&&\multicolumn{6}{c}{4800 MHz}&\multicolumn{5}{c}{1465 MHz}\\
\multicolumn{2}{c}{Source}  & Comp. 
&&\multicolumn{2}{l}{Total} &&\multicolumn{2}{l}{Polar-}&
&\multicolumn{2}{l}{Total} &\multicolumn{2}{l}{Polar-}
&\multicolumn{2}{l}{\hspace{2mm} $\alpha$} \\
   & & &&\multicolumn{2}{l}{Flux}&&\multicolumn{2}{l}{isation} 
&&\multicolumn{2}{l}{Flux}&\multicolumn{2}{l}{isation}
 &\multicolumn{2}{l}{ }\\
    &     & &&\multicolumn{2}{l}{(mJy)} &&\multicolumn{2}{c}{\%} 
&&\multicolumn{2}{l}{(mJy)} &\multicolumn{2}{c}{\%} 
&\multicolumn{2}{l}{ } \\ \hline 

      3C &65  &  W &&  524&0      &&19&3 &&1683&1 & 5&4 & -1&00\\
       && E   && 240&9 && 9&2 && 800&4  & 7&2 & -1&03 \\
      3C& 252  &  NW  && 178&7 && 6&4 && 592&2 & 5&9 & -1&00\\
        && SE  &&  80&0 && 14&1 &&  300&5 & 6&8 & -1&10\\
    & &Core &&1&98 &\multicolumn{3}{c}{--} &&1&33 &\multicolumn{2}{c}{--} & 0&33\\
       3C& 267  &  E && 184&0 && 8&9 && 745&7 & 6&8 & -1&17\\
    && Core &\multicolumn{3}{r}{ 1.87 $\pm 1$}&\multicolumn{3}{c}{--}  
&\multicolumn{3}{r}{ 1.05 $\pm 2$}& \multicolumn{2}{c}{--}  & 0&48 \\
    && W && 479&2  && 3&5 && 1294&6 & 3&3 & -0&83\\
      3C& 280  & E&& 326&0 && 8&0 && 1219&2 & 4&4 & -1&01\\
      && W && 1289&2 && 10&0 && 3191&6  &6&5& -0&76\\
      3C& 324  &  NE && 432&6 && 9&3 && 1525&2 & 5&6& -1&05\\
    & &SW &&  166&0 && 7&8 && 651&8 & 4&0 & -1&14 \\
      4C& 16.49 &  N  &&107&2	&& 8&3  && 307&2 & 13&0 & -0&88\\
    && Jet& \multicolumn{3}{r}{ 8.5 $\pm 4$}&&14&5&& 45&9  & 7&2 & -1&41\\
      &          &Core &\multicolumn{3}{r}{ 9.64 $\pm 2$}&& 2&7  && 58&2 &3&5  & -1&50\\
     && S  && 142&3 &&7&0	&&695&3  &6&4 & -1&32  \\
      3C& 68.1     & N  &&667&2 && 8&6	&& 1767&4&7&0 & -0&82\\
    & & S &&36&8 &&11&4   	&&134&6 &  6&6 & -1&08 \\
      3C& 265      & NW &&224&0&&10&0	&& 478&7&5&8& -0&62\\
    &   & SE&& 318&9	&&6&2  	&&835&0 &4&2 & -0&78\\
      3C &268.1    &  E   && 262&3	&& 5&0 	&&816&4 & 3&4 &-0&92\\
    & & W   &&2296&6 	&&4&7	&&4699&0&  2&7   & -0&58 \\ \hline
\end{tabular}
\caption[Properties of the sample B radio source components]{Properties of the sample B radio source components. Errors
are 5\% or less unless stated otherwise. The spectral indices are the
mean values for each component, calculated between approximately 4800
MHz and 1465 MHz.}\label{B}
\end{table}
\begin{table}
\centering
\begin{tabular}{ r@{ }l l l r@{.}l l r@{.}l r@{.}l l r@{.}l  r@{.}l  }\hline
&&&\multicolumn{6}{c}{4800 MHz}&\multicolumn{5}{c}{1452 MHz}\\
\multicolumn{2}{c}{Source}  & Comp. 
&&\multicolumn{2}{l}{Total} &&\multicolumn{2}{l}{Polar-}&
\multicolumn{2}{l}{Total} &&\multicolumn{2}{l}{Polar-}
&\multicolumn{2}{l}{\hspace{2mm} $\alpha$} \\
   & & &&\multicolumn{2}{l}{Flux}&&\multicolumn{2}{l}{isation} 
&\multicolumn{2}{l}{Flux}&&\multicolumn{2}{l}{isation}
 &\multicolumn{2}{l}{ }\\
    &     & &&\multicolumn{2}{l}{(mJy)} &&\multicolumn{2}{c}{\%} 
&\multicolumn{2}{l}{(mJy)} &&\multicolumn{2}{c}{\%} 
&\multicolumn{2}{l}{ } \\ \hline 
       3C& 42  & NW && 353&9 && 12&5 & 999&7 && 12&3 & -0&87\\ 
       & &SE && 450&6   && 7&5 &1266&3 && 7&4 & -0&86 \\
      4C& 14.27  &  NW  && 107&0 && 12&4 & 368&2 && 8&8 & -1&06\\
    && SE  &&  124&8&& 9&6  &  489&3 && 9&3 & -1&17\\
      3C& 46   & NE && 162&7 &&16&2 & 488&0 &&12&8 & -0&94\\
    &&SW && 173&7  &&13&7 & 506&4  && 12&0 & -0&91 \\
    & &Core &&2&54 & \multicolumn{3}{c}{--}&7&94 & \multicolumn{3}{c}{--}& -0&97\\
      3C& 457  &NE&& 208&6 &&15&8 & 692&0  &&15&7 &-1&02 \\
       &&SW && 290&5  && 11&9 & 898&9 && 10&0      & -0&94 \\
    &&Core && 3&03& \multicolumn{3}{c}{--}&2&63 & \multicolumn{3}{c}{--}  &0&12\\
      3C& 351  &  NE && 971&7 && 6&8 & 2327&9 && 8&1  & -0&73\\
    &&Diffuse && 122&2 && 26&4 & 421&3 && 13&7  & -1&03 \\
    &&Core && 18&3 && 18&6 & 46&2 && 8&2  &-0&77\\
    && SW &&  77&1 && 22&8 & 235&7 && 8&9 &-0&93\\
      3C &341   &  NE &&123&8	&&28&0  & 336&4 && 19&6 & -0&85\\
        && SW  && 265&9 &&26&4&862&6  &&17&2 & -1&00 \\
      3C& 299      & NE &&876&5& \multicolumn{3}{c}{ 0.79$\pm 0.5$}& 2592&4
& \multicolumn{3}{c}{0.43 $\pm 0.3$}& -0&93\\
    & & SW&&53&7 &&3&1   &122&6&&  2&6 & -0&71 \\
      3C& 16       & NE & \multicolumn{3}{c}{22.1 $\pm 3$} &&10&8  &60&9  && 4&9&-0&87\\
    && SW   && 484&9&& 14&7& 1510&9 &&8&2 & -0&97 \\ \hline
\end{tabular}
\caption[Properties of the sample C radio source components]
{Properties of the sample C radio source components. Errors are 5\% or
less unless stated otherwise. The spectral indices are the mean values
for each component, calculated between approximately 4800 MHz and 1452
MHz.}\label{C}
\end{table}

\begin{table}[!h]
\centering
\begin{tabular}{ l r@{.}l@{ $\pm$ }r@{.}l  r@{.}l@{ $\pm$ }r@{.}l  r@{.}l@{ $\pm$ }r@{.}l } \hline
Average &\multicolumn{12}{c}{Sample} \\
Prop.&\multicolumn{4}{c}{A}& \multicolumn{4}{c}{B}&\multicolumn{4}{c}{C}\\\hline 
z&1&06&0&04 			& 1&11& 0&05 		& 0&41 & 0&01  \\
P$_{151}$& (1&02&0&06)$\times10^{27}$ & (1&01&0&07)$\times10^{28}$&
(8&02&0&40)$\times10^{26}$\\
D&220&34&46&40 			& 262&69& 55&41 	& 399&57 & 122&50 \\
$\alpha$ & -0&87 &0&01 		& -0&92 & 0&06 		& -0&94 & 0&03 \\
 d$\alpha$& 0&13 &0&04 		&0&23 & 0&04 		& 0&10 & 0&03 \\
PF$_{1.4}$&6&29&0&86 		& 5&88& 0&64 		& 9&89 & 1&75 \\
PF$_{4.8}$&8&91&1&03 		& 8&76& 0&89 		&13&31 & 2&48 \\\hline 
\multicolumn{2}{c}{ }&\multicolumn{3}{l}{Average} &\multicolumn{8}{c}{Sample}\\
\multicolumn{2}{c}{ }&\multicolumn{3}{l}{Prop.}&\multicolumn{4}{c}{A+B}&\multicolumn{4}{c}{A+C}\\
\cline{3-13}
\multicolumn{2}{c}{ }&\multicolumn{3}{l}{z}&  1&08 &0&03 		&0&75 & 0&08 \\
\multicolumn{2}{c}{ }&\multicolumn{3}{l}{P$_{151}$}& (5&57&1&16)$\times10^{27}$ & (8&97&0&77)$\times10^{26}$\\
\multicolumn{2}{c}{ }& \multicolumn{3}{l}{D}& 241&51 &35&43 	& 304&69 & 64&50 \\
\multicolumn{2}{c}{ } &  \multicolumn{3}{l}{$\alpha$}& -0&90&0&03 			&-0&90 & 0&02 \\
\multicolumn{2}{c}{ }&  \multicolumn{3}{l}{d$\alpha$}& 0&18&0&01			& 0&12 & 0&02  \\
\multicolumn{2}{c}{ }& \multicolumn{3}{l}{ PF$_{1.4}$}&6&08 &0&52 			& 7&98 & 1&01 \\
\multicolumn{2}{c}{ }& \multicolumn{3}{l}{ PF$_{4.8}$}&8&83 &0&66 			&10&98 & 1&36 \\
\cline{3-13}
\end{tabular}
\caption[Mean spectral index and differential spectral index of the
sources averaged over the samples.]  {Mean properties of the sources
averaged over each sample with the associated error. Differential
spectral index (d$\alpha$) is derived by taking the difference in the
spectral index between the two lobes of each source and then averaging
this difference over each sample. Source size is in kpc and
radio-luminosity is in W/Hz.}\label{averagea}
\end{table}
The flux observed from both lobes of all sources was found to be
polarised at levels greater than 1\%, the only exceptions being 6C
0943+39 and 3C 299, see Tables \ref{A} to \ref{C}. At lower redshifts
the polarisation exhibits the largest range from 0.8\% in 3C 299 to
28.0\% in 3C 341. This range is not evident in either of the other two
samples. Statistically the sources at low redshift (sample C) are
slightly more polarised than those at high redshift, both at 1.4\,GHz
(sample A: 2.1$\sigma$, sample B: 2.3$\sigma$) and at 4.8\,GHz (sample
A: 1.8$\sigma$, sample B: 1.8$\sigma$). There is no difference between
the low radio luminosity sources (sample A) and the high luminosity
objects (sample B). This suggests that percentage polarisation
decreases for increasing redshift but is less dependent on radio
luminosity.

Table \ref{averagea} shows that there are no significant correlations
between spectral index, $\alpha$, and redshift or radio luminosity as
suggested by \citet{on89}, \citet{vv72} and \citet{athreya}, which use
much larger samples. This suggests the trend may be present, but may
be masked by my small sample size. However, there are trends found
with the difference in spectral index between the two lobes.  Sample B
shows a larger average difference in the spectral indices, d$\alpha$
between the two lobes of a given source than the other samples (sample
A: 2.5$\sigma$, sample C: 3.3$\sigma$). Sample B contains the sources
with the highest radio luminosity and so I find that in my samples the
difference in spectral index, d$\alpha$ increases with radio
luminosity rather than with redshift.  The trend of the difference in
the spectral index between the two lobes may be related to the extra
luminosity of the 3CRR hotspots compared to the 6C/7C hotspots (see
section \ref{obs}). On average I find that the hotspots of my sources
have shallower spectral indices. The average spectral index integrated
over the entire sources will therefore depend on the fraction of
emission from the hotspots compared to the extended lobe, thus
creating the observed trend.

\section{Rotation measure}
\index{rotation measure}
\label{rm}
Faraday rotation occurs when polarised light passes through a
plasma. AGN's emit synchrotron radiation which is almost completely
linearly polarised, this can be separated into
right-handed (RH) and left-handed (LH) circular polarisation which
have different phase velocities. At a distance $dl$ from a source the two
circular polarisations will be out of phase by $\Delta\theta$, which
is given by:
\begin{eqnarray}
\Delta\theta=\frac{\phi_{RH}-\phi_{LH}}{2} &=&\left(\sqrt(\epsilon_{RH})- \sqrt(\epsilon_{LH})\right)
\frac{\omega dl}{2c}\\
{\rm where} \hspace{1cm}\sqrt(\epsilon) &=& 1 - \frac{\omega_p^2}{2\omega^2}\left(1-\pm\frac{\omega_g}{\omega}\right)
\end{eqnarray}
where $\omega_p$ is the phase velocity of the polarisation, $\omega_g$
is the non relativistic gyro-frequency of the electrons in a magnetic
field, $\omega = 2\pi \nu$ where $\nu$ is the frequency of the
observations and $c$ is the speed of light. The total phase difference
is then
\begin{equation}\label{angle}
\Delta \theta = \frac{e^3}{2\omega^2 m_e ^2 c \epsilon_0}\int_0^z n_eB_{\|}dl
\end{equation}
where $B_\|$ is the line-of-sight component of the magnetic field,
$n_e$, the column density around the source, $dl$ is the path length
to the source, $e$ is charge on the electron, $m_e$ is the mass of an
electron and $\epsilon_\circ$ is the electric constant. From section
\ref{rmprobs} the observed rotation measure, RM, of a source depends
on the frequency of the observations, $RM=\Delta\theta/\lambda^2$,
hence equation \ref{angle} becomes:
\begin{equation}
RM = \frac{e^3}{8\pi^2 m_e ^2 c^3 \epsilon_\circ}\int_0^z n_eB_{\|}dl
\end{equation}
where RM is in rad m$^{-2}$.
\index{magnetic field}

Observationally it is not possible to measure the electron density or
the magnetic field strength without some prior assumptions, thus the
rotation measure must be calculated from the observed polarisation
angle. The rotation measure a source displays is related to the degree
of rotation of the polarisation position angle over a set frequency
range, see equation \ref{rmeq}. To gain an accurate value of the
rotation measure three frequencies (1.4 GHz, 1.6 GHz and 4.8 GHz) are
used to overcome the $n\pi$ ambiguities discussed in chapter
\ref{rmprobs} that arise when fitting to the observed polarisation angles
\citep{Simard,Rudnick}.

The observed RM and the degree of polarisation in a source may be
caused by the presence of plasma either inside the radio source itself
(internal depolarisation) or by a Faraday screen in between the source
and the observer (external depolarisation). In the latter case the
screen may be local to the radio source or within the Galaxy, or
both. Only in the case of an external Faraday screen {\it local} to
the radio source do the measurements contain any information on the
source environment.\index{Faraday screen}

\index{Galactic component in RM} \index{variations in RM} 

The average RM observed in my sources is consistent with a Galactic
origin \citep{l87}. This is also consistent with the absence of any
significant differences of RM between the samples (see Table
\ref{averageRM}). However, large variations of RM are observed within
individual lobes on small angular scales, e.g. dRM and $\sigma_{RM}$,
are probably caused by a Faraday screen local to the source
\citep{l87} and therefore must be corrected for the source redshift by
multiplying by a factor $(1+z)^2$. This allows a valid comparison between
sources. The variation of RM on angular scales of order 10s of
arcseconds, i.e. between the two lobes of a source, dRM, may still be
somewhat influenced by the Galactic Faraday screen. Nevertheless, the
large variations of RM found on arcsecond scales measured by $\sigma
_{RM}$ suggest an origin local to the source.
\begin{table}[!h]
\centering
\begin{tabular}{ l r@{.}l@{ $\pm$ }r@{.}l  r@{.}l@{ $\pm$ }r@{.}l  r@{.}l@{ $\pm$ }r@{.}l } \hline
Average &\multicolumn{12}{c}{Sample} \\
Prop.&\multicolumn{4}{c}{A}& \multicolumn{4}{c}{B}&\multicolumn{4}{c}{C}\\\hline 
z&1&06&0&04 				& 1&11& 0&05 		& 0&41 & 0&01  \\
P$_{151}$& (1&02&0&06)$\times10^{27}$ & (1&01&0&07)$\times10^{28}$&
(8&02&0&40)$\times10^{26}$\\
D&220&34&46&40 				& 262&69& 55&41 	& 399&57 & 122&50 \\
RM&  29&03 &12&81 			& 26&41 &7&74 		&16&24& 8&33\\
{\it dRM}&97&85 &36&41 			&115&96 &42&81 		&50&77&43&29\\
{\it $\sigma_{RM}$}&115&62 &33&43 	& 122&94 &28&48 	& 29&45 &9&98\\\hline
\multicolumn{2}{c}{ }&\multicolumn{3}{l}{Average} &\multicolumn{8}{c}{Sample}\\
\multicolumn{2}{c}{ }&\multicolumn{3}{l}{Prop.}&\multicolumn{4}{c}{A+B}&\multicolumn{4}{c}{A+C}\\
\cline{3-13}
\multicolumn{2}{c}{ }&\multicolumn{3}{l}{z}&  1&08 &0&03 		&0&75 & 0&08 \\
\multicolumn{2}{c}{ }&\multicolumn{3}{l}{P$_{151}$}& (5&57&1&16)$\times10^{27}$ & (8&97&0&77)$\times10^{26}$\\
\multicolumn{2}{c}{ }& \multicolumn{3}{l}{D}& 241&51 &35&43 	& 304&69 & 64&50 \\
\multicolumn{2}{c}{ } & \multicolumn{3}{l}{RM}&  27&64&7&05 			&23&55 &8&07\\
\multicolumn{2}{c}{ }&  \multicolumn{3}{l}{{\it dRM}}&107&44&27&62 			&77&68&27&54  \\
\multicolumn{2}{c}{ }&  \multicolumn{3}{l}{{\it $\sigma_{RM}$}}&119&49 &21&11	&78&69 &22&36\\
\cline{3-13}
\end{tabular}
\caption[Mean rotation measure, differential rotation measure and rms
variation of the rotation measure, averaged over the samples.]  {Mean
properties of the sources averaged over each sample with the
associated error. Differential rotation measure (dRM) is derived by
taking the difference of the rotation measure between the two lobes of
each source and then averaging this difference over each
sample. $\sigma_{RM}$ is the variation of the RM over the
source. Properties in italics are in the source frame of
reference.}\label{averageRM}
\end{table}
There is no statistically significant trend of dRM with respect to
redshift or radio luminosity as is seen in Table \ref{averageRM}. This
is again consistent with a Galactic origin of the RM properties of the
sources on large scales. On small angular scales the variation of RM
as measured by its rms variation, $\sigma _{RM}$, shows a significant
trend between the low redshift sample (C) and the high redshift
samples (A: 2.6$\sigma$, B: 3.3$\sigma$).  The exclusion of 3C 457 due
to $n\pi$-ambiguities in the rotation measure (see Section
\ref{rmprobs}) from the rotation measure averages could bias the
sample C results towards a small value of $\sigma _{RM}$. In
principle, similar ambiguities could also affect other sources in
sample C. However, as stated above the rotation measure is found to
vary smoothly across the other sources in this sample. The exclusion
of one source will not remove the trend found here. There is no
significant difference between the low and high radio luminosity
samples A and B, suggesting that at least the range of rotation
measures on small angular scales in a source {\it does} depend on the
source redshift and not the source radio luminosity.

\section{Depolarisation}\label{galactic}
\index{depolarisation}
The observed depolarisation depends on the amount of polarised flux at
two frequencies but also on the total intensity flux at the same two
frequencies. It is defined as:
\begin{equation}\label{dp}
DM^{4.8\, {\rm GHz}}_{1.4\, {\rm GHz}} = \frac{ PF_{4.8\, {\rm
      GHz}}}{PF_{1.4\, {\rm GHz}}}=
\frac{m_{6\, {\rm cm}}}{m_{20\, {\rm cm}}},
\end{equation}
where the $PF_{\nu}$ is the fractional polarisation, (polarised
flux)/(total intensity flux), at a given frequency, $\nu$, and
$m_{\lambda}$ is the percentage polarisation at an observed wavelength
$\lambda$.

\index{Faraday screen} \index{Burn's law} 

The observed depolarisation also depends on the distribution of the
Faraday depths covered by the projected area of the telescope beam,
i.e. the variation of RM on the smallest angular scales.
If the depolarisation is caused by a local but external Faraday screen
and the distribution of Faraday depths in this screen is Gaussian with
standard deviation $\Delta$, then \citep[e.g.][]{burn}

\begin{equation}\label{pp}
m_{\lambda} = m_0 \exp \left\{-2 \Delta ^2 \left[\lambda / \left( 1+z
\right) \right] ^4 \right\},
\end{equation}

\noindent where, $m_0$ is the initial percentage
polarisation before any depolarisation and $z$ is the redshift of the
source. Since I measure $m_{\lambda}$ at two observing frequencies
(1.4\,GHz and 4.8\,GHz), equation (\ref{pp}) can be solved for $\Delta$
as a function of the depolarisation measure,

\begin{equation}\label{delta}
\centering \Delta=\sqrt\frac{(1+z)^4\ln
DM^{4.8\,{\rm GHz}}_{1.4\,{\rm GHz}}}{2(\lambda_{1.4\,{\rm
      GHz}}^4-\lambda_{4.8\, {\rm GHz}}^4)} \hspace{0.5cm}{\rm
\,rad\,m}^{-2}.
\end{equation}

I now drop the GHz subscript.

For each source, it is possible to compare the value of $\Delta$ as
derived from the measured depolarisation with the observed rms of the
rotation measure, $\sigma_{RM}$. If the Faraday dispersion is less
than $\sigma_{RM}$, then the observations are consistent with an
external Faraday screen \citep{gc91a}.
\begin{figure}[!h]
\centerline{\psfig{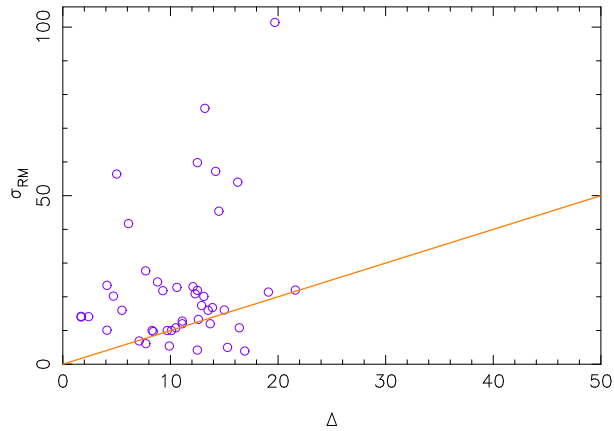}}
\caption[Plot of the Faraday dispersion, against the
rms of the rotation measure for each source.]
{Plot of the Faraday dispersion, $\Delta$, against the
rms of the rotation measure for each source.}\label{external}
\end{figure}\index{Faraday dispersion}
Figure \ref{external} displays the Faraday dispersion, $\Delta$, for
each lobe of the sources against the rms of the rotation measures
observed. It is evident from the plot that the value of $\sigma_{RM} >
\Delta$ for most components, see Tables \ref{ADM} to \ref{CDM} 
but there are a few sources where this is not the case.  However,
these components belong to sources where the depolarisation or
rotation measure is only determined reliably for a few pixels so an
accurate value is not obtainable for $\sigma_{RM}$ or $ \Delta$.
There is little correlation between $\Delta$ and $\sigma_{RM}$.  This
is a strong indicator that the Faraday medium responsible for
variations of RM on small angular scales, and thus for the
polarisation properties of the sources, is consistent with being
external but local to the sources'.
\begin{table}
\centering
\begin{tabular}{  r@{ }l cr@{.}l@{$\pm$}rcr@{.}l r@{.}l }\hline
     \multicolumn{2}{c}{Source}   & Component & \multicolumn{3}{c}{RM}  
& DM$^{4.8}_{1.4}$
& \multicolumn{2}{l}{Average}& \multicolumn{2}{l}{$\sigma_{RM}$} \\
&   & & \multicolumn{3}{c}{}  &  &\multicolumn{2}{c}{$\Delta$}
&\multicolumn{2}{c}{ }\\
&   & & \multicolumn{3}{c}{(rad m$^{-2}$)}& 
& \multicolumn{2}{c}{(rad m$^{-2}$)} 
& \multicolumn{2}{l}{(rad m$^{-2}$)}\\ \hline 
      6C& 0943+37  &  W & 1&6 &2  	& 2.19 		&15&01		&16&1\\
   	&	& E  & -19&1 &6   	& 3.86		&19&7  		&101&4\\
     6C& 1011+36  &  N  & 30&8&5 	&1.53		& 11&1		&12&8
    \\ 	&	& S   &12&6 &4 		& 1.06		&  4&1		&10&1\\
      6C& 1018+37  &  NE &0&85& 3 	& 1.23		&7&71		& 6&1
    \\ 		&& SW &11&2 & 2 		& 2.41 		& 15&9		 &2&9 \\
      6C& 1129+37  &  NW &-19&3 & 4	&2.51		& 16&3 		&54&0
    \\  	&& SE &0&08 &3 		& 5.09		& 21&61		&22&0\\
      6C& 1256+36  &  NE & 5&9 &3	 & 1.35		&  9&3		&21&8
     \\  & 	& SW &15&4 &3  		& 1.01 		& 1&7		&14&0\\
      6C& 1257+36  &  NW &-115&3 & 9 	& 1.27 		&  8&3		&10&0
    \\  	&& SE &  -115&6 & 10  	& 1.88		& 13&5		&16&0\\
      7C& 1745+642 &  N &\multicolumn{3}{c}{--} &1.85	& 13&3	&\multicolumn{2}{l}{--} 
     \\		&& Core &65&4 & 4 	&  1.69		& 12&3		&20&9
    \\ 		&& S &12&8 & 3		& 0.91		&\multicolumn{2}{l}{--} &3&1\\
      7C& 1801+690 &  N &44&8 &3		&1.11		&  5&5		&16&0
    \\		&& Core & 30&3 &3	& 0.45		&\multicolumn{2}{l}{--} &1&80\\
     		&& S &20&8&2		  & 1.39	&  9&7		&10&0\\
      7C& 1813+684 &  NE &13&9& 3 	& 0.81		&\multicolumn{2}{l}{--}&85&0
    \\ && SW & -68&4& 7 & 		1.14   		& 6&1  		&41&7  \\\hline
\end{tabular}
 \caption[Derived properties of sample A] {Properties of the sample A
radio source components. Errors are 5\% or less unless stated
otherwise. The depolarisation measures are the mean values of the
ratio of the fractional polarisation between approximately 4800 MHz
and 1465 MHz for each component. The rotation measures are the mean
values between approximately 4800 MHz, 1665 MHz and 1465 MHz and are
quoted in the observer's frame of reference.  The Faraday dispersion,
$\Delta$, is given in equation \ref{delta}. $\sigma_{RM}$ is the rms
in the rotation measure.  All mean values take into account pixels
above 5$\sigma_{rms}$ at 4.8 GHz and above 3$\sigma_{rms}$ at 1.4
GHz.}\label{ADM}
\end{table}
\begin{table}

  \centering 

\begin{tabular}{  r@{ }l cr@{.}l@{$\pm$}rcr@{.}l r@{.}l }\hline
     \multicolumn{2}{c}{Source}   & Component & \multicolumn{3}{c}{RM}  & DM$^{4.8}_{1.4}$
& \multicolumn{2}{l}{Average}& \multicolumn{2}{l}{$\sigma_{RM}$} \\

 &   & & \multicolumn{3}{c}{}  &  &\multicolumn{2}{c}{$\Delta$}&\multicolumn{2}{c}{ }\\
&         & & \multicolumn{3}{c}{(rad m$^{-2}$)}& & \multicolumn{2}{c}{(rad m$^{-2}$)} 
& \multicolumn{2}{l}{(rad m$^{-2}$)}\\ \hline 

3C& 65  &  W &-82&6 &7		&3.57	& 19&1	&21&4   \\
&       & E  &-86&1 & 6		& 1.28&  8&4 	& 9&7\\
3C& 252  &  NW  &15&7& 6	& 1.08 &  4&7	&20&2\\
&        & SE  &58&5 & 6	& 2.07 & 14&5 	&45&4\\
3C& 267  &  E &-9&6 & 3 	&1.31&  8&8	&24&4 \\
&    & W  &-21&5& 3 		& 1.06&  4&1	&23&4    \\
3C& 280  & E &-37&7& 8		& 1.82& 13&1	&20&1\\
&      & W&-7&5 &4 		& 1.54 & 11&1	&1&5 \\
3C& 324  &  NE& 22&1& 4		&  1.66&12&1	&23&0 \\
&    & SW &43&0& 5		& 1.95& 13&9	&16&8 \\
4C& 16.49 &  N &-4&3 &5		&  0.64& \multicolumn{2}{l}{--} 	&56&4\\
&     & Jet &30&1 &4		& 2.01& 14&2 	&57&2\\
&        &Core &  \multicolumn{3}{c}{--}  & 0.77& \multicolumn{2}{l}{--}  & \multicolumn{2}{l}{--}\\
&     & S  & 0&9 &3		& 1.09 & 5&0 	&56&4\\
3C& 68.1     & N  &-26&6& 5	&1.23&  7&7	&27&7 \\
&    &  S & 57&9 & 2 		& 1.73 & 12&5 	&59&8\\
3C& 265      & NW&42&2&6	&  1.72&12&5	&21&9\\
&    &    SE&32&8&3		&  1.48 &10&6 	&22&8  \\
3C& 268.1    &  E & 21&7 &5	&1.47& 10&5	&10&8\\
&    &  W   & 26&8 &6 		& 1.74     & 12&6 &13&3 \\\hline
\end{tabular}
 \caption[Derived properties of sample B]{As Table \ref{ADM} but for sample B.}\label{BDM}
\end{table}

\begin{table}
  \centering 
\begin{tabular}{  r@{ }l cr@{.}l@{$\pm$}rcr@{.}l r@{.}l }\hline
     \multicolumn{2}{c}{Source}   & Component & \multicolumn{3}{c}{RM}  & DM$^{4.8}_{1.4}$
& \multicolumn{2}{l}{Average}& \multicolumn{2}{l}{$\sigma_{RM}$} \\

 &   & & \multicolumn{3}{c}{}  &  &\multicolumn{2}{c}{$\Delta$}&\multicolumn{2}{c}{ }\\
&         & & \multicolumn{3}{c}{(rad m$^{-2}$)}& & \multicolumn{2}{c}{(rad m$^{-2}$)} 
& \multicolumn{2}{l}{(rad m$^{-2}$)}\\ \hline 

3C& 42  & NW &  -2&4& 5		& 1.02 & 2&4	&14&1\\
       && SE & 5&0& 5		& 1.01 & 1&7	&14&2\\
4C& 14.27  &  NW  &-13&0&4	&  1.41& 9&9	&5&4
    \\&& SE  &  -17&3 &5	& 1.03& 2&9	&4&8   \\
3C& 46   & NE &-4&8& 5		&  1.27&16&9	&3&9
    \\&&SW &-2&9 &1		& 1.14 & 12&5	&4&   \\
3C& 457  &NE& \multicolumn{3}{c}{--}  & 1.01 & 2&1  &   \multicolumn{2}{l}{--} \\
       &&SW & \multicolumn{3}{c}{--}            &1.19& 5&6   &  \multicolumn{2}{l}{--}  \\
3C& 351  &  NE & 1&00		& 5 & 0.84 &  \multicolumn{2}{l}{--}	&13&\\
    &&Diffuse & -8&7 & 1 	&  1.93 & 13&7	&12&\\
    &&Core & -0&53&2		&2.27 & 15&3	&5&\\
    && SW &  4&4 & 3		&2.56& 16&4	&10&8\\
3C& 341   &  NE&20&3&5		& 1.43& 10&1	&10&0
        \\ && SW  & 18&2&5	&1.53 & 11&1	&12&0\\
3C& 299      & NE &-126&3&8	& 1.84& 13&2	&75&9
    \\&&  SW&16&0 &5		& 1.19& 7&10	& 6&9\\
3C& 16       & NE&  \multicolumn{3}{c}{--} & 2.20 & 15&0	& \multicolumn{2}{l}{--} \\
    && SW   &-4&3 &3		& 1.79 &  12&9	& 17&4 \\ \hline
\end{tabular}
 \caption[Derived properties of sample C]{As Table \ref{ADM} but for sample C.}\label{CDM}
\end{table}
Comparing the average depolarisation of individual samples with each
other I find only very weak trends with redshift or radio luminosity. When
the samples are averaged together I find a trend in depolarisation
with redshift but none with radio luminosity. Samples A+C (low radio luminosity)
are statistically identical to sample B (high radio luminosity), suggesting
that there is no trend with radio luminosity in my sources. By considering
the averaged depolarisation of samples A+B (high redshift) compared
with that of sample C (low redshift) there is a weak trend,
$1.7\sigma$, with redshift. This trend is echoed in the dDM values
($1.9\sigma$). This may confirm the results of \citet{kcg72}: redshift
is the dominant factor compared to radio luminosity in determining the
depolarisation properties of a source but since the significance levels
are low it is not possible to say with any confidence.

\begin{table}[!h]
\centering 
\begin{tabular}{ l r@{.}l@{ $\pm$ }r@{.}l  r@{.}l@{ $\pm$ }r@{.}l  r@{.}l@{ $\pm$ }r@{.}l } \hline
Average &\multicolumn{12}{c}{Sample} \\
Prop.&\multicolumn{4}{c}{A}& \multicolumn{4}{c}{B}&\multicolumn{4}{c}{C}\\\hline 
z&1&06&0&04 		& 1&11& 0&05 		& 0&41 & 0&01  \\
P$_{151}$& (1&02&0&06)$\times10^{27}$ & (1&01&0&07)$\times10^{28}$&
(8&02&0&40)$\times10^{26}$\\
D&220&34&46&40 		& 262&69& 55&41 	& 399&57 & 122&50 \\
DM&1&82 &0&32 		& 1&61&0&13 		& 1&42& 0&12 \\
 dDM& 0&93 &0&26 	& 0&61 &0&22 		& 0&43 &0&18\\\hline
\multicolumn{2}{c}{ }&\multicolumn{3}{l}{Average} &\multicolumn{8}{c}{Sample}\\
\multicolumn{2}{c}{ }&\multicolumn{3}{l}{Prop.}&\multicolumn{4}{c}{A+B}&\multicolumn{4}{c}{A+C}\\
\cline{3-13}
\multicolumn{2}{c}{ }&\multicolumn{3}{l}{z}&  1&08 &0&03 		&0&75 & 0&08 \\
\multicolumn{2}{c}{ }&\multicolumn{3}{l}{P$_{151}$}& (5&57&1&16)$\times10^{27}$ & (8&97&0&77)$\times10^{26}$\\
\multicolumn{2}{c}{ }& \multicolumn{3}{l}{D}& 241&51 &35&43 	& 304&69 & 64&50 \\
\multicolumn{2}{c}{ } & \multicolumn{3}{l}{DM}&1&71 &0&17 		& 1&63 &0&18 \\
\multicolumn{2}{c}{ }& \multicolumn{3}{l}{dDM}&  0&77& 0&17 	& 0&69 & 0&17\\
\cline{3-13}
\end{tabular}
\caption[Mean depolarisation and differential depolarisation of the sources averaged over the samples.]
{Mean properties of the sources averaged over each sample with the
associated error. Differential depolarisation (dDM) is derived by
taking the difference of the depolarisation between the two lobes of
each source and then averaging this difference over each
sample.}\label{average}
\end{table}

\begin{figure}[!h]
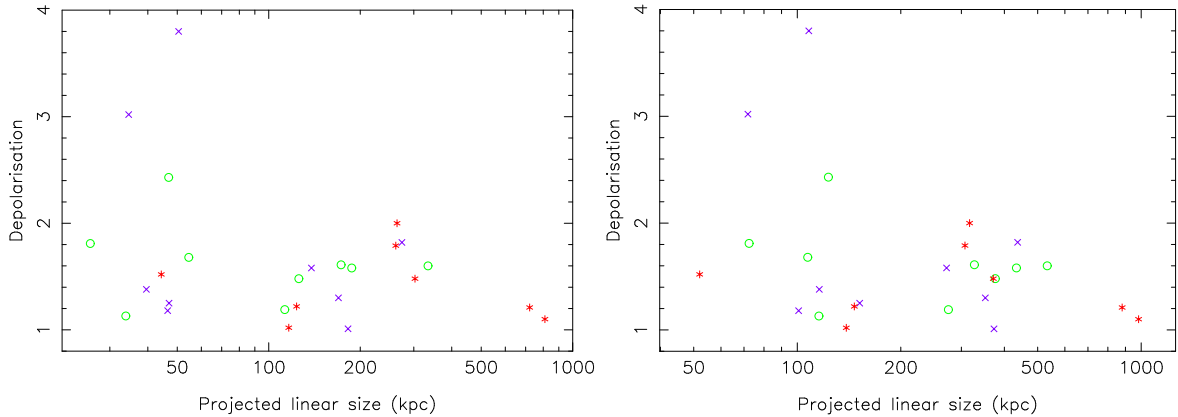

\centerline{\psfig{file=dep3_c.ps,width=7.8cm,angle=270}
\psfig{file=dep4_c.ps,width=7.6cm,angle=270}}
\caption[Average depolarisation against projected linear size
]{Average depolarisation against projected linear size
(kpc) for all 3 samples. Symbols as in Figure \ref{pzplot}. Figure (a)
assumes $H_o= 75$ kms$^{-1}$Mpc$^{-1}$, and $\Omega_{m} = 0.5$,
$\Omega_{\Lambda} = 0$. Figure (b) assumes $H_o= 75$
kms$^{-1}$Mpc$^{-1}$, and $\Omega_{m} = 0.35$, $\Omega_{\Lambda} =
0.65$.}\label{depolsize}
\end{figure}

\citet{strom,sj88,prm89,ics98} find an anti-correlation of
depolarisation with linear size.  Figure \ref{depolsize} shows that
there is a weak anti-correlation between physical source size and
depolarisation.  However, this trend is due to the two largest sources
in sample C, 3C 46 and 3C 457. Removing these sources yields an
average depolarisation measure of $1.50\pm0.15$ for sample C, which is
not significantly different from the average with these two sources
included. So I can rule out the possibility that the larger
depolarisation at high redshift is caused by selecting preferentially
smaller sources at high redshift.

\subsection{Laing--Garrington effect}\label{LGeffect}
\index{Laing-Garrington effect} Assuming that a radio source does not
lie flat to the observers line--of--sight and that both lobes are
embedded in the same Faraday medium then one lobe will lie further
away. The polarised flux from this lobe will have to travel through a
larger path length and will become more depolarised than the nearer
lobe. This creates an asymmetry in the depolarisation that is known as
the Laing--Garrington effect. 8 sources show dDM $\ge 1$ over the
source and a further 6 show $0.5 \le {\rm dDM} <1$. This implies that
14 sources out of 26 sources show a significant asymmetry in the
depolarisation of their lobes.  I would expect a proportion of the
sources to be observed at angles considerably smaller than
90$^{\circ}$ to my line-of-sight. Therefore it is not surprising that
so many sources are found to be candidates for the Laing-Garrington
effect.

\subsection{Depolarisation shadows}
\begin{figure}[!h]
\centerline{\psfig{file=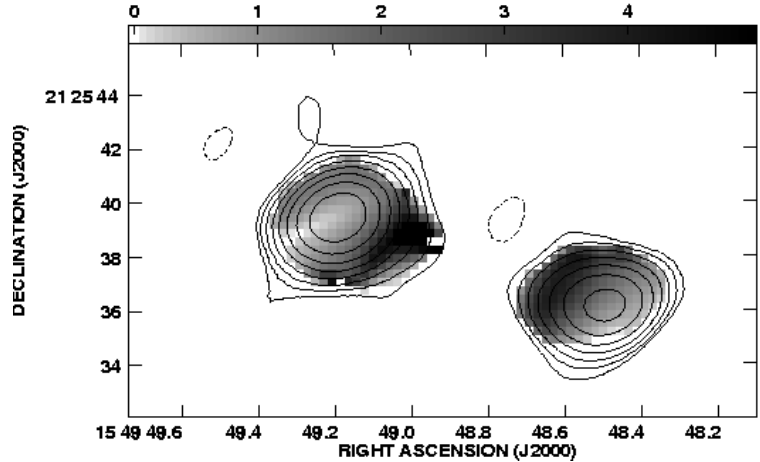,width=10cm}}
\caption[Example of a depolarisation shadow]{Depolarisation map of the radio source 3C 324 between 
4848 MHz and 1465 MHz. All contours are at $5 \sigma$
at 4848 MHz (1.5 mJy beam$^{-1}$)$\times$(-1, 1, 2, 4...,1024) with a beam
size of 2.2'' $\times$ 1.6''.}
\label{3c324dep}
\end{figure}

\index{depolarisation shadows}
Depolarisation shadows (regions where the depolarisation is
appreciably greater than in the surrounding area) are seen in a few
sources, e.g. Figure \ref{3c324dep}, and may be real features.  These were
first found by \citet{feb89} in a study of Fornax A. 
Depolarisation shadows can be caused by the parent
galaxy as in the case of 3C 324 or by an external galaxy in the
foreground of the source, causing a depolarising silhouette
\citep{blr97}. Several sources show signs of 
depolarisation shadows in at least one of the lobes. Of these,
6C 1256+36 was observed to lie in a cluster by \citet{reh98}, as were
3C 65 and 3C 324 \citep{b00}.  3C 324 has been observed by \citet{brb98}
in the sub-mm ($850\mu$m) and found a large dust mass centred around
3C 324. The host galaxy was shown to be the cause of the very strong
depolarisation, \citep{bcg98}.
\section{Using Burn's law}
\subsection{Polarisation}
\index{polarised flux} 
\index{Burn's law}
So far I have only considered the percentage polarisations of a source
measured in the observing frame. Variations of the RM on small angular
scales which determine the degree of polarisation are caused in
Faraday screens local to the sources. The trend with redshift may
therefore simply reflect the different shifts of the observing
frequency in the source rest-frame for sources at low and high
redshift. Using Burn's law in the form of equations (\ref{pp}) and
(\ref{delta}) it is possible to determine, for each source, the
percentage polarisation expected to be observable at a frequency
corresponding to a wavelength of 5\,cm in its rest-frame. This
wavelength was chosen as it lies close to the wavelengths already
observed in all the sources. The results for individual sources are
presented in Table \ref{hmm} and the sample averages are summarised in
Table \ref{averagez=1}. Again sources at low redshift (sample C) are
slightly more polarised than sources at high redshift (sample A:
1.9$\sigma$, sample B: 2.1$\sigma$).  As before, there is no trend
with radio luminosity, indicating that the trend with redshift is
dominant. This is not caused by pure Doppler shifts of the observing
frequencies.
\begin{table}[!h]
\centering 
\begin{tabular}{ l@{ }l ccr@{.}lcccr@{.}l}\hline
\multicolumn{2}{c}{Source} & z && \multicolumn{2}{c}{$\Delta_z$}	&DM$_{z=1}$& dDM$_{z=1}$
&\multicolumn{3}{c}{Polarisation}\\
&& &\multicolumn{2}{c}{ } && &&\multicolumn{3}{c}{$\lambda_{\rm rest}=5$\,cm}\\\hline
6C& 0943+37 &1.04 &	 &74&12 &	 3.31 & 1.98 	&&3&10 \\
6C& 1018+37 &0.81 & 	&42&95 &	 1.49 & 0.52		&&8&32\\
6C& 1011+36 & 1.04 &	 &36&11 & 1.33 & 0.66		&&5&63\\
6C& 1129+37 &1.06 & 	&83&07 &	 4.49& 3.43		&&3&09\\
6C& 1256+36 &1.07 & 	&29&53 & 	1.21 & 0.40	&&8&26\\
6C& 1257+36 &1.00 & 	&45&51 & 	1.58 & 0.61	&&9&39\\
7C& 1801+690 &1.27 & 	&41&24 &	 1.45 & 0.54		&&3&33\\
7C& 1813+684 &1.03 &  	&6&96 &	 1.01& 0.25		&&8&47\\ \vspace{3mm} 
7C& 1745+642 &1.23 & 	&47&81 &	 1.65& 1.36		&&7&32\\
 3C& 65 &1.18 & 	&75&33 & 	3.50 & 4.61&	&6&26\\
3C& 252 &1.11 & 	&50&66 & 	1.76 & 1.36	&&6&36\\
3C& 267 &1.14 & 	&32&34 & 	1.26 & 0.35	&&5&04\\
3C& 280 &1.00 & 	&48&79 & 	1.68 & 0.28	&&5&48\\
3C& 324 &1.21 &		& 54&34 &2.42 & 0.58		&&4&74\\
3C& 265 &0.81 & 	&35&56 & 	1.37 & 0.14	&&5&05\\
3C& 268.1 &0.97 &	 &42&13 & 	1.57 & 0.25	&&3&07\\
 3C& 68.1 &1.24 &	 &45&11 & 	1.85 & 0.98	&&10&20\\\vspace{3mm} 
 4C& 16.49 &1.29 &	 &31&04 & 1.23 & 0.47		&&6&76 \\
 3C& 42 &0.40 &  	&4&59 & 		1.00 & 0.01	&&9&85\\
 4C& 14.27 &0.39 & 	&14&34 &	 1.05 & 0.08		&&9&08\\
 3C& 46 &0.44 & 	&15&06 & 	1.05 & 0.03	&&12&45\\
3C& 457 &0.43 & 	&10&50 &		 1.03 & 0.04	&&12&88\\
3C& 351 &0.37 & 	&23&82 & 	1.14 & 0.15	&&8&59\\
3C& 341 &0.45 & 	&21&91 & 	1.11 & 0.02	&&18&57\\
 3C& 16 &0.41 & 	&27&54 & 	1.19& 0.06	&&6&64\\
3C& 299 &0.37 & 	&20&20 & 1.10 & 0.10		&&1&52\\\hline
\end{tabular}
 \caption[Recalculated polarisation parameters using Burn's law.]
{Recalculated average depolarisation and dDM for each source if it was
located at $z=1$ and average percentage polarisation of all sources if
it was emitted at 5 cm in the rest frame.}\label{hmm}
\end{table}
\subsection{Depolarisation}\label{sec:bdep}
\index{depolarisation} Cosmological Doppler shifts of the observing
frequencies influence the trend of DM with redshift, and in fact the
true trend is stronger than that naively observed. To demonstrate this
equation (\ref{delta}) can be used to derive the standard deviation of
Faraday depths, $\Delta$, for each source. By setting $z=1$ it is
possible to then rescale all the depolarisations, DM$^{4.8}_{1.4}$, to
the same redshift.  This allows all 3 samples to be compared without
any bias due to pure redshift effects (see Tables \ref{hmm} and
\ref{averagez=1}).  If there was no intrinsic difference between the
high-redshift and low-redshift samples then I would expect these
values to be consistent with each other. This is evidently not the
case. The high redshift samples are, on average, significantly more
depolarised (sample A: 2.2$\sigma$, sample B: 3.2$\sigma$) than their
low-redshift counterparts (sample C). Comparing sample A with sample B
I find no trend with radio luminosity. However, a note of caution must
be issued as the corrections applied use Burn's law and may actually
be too large. Considering the precorrected and the corrected values
together it is obvious that there is a connection between redshift and
depolarisation but there is no significant trend of depolarisation
with radio luminosity. There is also a connection between the
difference in the depolarisation, dDM, and the redshift of the source,
but no significant trend of the difference in the depolarisation with
the radio luminosity of the source. As noted in section \ref{rmprobs}
regions with signal-to-noise $< 3\sigma$ were blanked in the map
production. In individual sources blanking of low S/N regions in the
polarisation maps will cause the measured depolarisation to be
underestimated. Sources in sample A are more affected by this problem
than objects in the other samples. Therefore I probably underestimate
the average depolarisation in sample A implying that the trend with
redshift could be even stronger than my findings suggest.

\begin{table}[!h]
\centering
\begin{tabular}{ l r@{ $\pm$ }l r@{ $\pm$ }l r@{ $\pm$ }l r@{ $\pm$ }l  r@{ $\pm$ }l } \hline
Average & \multicolumn{10}{c}{Sample} \\
Prop.&	\multicolumn{2}{c}{A}&	\multicolumn{2}{c}{B} &	\multicolumn{2}{c}{C}&
	\multicolumn{2}{c}{A+B}&	\multicolumn{2}{c}{A+B}\\\hline 
DM$_{z=1}$&1.95 &0.39 & 1.85& 0.24 & 1.08 & 0.02 & 1.90 & 0.22 & 1.54 & 0.23 \\
 dDM$_{z=1}$& 1.08 &0.34 & 1.00 & 0.47 & 0.06 & 0.02& 1.04& 0.28 & 0.60 & 0.22 \\
PF$_{\lambda_{rest}=5\,{\rm cm}}$&6.32 & 0.86 & 5.88 &0.65 & 9.82 & 1.85 & 6.10 & 0.53 & 7.97 & 1.04 \\\hline
\end{tabular}
\caption[Shifted average DM and dDM.]  {Average DM and dDM now
shifted so that all the measurements are taken at $z=1$.  Average
polarisation of all the sources, shifted to a common rest frame
wavelength of 5 cm using equation \ref{delta}.}\label{averagez=1}
\end{table}

The trends of percentage polarisation and of depolarisation with
redshift are probably related in the sense that a lower degree of
depolarisation at low redshift also leads to a higher observed degree
of polarisation. Clearly a variation of the initial polarisation,
$m_0$, with redshift would lead to variations of the observed
$m_{\lambda}$ independent of the properties of any external Faraday
screen. Therefore both trends could also be caused by a significantly
higher level of $m_0$ for sources at low redshift (sample C). Using
equation (\ref{pp}) I find $m_0=9.3\pm1.1$ for sample A,
$m_0=9.1\pm1.0$ for sample B and $m_0=13.5\pm2.5$ for sample C. The
uncertainties associated with the use of Burn's law in extrapolating
from my observations to $\lambda =0$ are large. There is no difference
between the average initial polarisation of the sources in samples A
and B. The difference found for $m_0$ at low redshift (sample C)
compared to high redshift (samples A and B) is small compared to the
difference found for the depolarisation comparing the same
samples. This suggests that the differences in percentage polarisation
and depolarisation are due to variation with redshift of the Faraday
screens local to the sources rather than to differences in the initial
degree of polarisation. However note, that the variation of $m_0$ is
not significantly smaller than the trend of percentage polarisation
with redshift.

Burn's law predicts a steep decrease of percentage polarisation with
increasing observing frequency. Although the decrease may well be
`softened' by geometrical and other effects in more realistic source
models \citep{laing}, the value of DM may be small for sources in
which the observing frequencies are lower than the frequency at which
strong depolarisation occurs. For such sources I would expect to
measure a low value of DM associated with a low percentage
polarisation. \citet{ag95} show that one of my sources, 3C 299 (sample
C), is strongly depolarised between 1.6\,GHz and 8.4\,GHz with almost
all of the depolarisation taking place between 4.8\,GHz and
8.4\,GHz. I measure only a very low percentage polarisation for this
source and the average value of DM$^{4.8}_{1.4} = 1.5$ is also lower
than DM$^{8.4}_{1.6} \sim 4$ as measured by \citet{ag95}. If many of
the sources at low redshift are affected by strong depolarisation at
frequencies higher than 4.8\,GHz, then this may cause the trend of DM
with redshift noted above. The absence of any other sources with very
low degrees of polarisation combined with a low value for DM, at least
in the low redshift sample C, argues against this bias. In fact,
\citet{tab} show that the sources 3C 42, 3C 46, 3C 68.1, 3C 265, 3C 267,
3C 324 and 3C 341 depolarise strongly only at frequencies lower than
my observing frequencies. 3C 16, 3C 65, 3C 252, 3C 268.1 and 3C 280
do depolarise strongly between 1.4 GHz and 4.8 GHz. In all the sources
mentioned in \citet{tab} none have strong depolarisation at
frequencies higher than 4.8 GHz.  There is no information for 4C
14.27, 4C 16.49, 3C 351 and 3C 457. Those sources which do
depolarise strongly between my observing frequencies, 1.4 and
4.8\,GHz, could in principle have inaccurate values of the rotation
measure because of this. However, the pixels containing most
depolarised regions of the source are likely to have been blanked
because of insufficient signal--to--noise in their polarisation at
1.4\,GHz; the rotation measure is determined from the unblanked (less
depolarised) regions of the source, and will therefore be reliable.

\section{Discussion}
In this chapter I present the complete data set of the three samples
of radio galaxies and radio-loud quasars. The three samples were
defined such that two of them overlap in redshift and two have similar
radio luminosities. Allowing the effects of redshift and radio luminosity on
various source properties to be analysed.

There is little correlation between $\Delta$ and $\sigma_{RM}$,
suggesting that the Faraday medium responsible for variations of RM on
small angular scales, is consistent with being external but local to
the sources. This implies that any trends with radio luminosity and/or
redshift reflect changes of the source environment depending on these
quantities. There is also little correlation between rotation measure
and redshift or radio luminosity which is consistent with a Galactic origin
of the RM properties of the sources on large scales. However, I find
that the rms fluctuations of the rotation measure correlate with
redshift but not radio luminosity to a confidence level of $> 99.9\%$,
determined in the sources' frame of reference.

I find that the polarisation of a source anti-correlates with its
redshift but is independent of its radio luminosity, resulting in the
low redshift sample having much higher degrees of polarisation, in
general. I also detect higher degrees of depolarisation in the high
redshift samples (A and B) compared to sources at lower redshift
(sample C). This suggests that depolarisation is correlated with
redshift. These two results are probably related in that lower
depolarisation at low redshift leads to both lower depolarisation
measurements and also higher degrees of observed
polarisation. According to Burn's law \citep{burn}, this implies an
increase in the source environments of either the plasma density or
the magnetic field strength or both with redshift. Such an
interpretation is also supported by the increased depolarisation
asymmetry of sources at high redshift compared with their low redshift
counterparts.

The findings on the rotation measurements and polarisation properties
of my sources are indicative of an increase of the density and/or the
strength of the magnetic field in the source environments with
increasing redshift.

I do not find a strong correlation between the projected sizes of the
sources and their depolarisation measure as found by
\citet{strom,sj88,prm89,ics98}. 

I find no correlation between the spectral index and redshift or radio
luminosity. However, I do find the difference in the spectral index,
across individual sources, increases for increasing radio luminosity of the
source and also with increasing redshift of a source.

The correlations found are only general trends within the samples. To
completely break the z-P degeneracy a much more rigorous and detailed
approach to the statistical analysis is needed. In the next chapter I
will investigate the observational properties using Principal
Component Analysis and Spearman Rank and determine the significance of
the results presented here.

\newpage
\thispagestyle{empty}
\chapter{Analysing the observations}
\index{analysis of the observations}
\label{statsobs}
\index{Faraday screen}

In this chapter I investigate the trends and correlations of a number
of observational parameters (spectral index, rotation measure and
depolarisation measure) with the `fundamental' parameters, low
frequency radio luminosity, redshift and the physical size of the
radio structure. The observational parameters mainly constrain the
properties of the gas in the vicinity of the radio structure which
acts as a Faraday screen, see sections \ref{rm} and \ref{galactic}.

In my statistical analysis I use Spearman Rank as well as Partial
Spearman Rank and Principal Component Analysis techniques. Thus I am
able to assess whether any observed correlations of parameters with
redshift and radio luminosity are significant in their own right or
whether they simply arise from a Malmquist bias.

The 3 samples can also be used to determine if there are any
asymmetries present between the lobes of sources and if so, what
causes these asymmetries. Traditionally this has been done with
sources in which jets are detected so that the asymmetries between the
jet side and counter--jet side can be studied,
\citep[e.g.][]{gc91a,dtb97,l87}. This type of study allows a direct
insight into the orientation of a source. In my sample only a few
sources have a well defined jet at either 4.8 GHz or 1.4 GHz, so this
type of analysis is not possible. However, I compare the brighter lobe
to the fainter lobe to determine if there are any asymmetries present
in the samples and if possible, explain the underlying physical
cause(s) of the asymmetries.
\section{Statistics}\label{statistics}
\index{Spearman Rank}
\subsection{Spearman Rank}
The Spearman Rank test is a non-parametric correlation test assigning
a rank to given source properties, X and Y and then performing a correlation on
the rank. The statistic is given by
\begin{eqnarray}\label{rs}
r_{\rm XY} = \frac{\Sigma x^r_i y^r_i}{\sqrt{\Sigma x^{r2}_i \Sigma y^{r2}_i}}\\ \nonumber
 = 1-\frac{6\Sigma d^2}{n^3-n}
\end{eqnarray} 
where $d$ is the difference between the rank in the X direction and the
rank in the Y direction and $r_{\rm XY}$ has a Student--t distribution:
\begin{equation}
D_{\rm XY} = \frac{r_{\rm XY}\sqrt{n-2}}{\sqrt{1-r_{\rm XY}^2}}
\end{equation}
\noindent where n is the number of sources.

The null hypothesis states that no correlation is present between two
properties X and Y, corresponding to $r_{\rm XY}$=0 and a (anti--)
correlation corresponds to (-)1. A correlation with a confidence level
above 95\% will show a Student-t value greater than 2 for my sample
size \citep{book_stat}.

Partial Spearman Rank is used to determine if a correlation between
two variables (X, Y) depends on the presence of a third, A
\citep{Macklin}. The null hypotheses are:
\begin{itemize}
\item the A-X correlation is entirely due to the X-Y and the Y-A
independent correlations, and
\item the Y-A correlation is entirely due to the X-Y and the X-A
independent correlations
\end{itemize}
\index{Partial Spearman Rank}
The Partial Spearman Rank correlation coefficient is given by:
\begin{equation}
r_{\rm AX, Y} =
\frac{r_{\rm AX}-r_{\rm XY}r_{\rm AY}}{\sqrt{\left[(1-r_{\rm
	XY}^2)(1-r_{\rm AY}^2)\right]}}
\end{equation}
where $r_{\rm AX}$ is the normal Spearman Rank correlation coefficient
described above.  The significance level for the A-X correlation,
independent of Y is
\begin{equation}
D_{\rm AX,Y}=\frac{1}{2}(n-4)^{\frac{1}{2}}ln\left(\frac{1+r_{\rm
    AX,Y}}{1-r_{\rm AX,Y}}\right).
\end{equation}
To fully determine the relationships between X,Y and A, the triplet
$r_{\rm AX,Y}$, $r_{\rm AY,X}$ and $r_{\rm XY,A}$ must be calculated and their
corresponding significance levels found. By comparing this triplet it
is possible to determine which relationship, if any, between X, Y and
A dominates.
\subsection{Principal Component Analysis, PCA}
\index{Principal Component Analysis}
Principal Component Analysis, PCA, is a multi-variate statistical test
which is described in detail by \cite{d64} and \citet{ef84}. PCA is a
linear self--orthogonal transformation from an original set of $n$
objects and $m$ attributes forming a $n \times m$ matrix, with zero mean
and unit variance, to a new set of parameters, known as principal
components. The principal components of the new dataset are all
independent of each other and hence orthogonal. The first component
describes the largest variation in the data and has, by definition,
the largest eigenvalue of the $n \times m$ matrix. In physical terms
the magnitude of the eigenvalue determines what fraction of the
variance in the data any correlation describes. Thus a strong
correlation will be present in the first eigenvector and will only
strongly reverse in the last. PCA in its simplest terms is an
eigenvector--eigenvalue problem on a transformed, diagonalised and
standardised set of variables. $n$ objects will create an
eigenvector--eigenvalue problem in $n-1$ dimensions. To avoid problems
with the interpretation of my results I do not use more than 4
parameters in any part of the analysis.

\section{Fundamental Parameters}
\index{fundamental parameters}\index{z-P degeneracy}
\label{fundamentals}
In the following I investigate the relationship of the observational
properties of my sources with the `fundamental' properties redshift,
z, radio luminosity at 151 MHz, P$_{151 }$, and physical size,
D$_{source}$. For this I first need to understand the relations
between these three fundamental parameters.  Table \ref{zpd} contains
the associated Spearman Rank results.  Redshift and radio luminosity
are highly correlated, to a significance of $\approx 99.9\%$ but this
is simply due to the way the samples were selected. Using the
statistical techniques described above this correlation can be
isolated from other correlations. In fact, it provides a `bench mark'
for other correlations.
\begin{table}[!h]
\centering
\begin{tabular}{ r@{ }l r@{.}l r@{.}l }\hline
\multicolumn{2}{c}{Parameters}&
\multicolumn{2}{c}{rs value}&\multicolumn{2}{c}{ t value}\\ \hline
z& P$_{151}$&0&69846& 4&78\\ 
z& D$_{source}$&-0&22940&-1&15\\
 P$_{151}$&D$_{source}$&-0&13026&-0&64\\\hline
\end{tabular}
\caption{Spearman rank values for z, P$_{151}$ and D$_{source}$.}\label{zpd}
\end{table}

\begin{figure}[!h]
\centerline{\psfig{file=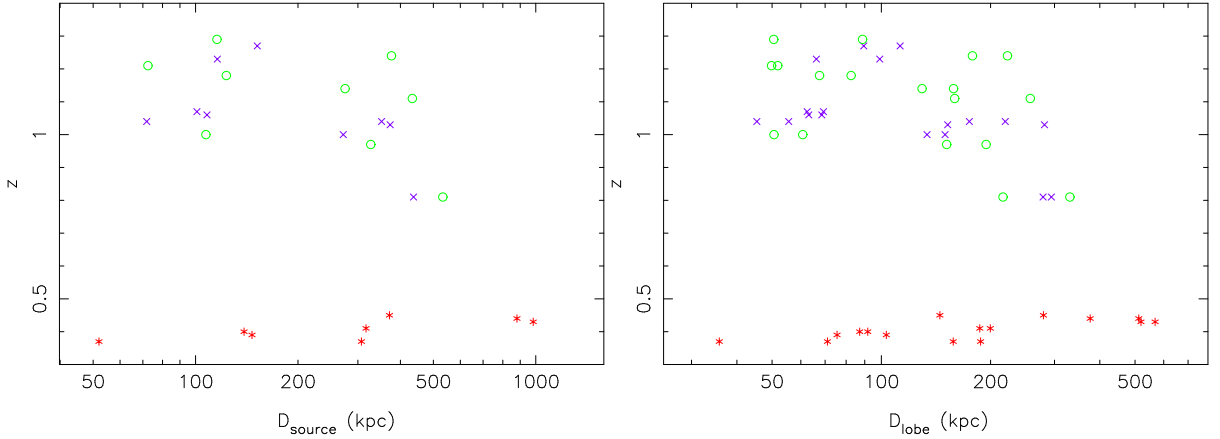,width=16cm}}
\caption[Redshift-size relation] {(a - left) Redshift against average source 
sizes in kpc and (b - right) with individual lobe sizes in kpc. 
Sample A is represented by `$\times$', sample B by `$\circ$' and
sample C by `*'.}\label{zd}
\end{figure}

There is also a weak z - D$_{source}$ anti-correlation, but as Table
\ref{zpd} demonstrates this anti-correlation is not significant. Figure
\ref{zd}(a) shows how the source size is distributed over the redshift
range of my samples. The presence of the two low redshift giants, 3C 46
and 3C 457 are the cause of the weak correlation. By removing these two
giants the anti-correlation completely disappears. However, as the
anti-correlation is below 95\% significant, even with the two giants
included in the sample, any correlations found with size will not be
biased towards lower redshift sources. Figure \ref{zd}(b)
demonstrates that the sizes of individual lobes are evenly distributed
over the redshift range of the three samples and is also not unduly
affected by the presence of the two giants.

Partial Spearman Rank was used to determine the dependencies between
z, P$_{151}$ and D$_{source}$.
\begin{eqnarray}
 r_{z D_{source},P_{151}}=&\hspace{0.3cm} 0.04,\hspace{2cm}& D = 0.20
\nonumber\\ r_{z P_{151},D_{source}} =&\hspace{0.3cm}
0.69,\hspace{2cm} & D = 4.00\nonumber \\ r_{P_{151} D_{source},z}=
&-0.20 ,\hspace{2cm} &D = 0.93 \nonumber
\end{eqnarray}
Using Partial Spearman Rank I find no independent z - D$_{source}$
anti--correlation, proving that my sample is not biased towards
larger sources at lower redshifts, allowing a fair comparison with
higher redshift sources. Interestingly Table \ref{zpd} indicates that
there is no relationship between a sources' radio-luminosity and its
size and this result is confirmed by the Partial Spearman Rank results
at any fixed redshift. In a study by \citet{k89} it was found that
D$_{source} \propto (1+z)^{-3\pm0.5}$ using $\Omega = 1$, but at
constant redshift it was found that D$_{source} \propto P^{0.3\pm
0.1}$. \citet{s93} also found that for his sample of 789 sources, size
predominantly correlated with radio-luminosity and was only a very
weak function of redshift. However it is interesting to note that
\citet{bam88} found that D $\propto (1+z)^{-1.5\pm1.4}P^{-0.03\pm0.3}$
using a sample of steep spectrum radio quasars. My results are
consistent with the results from \citet{bam88}. These results
demonstrate that it is very complicated to disentangle the redshift,
radio-luminosity and size correlation and my sample is simply too
small to add anything significant to this well studied field.

In Tables \ref{rsa} to \ref{pcaRM} the relationships
between the observational parameters with the fundamental
parameters are given.

\section{Observational Parameters}
In order to compare the differential observational parameters
e.g. dDM$_z$\footnote{dDM$_z$ = $|DM_{z_{lobe 1}}-DM_{z_{lobe 2}}|$, see
also the previous chapter.} with averaged properties like DM, all
averaged properties are taken over the entire source as opposed to
individual lobes. This is the only method possible as for each source
I only have one measurement of the differential properties and two
measurements of the average properties (taken from each lobe). In the
case of the RM properties (RM, dRM$_z$ \& \sig) I only use 23 of the
26 sources in the analysis. This is simply due to the fact that
7C 1745+642, 3C 16 and 3C 457 do not have reliable RM information in at
least one of their lobes.
\subsection{Spectral index}
\index{spectral index}
\begin{table}[!h]
\centering
\begin{center}
\end{center}
\begin{tabular}{ l@{ }l r@{.}l r@{.}l c l@{ }l  r@{.}l r@{.}l }
\cline{1-6}\cline{8-13}
\multicolumn{2}{c}{Parameters}&
\multicolumn{2}{c}{rs value}&\multicolumn{2}{c}{ t value}&&
\multicolumn{2}{c}{Parameters}&
\multicolumn{2}{c}{rs value}&\multicolumn{2}{c}{ t value}\\
\cline{1-6}\cline{8-13}
$\alpha_s$&z&-0&12615&-0&62&&d$\alpha$&z&0&15487& 0&77\\
$\alpha_s$&D$_{source}$&0&20889&1&05&&d$\alpha$&D$_{source}$&-0&20957&-1&05\\
$\alpha_s$&P$_{151}$&0&03932& 0&19&&d$\alpha$&P$_{151}$&0&26632&1&35\\
\cline{1-6}\cline{8-13}
\end{tabular}
\caption[Spearman rank values for spectral index parameters]
{Spearman rank values for all spectral index parameters where the
subscript {\it s} indicates that the spectral index is averaged over
both lobes in a source.}\label{rsa}
\end{table}
There are no strong correlations between spectral index, $\alpha_s$,
and any of the fundamental parameters as the Spearman Rank results in
Table \ref{rsa} demonstrates. The PCA analysis confirms these findings
(Table \ref{pcaa}). The difference in the spectral index between the
lobes, d$\alpha$, also shows little correlation with any of the
fundamental parameters. \citet{vv72} and \citet{on89} find a strong
spectral index--radio luminosity correlation, but no corresponding
correlation with redshift. Interestingly \citet{athreya} find a strong
correlation of spectral index with redshift, but no significant
correlation with radio luminosity when they analysed the MRC/ 1Jy
sample. Although I do not find strong support for either trend, this
could simply be due to my small sample size.

A recent study by \citet*{brw99} using the 3CRR, 6CE and 7C samples
found that $\alpha$ anti-correlated not only with radio-luminosity,
but also with source size. They found that in general, larger sources
had a steeper spectral index. Again the fact that I do not find this
trend could simply be due to my much smaller sample size.
\begin{table}
\centering
\begin{center}
\end{center}
\begin{tabular}{l  r@{.}l r@{.}l r@{.}l  r@{.}l }\hline
Parameter  & \multicolumn{2}{c}{1} &  \multicolumn{2}{c}{2}    &  
 \multicolumn{2}{c}{3}  &   \multicolumn{2}{c}{4}     \\\hline
    z & 0&6354 &  -0&2475& 0&0568& -0&7292  \\
    P & 0&5781&  -0&1595& 0&5304&  0&5992  \\
    D & -0&4690&  -0&1168&0&8206&  -0&3051 \\
$\alpha_s$ &-0&2053& -0&9485&-0&2050&   0&1271 \\\hline
Eigenvalue&44&6\% &  24&6\% &20&5\% & 10&2 \%   \\ \hline
    z & 0&5565& -0&0810& 0&5422&-0&6243\\
    P & 0&5487&  0&4404& 0&2629& 0&6602\\
    D &-0&4038&  0&8418& 0&1920&-0&3024\\
    d$\alpha$ &  0&4756& 0&3014&-0&7746&-0&2879\\\hline
Eigenvalue& 51&4\%&21&3\% & 17&7\% &  9&5 \%  \\ \hline
\end{tabular}
\caption{Eigenvectors and Eigenvalues for spectral index.}\label{pcaa}
\end{table}

\subsection{Depolarisation}\label{4:dep}
\index{depolarisation} \index{Burn's law} In the previous chapter I
motivated the adjustment of the measured depolarisation to a common
redshift, $z=1$, using the Burn\footnote{As noted in the previous
chapter the Burn method for shifting the depolarisation may result in
an over-estimation of the depolarisation. In the next chapter I
discuss other methods for shifting the depolarisation and determine
what effect this has on the correlations presented here.} theory of
depolarisation \citep{burn}, for all sources. There is little
difference between the redshift and the radio luminosity correlations
with the average adjusted depolarisation, taken over the source as a
whole, DM$_z$, (see Table \ref{rsDM}) as both correlations show
significance levels greater than 99.9\%. There is also an indication
of a DM$_z$-D$_{source}$ anti-correlation, but this is found not to be
significant. The Partial Spearman Rank test is used to determine if
the P$_{151}$-DM$_z$ correlation can be explained by the independent
z-DM$_z$ and z-P$_{151}$ correlations.
\begin{eqnarray}
r_{z DM_z, P_{151}} = 0.40 ,\hspace{2cm} D = 2.00 \nonumber\\ 
r_{z P_{151}, DM_z} = 0.44 ,\hspace{2cm} D = 2.21 \nonumber \\ 
r_{P_{151} DM_z, z} = 0.41 ,\hspace{2cm} D = 2.04  \nonumber
\end{eqnarray}
\begin{figure}[!h]
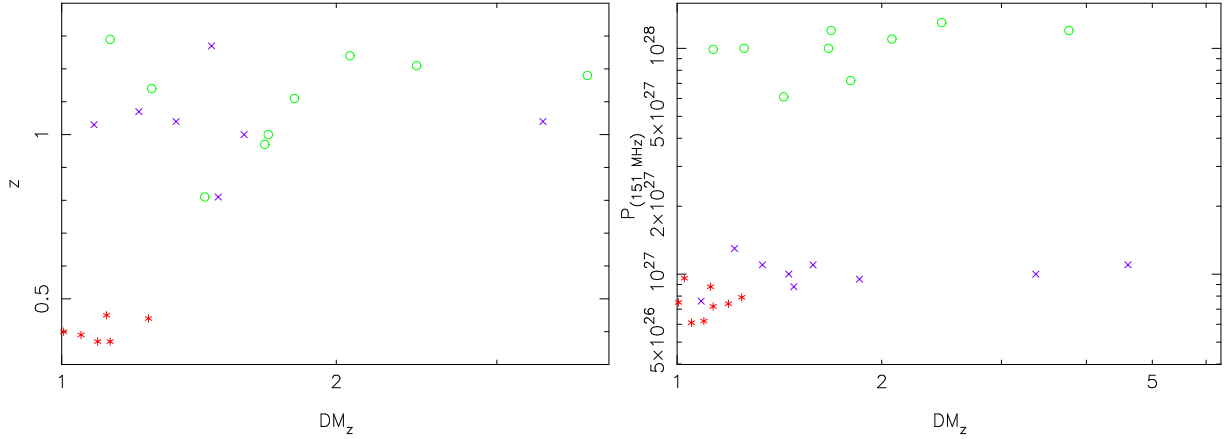

\centerline{
\psfig{file=ABCdmzobs2_c.ps,width=8cm,angle=270}
\psfig{file=DMP_c.ps,width=8cm,angle=270}
}
\caption[Redshift and radio-luminosity against DM$_z$]
{(a - left) Redshift against the average depolarisation of a
source. (b - right) Radio-luminosity against the average
depolarisation.  The depolarisation has been shifted to a common
redshift, z=1.  Symbols are as in Figure \ref{zd}.}\label{zDM}
\end{figure}
\begin{table}[!h]
\centering
\begin{center}
\end{center}
\begin{tabular}{ l@{ }l r@{.}l r@{.}l c l@{ }l  r@{.}l r@{.}l }
\cline{1-6}\cline{8-13}
\multicolumn{2}{c}{Parameters}&
\multicolumn{2}{c}{rs value}&\multicolumn{2}{c}{ t value}&&
\multicolumn{2}{c}{Parameters}&
\multicolumn{2}{c}{rs value}&\multicolumn{2}{c}{ t value}\\
\cline{1-6}\cline{8-13}
DM$_z$&z&0&68955& 4&66&&dDM$_z$&z&0&74017& 5&39\\
DM$_z$&P$_{151}$&0&68821& 4&65&&dDM$_z$&P$_{151}$&0&54462& 3&18\\
DM$_z$&D$_{source}$&-0&31145&-1&61&&dDM$_z$&D$_{source}$&-0&31009&-1&60\\
DM$_z$&$\sigma_{RM_z}$&0&45257&  2&33&&dDM$_z$&$\sigma_{RM_z}$&0&55929&  3&09\\
DM$_z$&dDM$_z$&0&89197& 9&67&&dDM$_z$&dRM$_z$&0&40514&  2&03\\
\cline{1-6}\cline{8-13}
\end{tabular}
\caption{Spearman rank values for all shifted depolarisation parameters.}\label{rsDM}
\end{table}
It is evident that there is still little difference between the redshift
and radio-luminosity correlations. This is confirmed using the PCA
results in Table \ref{pcaDM}(top). Figure \ref{zDM}(a) shows that the
higher redshift sources display, on average, a larger degree of
depolarisation and also a much larger spread of depolarisation at any
given redshift compared to their low redshift counterparts. There is
also very little difference between the two high redshift samples
indicating that the difference in the radio-luminosity of the two
samples is unimportant, see Figure \ref{zDM}(b). Interestingly
\citet{mt73} found depolarisation to correlate with radio--luminosity
whereas \citet{kcg72} found the correlation was predominantly with
redshift. My results support the findings of both groups.

The trend with redshift is more pronounced when the difference in the
depolarisation over the source, dDM$_z$, is considered (Tables
\ref{rsDM} and \ref{pcaDM}(bottom)). Using Partial Spearman Rank I
find
\begin{eqnarray}
r_{z dDM_z,  P_{151}} &= 0.60 ,\hspace{2cm}& D = 3.25 \nonumber\\ 
r_{z P_{151}, dDM_z} &= 0.53 ,\hspace{2cm}& D = 2.77 \nonumber \\ 
r_{P_{151} dDM_z, z} &= 0.05 ,\hspace{2cm}& D = 0.23 \nonumber
\end{eqnarray}
This shows a definite correlation between redshift and dDM$_z$ at any
given radio luminosity, but the corresponding P$_{151}$-dDM$_z$ correlation
is considerably weaker at a given redshift.

\begin{figure}[!h]
\centerline{\psfig{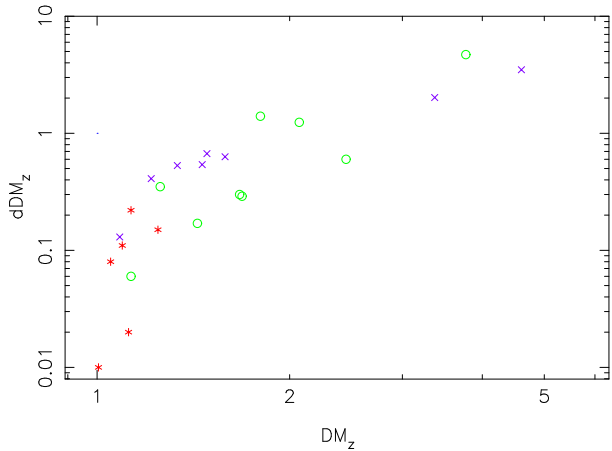}}
\caption[DM$_z$ against the difference in depolarisation, dDM$_z$.]
{DM$_z$ against the difference in depolarisation, dDM$_z$. dDM$_z$ is the
difference in the average depolarisation of each lobe. The
depolarisation has been shifted to a common redshift, $z=1$. Symbols are
as in Figure \ref{zd}.}
\label{DMdDM}
\end{figure}

\index{Laing-Garrington effect}
It is not surprising that DM$_z$ and dDM$_z$ are highly correlated,
with a significance greater than 99.9\% (Figure \ref{DMdDM}). A source
with a high average DM$_z$ will be located in a region with a dense
Faraday medium, which also causes a significant difference between the
lobes and hence the correlation. The Laing Garrington effect discussed
in section \ref{LGeffect} could also cause the observed correlation
between \DMz\, and dDM$_z$, assuming that the sources are not angled
flat to the line-of-sight.

\begin{table}
\centering
\begin{tabular}{l  r@{.}l r@{.}l r@{.}l  r@{.}l }\hline
Parameter  & \multicolumn{2}{c}{1} &  \multicolumn{2}{c}{2}    &  
 \multicolumn{2}{c}{3}  &   \multicolumn{2}{c}{4}     \\\hline
    z  &  0&5837&  0&2185& -0&0001&  -0&7820\\
    P  & 0&4699& 0&6727& -0&1909&  0&5387  \\
    D  &-0&4397& 0&6135 & 0&6368 & -0&1570 \\
  DM$_{z}$&  0&4951& -0&3511& 0&7470&   0&2713 \\\hline
Eigenvalue& 52&7\%& 22&0\% & 15&5\%& 9&8 \%   \\ \hline
    z & 0&5957&   0&1851 & 0&0132 & 0&7815 \\
    P & 0&4760&   0&6755 &-0&2183 &-0&5192 \\
    D &-0&4281 &  0&6314 & 0&6249 & 0&1662 \\
    dDM$_{z}$ & 0&4851&  -0&3329 & 0&7495& -0&3036 \\\hline
Eigenvalue& 52&0\%&21&6\%&16&9\% &  9&5 \%  \\ \hline
\end{tabular}
\caption{Eigenvectors and Eigenvalues for depolarisation}\label{pcaDM}
\end{table}

Table \ref{rsDM} shows a very weak DM$_z$-D$_{source}$
anti-correlation. This detection is marginal and as the PCA results in
Table \ref{pcaDM}(top) demonstrate this anti-correlation is not very
significant. Interestingly this trend has been found by
\citet{strom,sj88,prm89,bel99} and \citet{ics98}, but was not found by
\citet{jane}[DT96]. The observations by \citet{strom}, \citet{sj88} and
\citet{ics98} are at a lower resolution than my measurements and thus
their depolarisation measurements may be affected by this resolution
difference. The observations by \citet{bel99} are at a higher
resolution, but also at higher frequencies (4.8 GHz to 8.4 GHz) and so
I would expect any depolarisation trends to be stronger in this
sample. \citet{prm89}(P89) however observe at the same frequency range
and resolution as my samples and they find an anti-correlation with
size. DT96 also observes at the same frequency and resolution as P89,
but also does not find any trend with size. Even when my results from
sample B, which are taken at the most similar redshift and
radio-luminosity range to those of P89, are analysed separately, I
still find no trend with size. There is no difference in the selection
of the P89 sample and my sample B except that P89 chose their sample
from sources which had strong emission lines. By preferentially
selecting sources with strong emission lines P89 has chosen sources
that would have strong depolarisation asymmetries and hence large
depolarisation measurements overall. This could strengthen any
D$_{source}$-DM correlation found. However, there does not seem to be
any convincing physical explanation why one set of sources should show
such a strong size-depolarisation anti-correlation and another set of
sources should show no trend.

\index{Faraday screen}\index{density of source environment}
It is usually assumed that radio sources are located in stratified
atmospheres. Therefore a small source will be embedded in denser gas
which acts as a more efficient Faraday screen. As the source expands
the lobes have a higher probability of extending beyond the denser
inner atmosphere, thus reducing the amount of depolarisation observed.
In section \ref{fundamentals} it was shown that there is no
significant z-D$_{source}$ anti-correlation in my sample. Therefore a
lack of any DM$_z$-D$_{source}$ anti-correlation suggests that there
is no significant difference in the source environments at any
redshift, of my samples, as the sources become larger. This may be
evidence that on scales of up to a Mpc the environment is relatively
homogeneous and does not show evidence of stratification. The lack of
any \Ds-dDM$_z$ could simply be caused by the Laing-Garrington effect
(see section \ref{LGeffect}). \index{Laing-Garrington effect}

\subsection{Rotation measure}
\index{rotation measure}
The Spearman Rank results in Table \ref{rsRM} show that there are no
strong correlations of the fundamental parameters with rotation
measure, RM. In every case the correlation is below 95\%
significant. The PCA results in Table \ref{pcaRM}(top) also
demonstrate that the RM of a source does not depend on the size,
redshift or the radio luminosity of the source. This is consistent
with the idea that the observed RM is due to Galactic variations and
not to a local Faraday screen (see section \ref{galactic}) and would
not be expected to vary in response to any change in the source
properties.

\begin{figure}[!h]
\centerline{\psfig{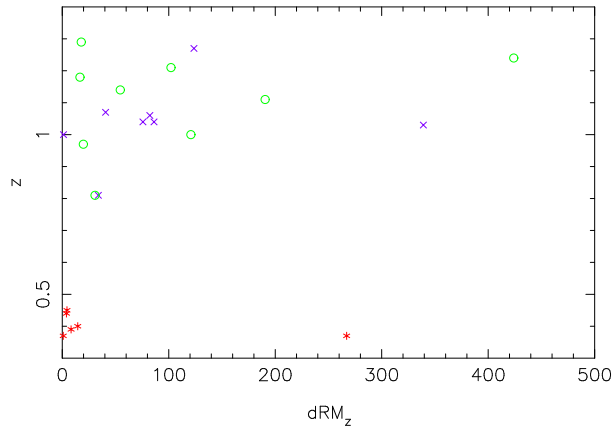}}
\caption[dRM$_z$ against redshift]
{Redshift against difference in rotation measure, dRM$_z$ across the
lobes of a source. dRM$_z$ has been shifted to the sources' frame of
reference. Symbols are as in Figure \ref{zd}.}
\label{zdrm}
\end{figure}

\index{variations in RM}
The difference in the rotation measure, dRM$_z$, between the two lobes
of a given source and the rms variation in the rotation measure,
$\sigma_{RM_z}$, are local to the source and have been corrected to
the sources' frame of reference. There is a dRM$_z$-z correlation
present in the Spearman Rank results, (see Table \ref{rsRM}), but it
is weaker in the PCA results where the correlation is strongly
reversed in the 3rd eigenvector, see Table \ref{pcaRM}
(middle). Figure \ref{zdrm} shows that in principal I should find a
strong dRM$_z$-z correlation. However, the presence of the low
redshift - high dRM$_z$ source (3C 299) weakens this trend. 3C 299 has a
large dRM$_z$ produced by a lack of polarisation information in one
lobe. Removing this source strengthens the correlation giving
r$_{dRM_z,z} = 0.632$ which corresponds to a significance of 99.98\%.

\begin{figure}[!h]
\centerline{\psfig{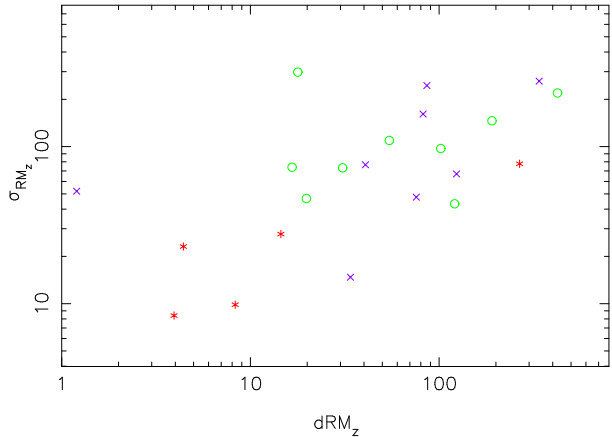}}
\caption[Rotation measure variations]
{The rms variation in the rotation measure,\sig against the difference
in rotation measure across the lobes of a source. Symbols are as
in Figure
\ref{zd}.}
\label{sdrm}
\end{figure}

As previously noted, $\sigma_{RM_z}$ measures variations of the
Faraday screen on scales smaller than dRM$_z$, which may still be
influenced by the Galactic Faraday screen. However, it is interesting
to note that dRM$_z$ correlates strongly with $\sigma_{RM_z}$, see
Figure \ref{sdrm}. So the two parameters may sample the same, local
Faraday screen even if dRM$_z$ may still contain some Galactic
contribution.
\begin{table}
\centering
\begin{tabular}{ l@{ }l r@{.}l r@{.}l c l@{ }l  r@{.}l r@{.}l }
\cline{1-6}\cline{8-13}
\multicolumn{2}{c}{Parameters}&
\multicolumn{2}{c}{rs value}&\multicolumn{2}{c}{ t value}&&
\multicolumn{2}{c}{Parameters}&
\multicolumn{2}{c}{rs value}&\multicolumn{2}{c}{ t value}\\
\cline{1-6}\cline{8-13}
RM$_a$& z &0&16996&  0&79&&	$\sigma_{RM_z}$&z&0&67885&  4&24\\
RM$_a$&P$_{151}$&0&27470&  1&31&& $\sigma_{RM_z}$&P$_{151}$&0&34881&  1&71\\
RM$_a$&D$_{source}$&-0&05040&-0&23&&$\sigma_{RM_z}$&D$_{source}$&-0&30435&-1&46\\
 dRM$_z$&z&0&50000&  2&65&&$\sigma_{RM_z}$&dDM$_z$&0&55929&  3&09\\
dRM$_z$&P$_{151}$&0&22530&  1&06&&$\sigma_{RM_z}$&DM$_z$&0&45257&  2&33\\
dRM$_z$&D$_{source}$&-0&22332& -1&05&
&$\sigma_{RM_z}$&dRM$_z$&0&68874&  4&35\\
\cline{8-13}
dRM$_z$&dDM$_z$&0&40514&  2&03&&\multicolumn{6}{c}{ }\\ 
\cline{1-6}
\end{tabular}
\caption[Spearman rank values for the rotation measure parameters]
{Spearman rank values for all rotation measure parameters. The
subscript {\it a} indicates that the RM has not been shifted to the
sources' frame of reference.}\label{rsRM}
\end{table}
\begin{table}
\centering
\begin{tabular}{l  r@{.}l r@{.}l r@{.}l  r@{.}l }\hline
Parameter  & \multicolumn{2}{c}{1} &  \multicolumn{2}{c}{2}    &  
 \multicolumn{2}{c}{3}  &   \multicolumn{2}{c}{4}     \\\hline
    z&  0&6398&  0&2390& -0&0491&     0&7288\\
    P &  0&5968&  0&4094& -0&1678& -0&6694\\
    D  & -0&3781&  0&4402& -0&8034& 0&1335\\
    RM &  0&3026& -0&7625& -0&5693&-0&0540\\ \hline
Eigenvalue& 43&8\%& 24&1 \%& 21&5\% &10&6\%   \\ \hline
    z&  0&6293&  -0&0201&  0&2213&  0&7447\\
    P&  0&5820&  -0&1197&  0&4874& -0&6399\\
    D& -0&3224&  -0&8641&  0&3594&  0&1423\\
   dRM$_z$&  0&4017& -0&4885& -0&7644& -0&1255\\ \hline
Eigenvalue& 45&9\%&23&3\% & 20&3\% & 10&5 \%  \\ \hline
   z   &  0&6215& 0&1765&  0&0623&  0&7607\\
    P  &  0&4923& 0&4210& -0&6151& -0&4495\\
    D  & -0&3212& 0&8876&  0&3288&  0&0296\\
$\sigma_{RM_z}$&  0&5178& -0&0615&  0&7139& -0&4673\\ \hline
Eigenvalue& 51&3\%&22&4\% & 18&6\% &  7&6 \%  \\ \hline
\end{tabular}
\caption{Eigenvectors and Eigenvalues for rotation measure.}\label{pcaRM}
\end{table}
$\sigma_{RM_z}$ shows a strong correlation with redshift, $>99.9\%$
significance, and a weak anti-correlation with D$_{source}$. The
$\sigma_{RM_z}$--z correlation also strengthens with the removal of
3C 299, becoming r$_{\sigma_{RM_z},z} = 0.767$ which corresponds to a
significance exceeding 99.99\%. These correlations are more obvious
when the PCA results are considered, see Table \ref{pcaRM}
(bottom). The first eigenvector contains 51.3\% of the variation in
the data and shows a strong z-$\sigma_{RM_z}$ correlation and a weaker
D$_{source}$-$\sigma_{RM_z}$ anti-correlation. The
D$_{source}$-$\sigma_{RM_z}$ anticorrelation reverses in the third
eigenvector indicating that it is a considerably weaker trend compared
to the z-$\sigma_{RM_z}$ correlation which only reverses in the last
eigenvector. The Partial Spearman Rank results below confirm that
there are no significant correlations with radio--luminosity for both
dRM$_z$ and $\sigma_{RM_z}$,
\begin{eqnarray}
r_{z dRM_z P_{151}} =&\hspace{0.3cm} 0.49 ,\hspace{2cm}& D = 2.40
\nonumber\\ r_{z P_{151}, dRM_z}= &\hspace{0.3cm} 0.69 ,\hspace{2cm}&
D = 3.79 \nonumber \\ r_{P_{151} dRM_z, z} =&-0.19 ,\hspace{2cm} &D =
0.86 \nonumber
\end{eqnarray}
\begin{eqnarray}
r_{z \sigma_{RM_z} P_{151}} =&\hspace{0.3cm} 0.65 ,\hspace{2cm} &D =
3.47 \nonumber\\ r_{z P_{151}, \sigma_{RM_z}} =&\hspace{0.3cm} 0.67
,\hspace{2cm} &D = 3.63 \nonumber \\ r_{P_{151} \sigma_{RM_z}, z}
=&-0.24 ,\hspace{2cm} &D = 1.09 \nonumber
\end{eqnarray}

In a survey of 27 high redshift sources \citet{pent}[P02] found that
the Faraday rotation was independent of size or radio--luminosity but
they also found that the number of sources with high levels of Faraday
rotation increased with redshift. The results from P02 use the RM of a
source in the sources' frame of reference. As noted in section
\ref{rm} the RM from my sources is taken to be dominated by
contributions from the Galaxy thus I use \sig\, and dRM$_z$ as
indicators of the environment instead of RM. \,\sig, dRM$_z$ and RM
sample the same medium and both sets of results find a trend with
redshift. The P02 trend indicates that the strength of the magnetic
field or the density of the environment (or both) is increasing with
redshift but by using only RM they are unable to determine how the
disorder of the environment changes with redshift. Conversely by using
only \sig\, and dRM$_z$ I have no information on the density of the
environment or the magnetic field but I find that the disorder in the
environment increases with redshift.

The Spearman Rank results in Table \ref{rsRM} show that there is a
strong $\sigma_{RM_z}$-DM$_z$ correlation, indicating that
$\sigma_{RM_z}$ is sampling the same medium as the depolarisation
measurements and thus either parameter can be used as a test of the
Faraday screen.\label{sDM}

\subsection{Summary}
Depolarisation, DM$_z$, and the differential rotation measure
properties, dRM$_z$ and $\sigma_{RM_z}$, correlate with redshift,
indicating that there is a change in the source environments as
redshift increases. Only depolarisation shows a trend with
radio--luminosity. By analysing the rotation measure down to its rms
variations ($\sigma_{RM_z}$) and the depolarisation measure, I am
probing smaller and smaller variations in the Faraday screen. I find
that there is an increasingly strong correlation with redshift over
radio--luminosity which suggests that the environment has no direct
link to the radio--luminosity of a source. The fact that there are no
spectral index correlations found also corroborates this view.

\section[Source asymmetries]{Source asymmetries}\label{asym:intro}
\index{source asymmetries} 

The previous sections have allowed me to analyse the bulk trends of
the source properties with the fundamental properties, but also with
themselves. However, it is also interesting to look at the asymmetries
of the sources and to determine their underlying physical cause.

The angle at which a source is orientated to the line--of--sight,
$\theta$, can affect the projected lobe length. Sources orientated at
a small angle will have lobes that appear more asymmetrical than
sources at a large angle, assuming there are no environmental
differences between the lobes. The orientation of a source will also
affect the observed depolarisation \citep{gc91a}, the spectral index
\citep{lp91} and the rotation measure.  In a simple orientation model
emission from the lobe pointing away from the observer will have a
longer path length through the local Faraday screen and thus any
depolarisation measurements of this lobe would be larger than the lobe
pointing towards the observer. This is known as the Laing-Garrington
effect, see section \ref{LGeffect}. \citet{lp91} also find that the
least depolarised lobe has a flatter spectrum.
\index{Laing-Garrington effect} \index{spectral index}\index{beaming}
\index{Liu-Pooley effect}

Beaming can also cause asymmetries in a source, independent of path
length through the Faraday screen. Assuming a source is angled at
$\theta$ to the line--of--sight, then the ratio of lobe lengths,
$D_1/D_2$, is given by
\citep{lr79}
\begin{equation}
\frac{D_1}{D_2}=\frac{1+v_\circ/c\; {\rm cos} \theta}{1-v_\circ/c\;
{\rm cos} \theta}
\end{equation}
where $v_\circ$ is the velocity at which the hotspot \citep{lr79} is
moving away from the nucleus and $c$ is the speed of light. As the
hotspot becomes increasingly more beamed (i.e. $\theta\rightarrow0$)
the lobe ratio increases. The hotspot will begin to dominate the flux
and spectral index in the beamed lobe. Thus as beaming becomes more
dominant in a source, the beamed lobe will become brighter, longer and
its spectrum flatter than the receding lobe.

Half of my sources do not contain any detection of a core at either
4.8 GHz or 1.4 GHz. This was a problem as I needed to define where the
core was located to determine an accurate estimation of the lobe
length. In the cases were there was no core detection I used
previously published maps to determine where a core is likely to be
located. However in several cases, most notably 3C 16, there was no
core to be found. In these cases I estimated the location of the core
position by using the most likely core position from the literature
\citep{lp91a}. This may seem a rather drastic approach, but as the
maps are not of very high resolution, the values calculated for the
lobe length can only be an approximation to the true length. The lobe
volume is calculated by assuming cylindrical symmetry and using the
measured lobe length, D$_{lobe}$, but also the parameter R, which
determines the ratio of the lobe length to the width at half the lobe
length,

\beg
 V_{lobe} =\frac{\pi D_{lobe}^3}{4 R^2} 
\ee

In most cases R $>$ 1 but in extreme cases like the southern lobe of
7C 1745+642 (see Figure \ref{1745.1}) R is less than unity, which
causes the large volume ratio ($>$ 10) for this source apparent in
Figure \ref{Dratio}(a).

In the following sections the subscripts {\it b} and {\it f} are used
to denote the value from the brighter lobe and the fainter lobe of an
individual source, respectively.

In the previous section I used the corrected $z=1$ depolarisation in
my analysis of the trend of depolarisation with the fundamental
parameters. However, by correcting the depolarisation to a common
redshift I am assuming a specific underlying structure to the Faraday
screen which may mask any underlying asymmetries between lobes. In
previous studies \citep[e.g.][etc.]{prm89,lp91b,gc91} the observed
depolarisation is used as an asymmetry indicator. To ensure my results
are comparable to those published I use only the {\it observed}
depolarisation values given in Tables \ref{A} to \ref{C} in my
analysis of any depolarisation asymmetries.\index{depolarisation}

\subsection{Flux asymmetries}\label{fluxasym}
\begin{figure}[!h]
\centerline{\psfig{file=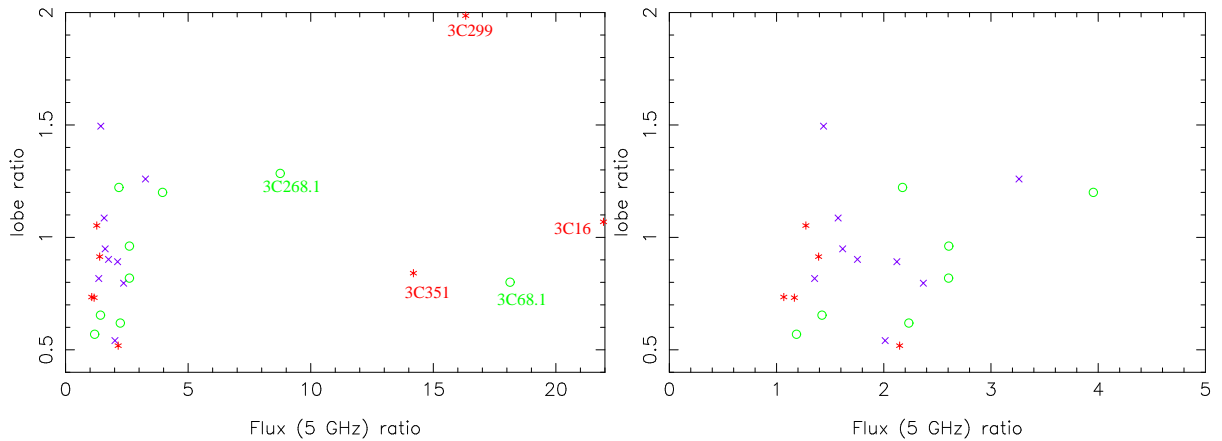,width=16cm}}
\caption[Flux ratio against lobe ratio]
{Plot of the flux ratio at 4.8 GHz against the lobe ratio, (a - left)
with all sources and (b - right) zoomed in on the more `normal'
sources.  Symbols are as in Figure \ref{zd}.}
\label{FLr}
\end{figure}

\begin{figure}[!h]
\centerline{\psfig{file=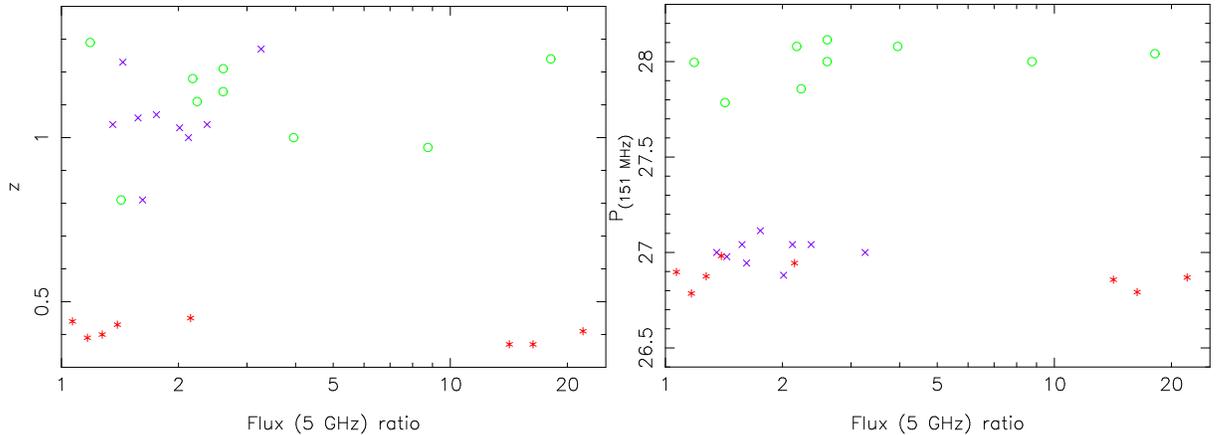,width=16cm}}
\caption[Flux ratio against the fundamental parameters]
{(a - left) Plot of the flux ratio at 4.8 GHz against redshift,  (b -
right) and against radio-luminosity.  Symbols are as in Figure \ref{zd}.}
\label{Flux}
\end{figure}

\index{flux asymmetries}
\begin{figure}[!h]
\centerline{\psfig{file=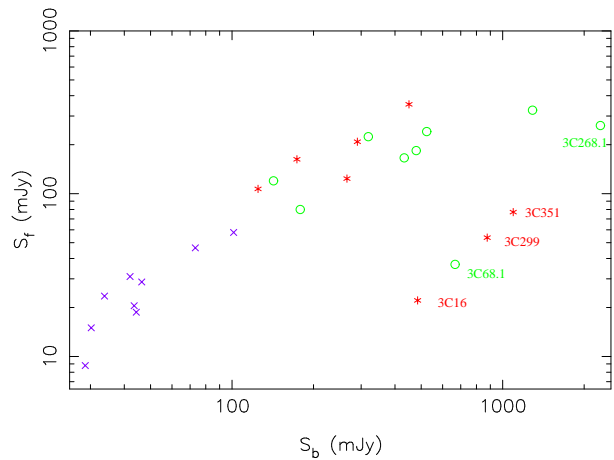,width=8cm}}
\caption[Fluxes of the brightest to faintest lobe]
{Plot of the flux of the brighter lobe against the fainter lobe at 4.8
GHz.  Symbols are as in Figure \ref{zd}.}
\label{F1F2}
\end{figure}
Figures \ref{F1F2} to \ref{Flux} use the 4.8 GHz flux data as a test for
flux asymmetry over the source. Tables \ref{A} to \ref{C} show that
the 1.4 GHz flux is much larger compared to the 4.8 GHz flux for each
source. Although this suggests that the 1.4 GHz flux data would be a
better indicator of any source asymmetries than the 4.8 GHz flux data,
since fainter lobes such as the northern lobe of 7C 1745+642 are much more
prominent, but the noise level in each source is also much higher. To
ensure that my analysis was not unduly affected by high noise levels I
chose to use the 4.8 GHz flux instead of the 1.4 GHz flux, in my
analysis of the asymmetries.

Figure \ref{F1F2} demonstrates that in all my sources there is a tight
correlation between the flux of the brightest lobe when compared to
the flux of the fainter lobe. In several cases the fainter flux is
much smaller than the brighter flux. 3 of these sources are quasars,
3C 351, 3C 68.1 and 3C 16 but interestingly 3C 299 and 3C 268.1 also
display this atypical behaviour. \citet{msb95} noted that 3C 299 showed
a strong asymmetry and was more similar to a high redshift ($z>$1)
radio galaxy than a low redshift galaxy. \citet{hs02} also noted that
this source has lobes of very different lengths. I find 3C 299 to have a
\sig\, comparable to the high redshift samples.

Figure \ref{FLr}(a) shows that 3C 299, 3C 268.1 and 3C 16 do show
larger ($>$1) lobe ratios but 3C 351 and 3C 68.1 do not. However 3C
351's position in this plot is uncertain. The estimation of the lobe
length requires an estimation of where the lobe terminates but in the
case of 3C 351's northern lobe it extends far to the west of the
hotspots whereas the southern lobe is much more well defined (see
Figure \ref{3c351.1}). Thus the estimation of the length of the
northern lobe may be incorrect and hence the calculated lobe ratio may
be smaller than it really is. 3C 68.1 and 3C 16 have no detected cores
either by me or previously published data\footnote{\citet{bhl94} do
marginally detect a core in deeper observations of 3C 68.1.}. This
means that the lobe length estimation in these sources may also be
flawed. To further aggravate this problem both of these sources have
very faint lobes (see Figures \ref{3c68.1} \& \ref{3c16.1}) and so it
is hard to determine where exactly the lobe terminates. However, it is
worth noting that when these 5 sources are removed from the analysis
(Figure \ref{FLr}(b)) there is no significant trend for the brighter
flux ratio sources to have a larger lobe ratio.

Figures \ref{Flux}(a) and \ref{Flux}(b) show that the trends noted are
intrinsic to the sources and not due to any redshift and
radio-luminosity differences in the samples. It is worth noting that
sources in the high-redshift, low radio-luminosity sample, sample A
(`$\times$'), always seem to have a smaller flux ratio than sources in the
other two samples.

\subsection{Asymmetries in spectral index}
\index{spectral index asymmetries}
\index{Liu-Pooley effect}
\subsubsection{Liu-Pooley effect}\label{LPe}
The Liu--Pooley (LP) effect predicts that the lobe with the flatter
spectrum in a source is also the least depolarised lobe
\citep{lp91b,ics01}. \citet{lp91b} found that 12 out of their 13
sources demonstrated this effect. As noted in section \ref{asym:intro},
to allow a fair comparison between my results and those of
\citet{lp91b} I use only the {\it observed} depolarisation in the
following analysis.

\begin{figure}[!h]
\centerline{\psfig{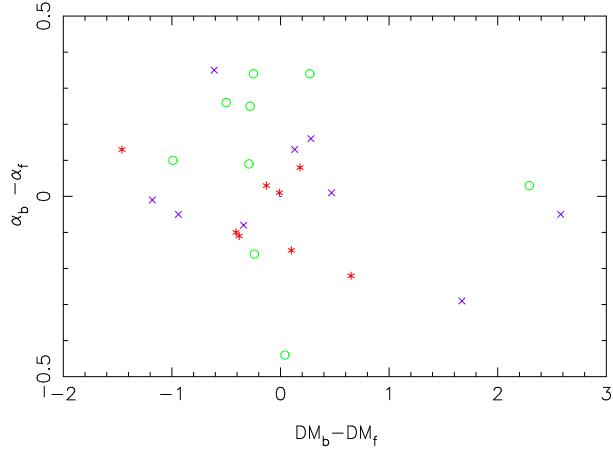}}
\caption[Liu - Pooley effect I]
{The difference in the spectral index of the bright lobe to the faint
lobe against the difference in the depolarisation of the bright lobe
to the faint lobe. Depolarisation is taken to be the observed
(unshifted) value. Symbols are as in Figure \ref{zd}.}
\label{LP1}
\end{figure}

\begin{figure}[!h]
\centerline{\psfig{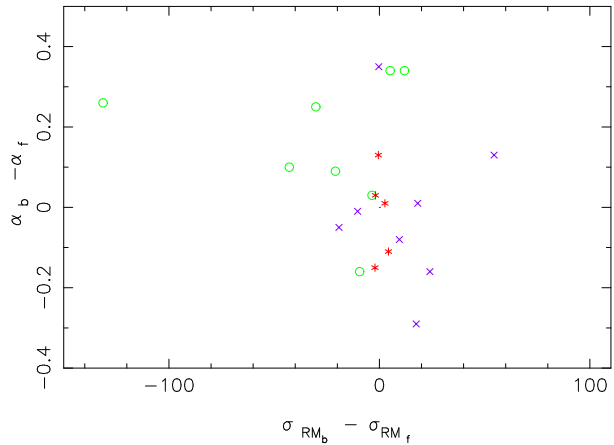}}
\caption[Liu - Pooley effect II]
{The difference in the spectral index of the bright lobe to the faint
lobe against the difference in the $\sigma_{RM_z}$ of the bright lobe
to the faint lobe. $\sigma_{RM_z}$ is the value {\it local} to the
source. Symbols are as in Figure \ref{zd}.}
\label{LP2}
\end{figure}

As Figure \ref{LP1} shows I find no significant Liu-Pooley (LP) effect
in my sample. As another comparison I used the rms variations in the
rotation measure in the sources' reference frame, \sig, instead of
depolarisation measure. Clearly Figure \ref{LP2} shows that with
\sig\, I find no significant evidence of any LP effect. Even by
plotting the difference in depolarisation or the difference in
$\sigma_{RM_z}$ against the spectral index difference, insuring that
the difference in DM or \sig\, were always positive, there was no
significant LP effect. In all cases I find only a maximum of 14
sources out of the 26 sources (12 out of 23 in the case of \sig) to
show definite signs of the LP effect.

\citet{ics01}[IC0] have shown that the LP effect is stronger
for smaller sources and shows no dependence on redshift. They also
find that the LP effect is not affected by the choice of quasars or
radio galaxies in a sample and is equally significant in both
species. They also find that for their large sample (comprised of the
\citet{lp91b,prm89,gcl91} samples and the Molongo Reference
Catalogue/1 Jy sample) only 58\% of radio galaxies and 59\% of quasars
show the LP effect which is consistent with my results. The inherent
difference between sources chosen in samples as well as orientation effects
has been suggested by IC0 as an explanation to the number of sources
showing the LP effect, in any given sample.

\subsubsection{Effect of changes in the lobe length}\label{3C16}
\index{lobe length asymmetries}
\begin{figure}[!h]
\centerline{\psfig{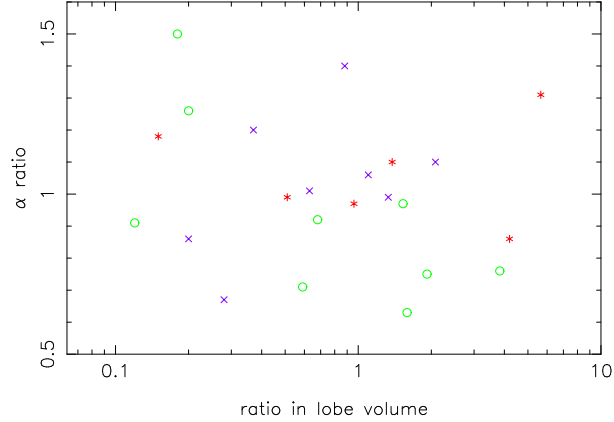}}
\caption[The ratio of spectral indices against the lobe ratio.]
{The ratio of spectral indices against the lobe ratio.  Symbols are as
in Figure \ref{zd}.}
\label{ar}
\end{figure}

Figure \ref{ar} shows that the ratio of spectral indices of the
sources' lobes is insensitive to changes in the lobe volume. A source with a
large lobe volume ratio ($>$ 1) is equally likely to have a low spectral
index ratio as a high spectral index ratio. This suggests that the
environment, which affects how large a volume a source can grow to, has
no direct effect on the spectral index of a lobe. Thus spectral index
should not be used as a tracer for the environment.

\begin{figure}[!h]
\centerline{\psfig{file=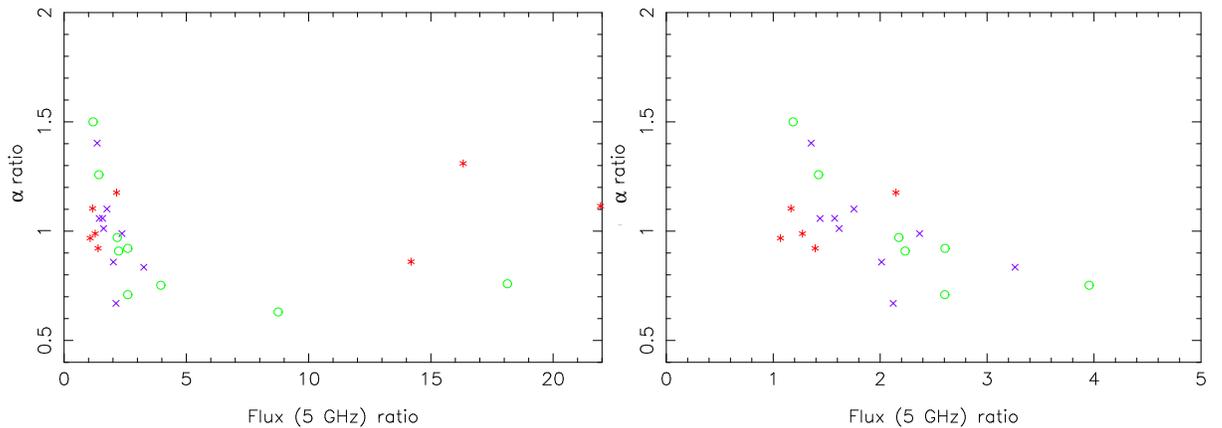,width=16cm}}
\caption[$\alpha$ ratio against flux ratio]{The spectral index ratio
against the flux ratio,  (a - left) all sources and (b - right) zoomed
in on sources with a flux ratio $<$ 5. Symbols are as in Figure
\ref{zd}.}
\label{Far}
\end{figure}

Figure \ref{Far}(a) shows that in general, a source with a higher flux
ratio has a smaller spectral index ratio. This means that the brighter
lobe becomes progressively flatter (i.e. $\alpha_b/\alpha_f
\rightarrow 0$) as the difference in the flux ratio's increases. This
is consistent with the theory that beaming is present in these
sources. 3C 16 and 3C 299 are notable exceptions to this trend. 3C 299
has already been discussed above and possess exceptional properties
when compared to other radio galaxies. 3C 16 is a quasar and I would
expect this source to have shown more evidence for beaming but it has
almost similar spectral indices in both lobes. \citet{hh98} found
evidence that 3C 16 maybe a radio source that is just beginning to
restart and hence the fainter lobe will not have a well defined
hotspot, therefore beaming is currently not important in this
source. Figure \ref{Far}(b) shows that when the more extreme sources
are removed the anti-correlation is even more evident.

\begin{figure}[!h]
\centerline{\psfig{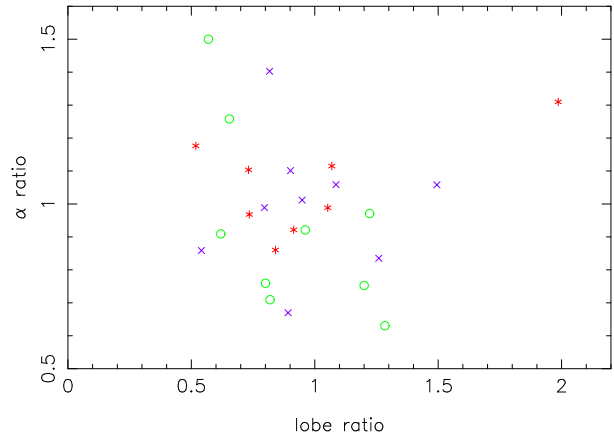}}
\caption[$\alpha$ ratio against lobe ratio]
{The ratio of spectral indices against the lobe ratio.  Symbols are as
in Figure \ref{zd}.}
\label{ad}
\end{figure}

Figure \ref{ad} shows that as the lobe length ratio increases the
spectral index ratio decreases. This means that in the more length
asymmetric sources the larger lobe has the flatter spectral
index. Once again 3C 299 again contradicts this result. It has a high
ratio of spectral indices and a high lobe length ratio. The trend
between lobe ratio and the spectral index ratio adds further evidence
to the presence of beaming in many of my sources.

Although Figures \ref{Far} and \ref{ad} show evidence for beaming,
another possibility would be that the observed asymmetries are caused by a
difference in the density of the environment between both lobes. A
denser environment would mean that the lobe has a higher minimum
magnetic field. Thus any asymmetries in the \bmin\, would indicate
that it is the environment and not beaming that is causing the observed
asymmetries with spectral index.
\subsubsection{Minimum energy of the magnetic field}
\index{magnetic field asymmetries}
\label{bmin:theory}\index{minimum energy}
A simple method to estimate the magnetic field of a radio source is to
use the minimum energy argument \citep[e.g][]{longair}. For any given
radio source of a given volume and radio-luminosity the \bmin\, argument
calculates the minimum energy that must be present in the emitting
volume. The minimum energy argument assumes that $W_{{\rm
mag}}=\frac{4}{3}W_{{\rm particles}}$, i.e. a source has approximately
equal energy in its relativistic particles as it has in its magnetic
field.
 
The minimum energy approach assumes a perfectly uniform magnetic field
throughout the lobe. There are no energy losses from the
electrons. Therefore, the energy distribution of the relativistic
electrons is a perfect power--law throughout the life--time of the
source. This is a simplistic model but it can be used to model the
magnetic field, B$_{min}$, in the lobe with few assumptions
required. For full details see
\citet{longair} which gives,\index{magnetic field}

\begin{eqnarray}\label{eq:bmin}
B_{min} &=&
\left(\frac{9\sqrt{3}c\mu_0^2L_\nu\bar{\nu}^{-\frac{1}{2}}e^{\frac{1}{2}}m_e^{\frac{1}{2}}\kappa}{2^3\sqrt{\pi}\sigma_TV}\right)^{\frac{2}{7}},\hspace{1cm}{\rm
(Tesla)}\\ {\rm where}&&\nonumber \\
\bar{\nu}^{-\frac{1}{2}}&=&\left[\left(\frac{3-x}{2-x}\right)\left(\frac{\nu_1^{\frac{2-x}{2}}-\nu_2^{\frac{2-x}{2}}}{\nu_1^{\frac{3-x}{2}}-\nu_2^{\frac{3-x}{2}}}\right)\right]
\end{eqnarray}
and $x=1-2\alpha$. The volume of a lobe, V, is assumed to be
cylindrical with $V = \frac{\pi}{4}D_{lobe}^3/R^2$ as before. $\nu_1 =
10$ MHz and $\nu_2 = 100$ GHz are taken to be the lower and upper
limits of the synchrotron spectrum and the luminosity, L$_{\nu}$, is
calculated using the 4.8 GHz flux of each lobe.

\begin{figure}[!h]
\centerline{\psfig{file=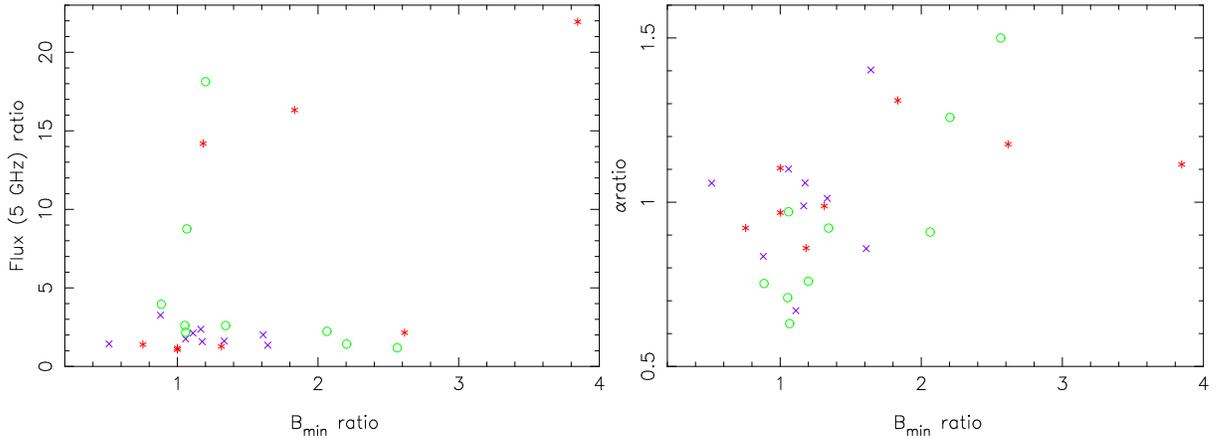,width=16cm}}
\caption[The \bmin\, ratio against flux and $\alpha$ ratios]
{(a - left) The \bmin\,
ratio against the flux ratio and (b - right) against the spectral
index ratio. Symbols are as in Figure \ref{zd}.}
\label{Bminp}
\end{figure}
As Figure \ref{Bminp}(a) demonstrates, there is no relationship between
the \bmin\, ratio and the flux ratio. This indicates that there is no
significant environmental difference causing the flux differences
between two lobes of a source.

Correspondingly Figure \ref{Bminp}(b) shows that there is a weak
correlation in the \bmin\, ratio when compared with the spectral index
ratio. A beamed, flat spectrum hotspot will have a stronger magnetic
field which could explain the weak trend in Figure \ref{Bminp}(b).
However, the asymmetries in the \bmin\, could also be caused by
differences in the density of the environmental across the lobes.
Figure \ref{Bminz} shows that there is no trend with redshift and so
the \bmin\,- $\alpha$ ratio is intrinsic to the local source
environment and is not affected by the redshift of a source.
\begin{figure}[!h]
\centerline{\psfig{file=Bminz_c.ps,width=8cm,angle=270}}
\caption[The \bmin\, ratio against z]{The \bmin\, ratio against the redshift
of the source. Symbols are as in Figure \ref{zd}.}
\label{Bminz}
\end{figure}
\begin{figure}[!h]
\centerline{\psfig{file=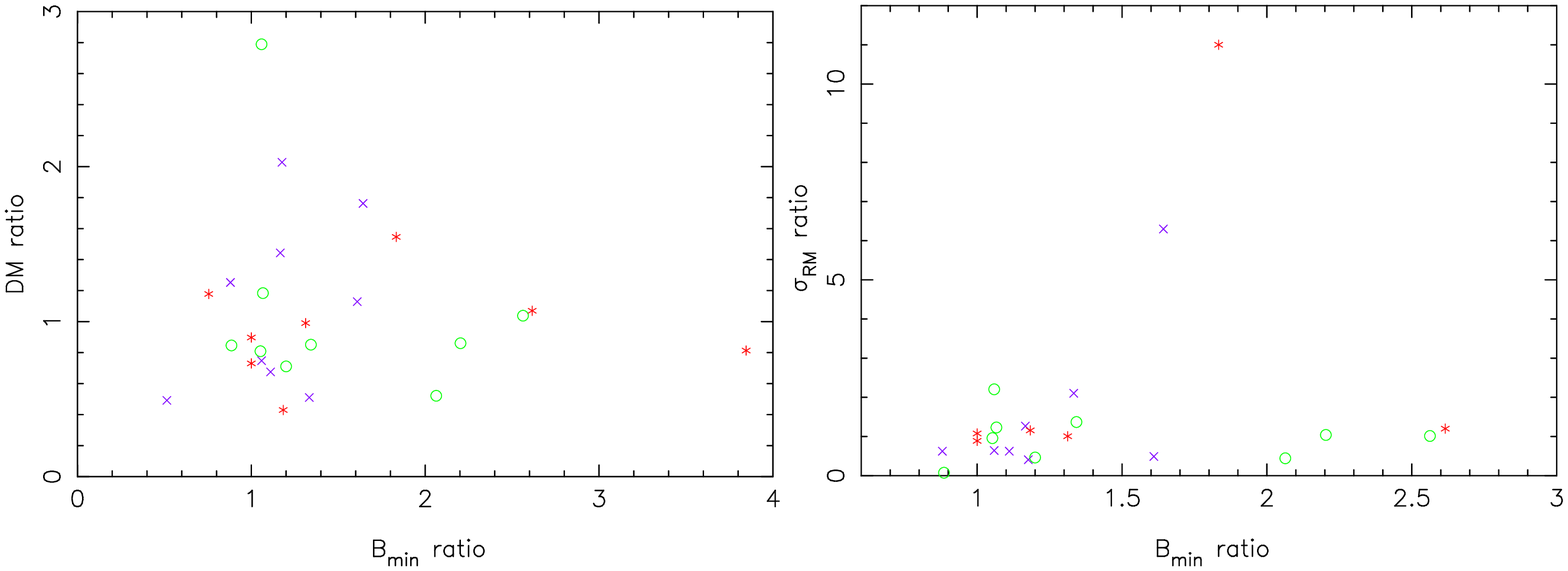,width=16cm}}
\caption[The \bmin\, ratio against DM \& \sig\, ratios]
{The \bmin\, ratio against (a - left) the depolarisation ratio and (b -
right) \sig\, ratio. Symbols are as in Figure \ref{zd}.}
\label{bmindep}
\end{figure}

\index{depolarisation asymmetries} \index{rotation measure
asymmetries} \index{density of source environment}

It is interesting to note that the asymmetries in depolarisation and
\sig\, are not related to asymmetries in the \bmin\, (see Figure
\ref{bmindep}).  Depolarisation and \sig\, are both thought to be
observational indicators of density\footnote{Rotation measure is a
direct function of density and the magnetic field of a source (section
\ref{rm}) and since \sig\, and the shifted depolarisation have been
shown to be correlated (section \ref{sDM}) then both can be said to be
tracers of the density of the environment.}, in the source environment
but neither show any relationship with the \bmin\, ratio. This
suggests that the environment has no direct impact on either
depolarisation or \sig\, asymmetries.

\subsection{Asymmetries in depolarisation}\index{depolarisation asymmetries}
Unlike the flux asymmetry (Figure \ref{F1F2}) the depolarisation shows
no obvious correlation between the depolarisation of the brighter and
fainter lobes, see Figure \ref{DM12}. It is worth noting that in most
cases the depolarisation of the brighter lobe is smaller than the
depolarisation of the fainter lobe. This is again indicative of
beaming in these sources. In the case of beaming, the fainter lobe is
assumed to be pointing away from the observer. The radiation of the
fainter lobe will then have a larger path length to traverse through
the local Faraday screen compared to the radiation of the brighter
lobe and will appear more depolarised.
\bfi[!h]
\centerline{\psfig{file=DM1DM2_c.ps,width=8cm,angle=270}}
\caption[Depolarisation of the brightest to faintest lobe]
{Depolarisation of the brighter lobe against the fainter lobe.
Symbols are as in Figure \ref{zd}.}
\label{DM12}
\efi
\bfi[!h]
\centerline{\psfig{file=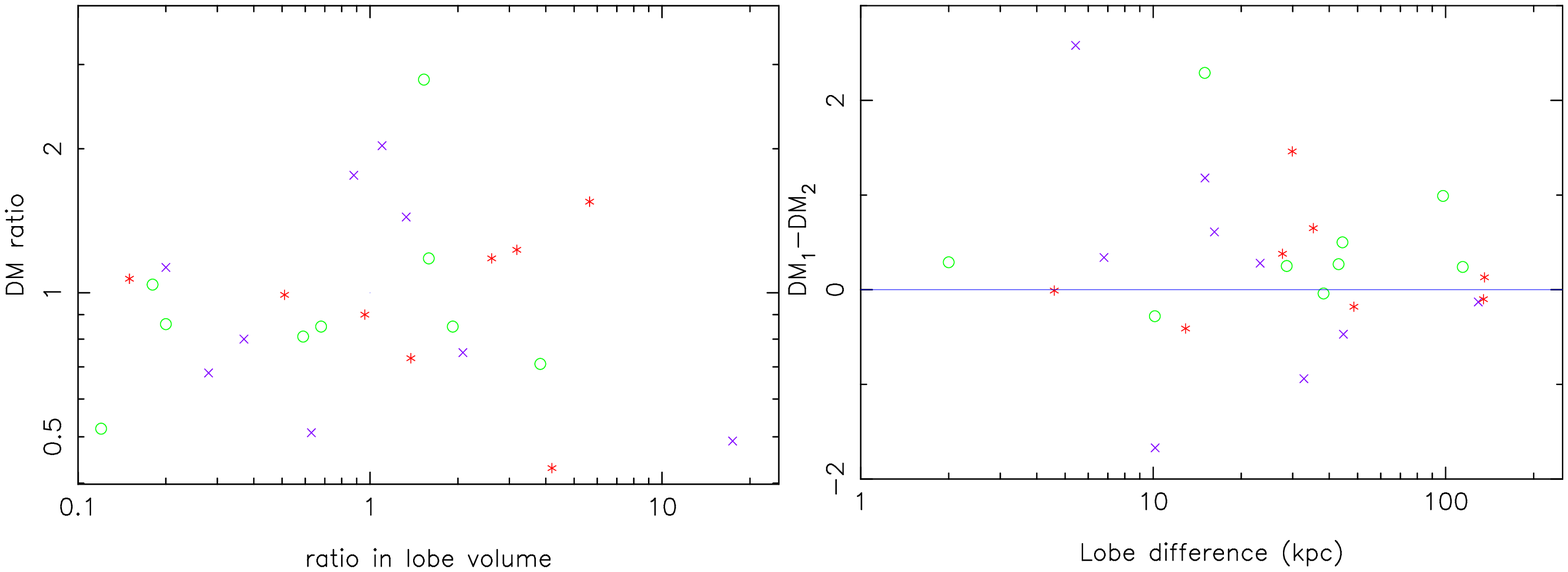,width=16cm}}
\caption[Depolarisation asymmetries]
{(a - left) Depolarisation ratio against the lobe volume ratio.  (b -
right) Difference in depolarisation against the difference in size of
the lobes. Plotted so that \DLa-\DLb $>$ 1 in all cases.  Symbols are
as in Figure \ref{zd}.}
\label{Dratio}
\efi

The flux asymmetry is independent of any depolarisation asymmetry as
3C 299, 3C 16, 3C 68.1, 3C 351 and 3C 268.1 (the high flux ratio sources)
show similar spreads in their depolarisation ratios when compared to
all the other sources. Figure \ref{Dratio}(a) at first sight indicates
that the volume of the lobe also plays no part in the magnitude of the
depolarisation ratio. However, on closer inspection it is worth noting
that in 16 out of the 26 sources, the lobe with the larger volume is the
more depolarised. In fact, only 4C 14.27, 3C 68.1, 3C 351, 6C 1256+36 and
7C 1745+642 vary significantly from this trend.

By plotting the depolarisation ratio against the lobe volume ratio in
Figure \ref{Dratio}(a) I assume that all the lobes in the three
samples can be described by a cylinder. This is not a good
approximation for extremely distorted sources like 3C 351 and 7C
1745+642. As Figure \ref{Dratio}(b) shows the trend still exists when
the difference in the lobe lengths is compared with the difference in
the depolarisations of each lobe. In this case the trend is much more
pronounced with only 6C 0943+39 and 7C 1745+642 showing significant
disagreement with the findings. Both of these sources show the same
trends in (a) and (b). This disagrees with the results of
\citet{prm89} and \citet{lp91b}, who find that the shorter lobe is the
more depolarised lobe. However DT96 also finds this reverse
correlation with only 3C 351 in common between her sample and my
sample. The DT96 correlation was found to be strong in the quasars,
but the depolarisation of the radio galaxies was too low to be able to
determine if the correlation was significant. This may indicate that
there is an inherent difference in the \citet{prm89} and \citet{lp91b}
samples and the DT96 sample and mine. This may be due to the fact that
the \citet{prm89} sample was selected from sources with strong
emission lines and the \citet{lp91b} sample only contains sources
smaller than 23$''$.

\subsection{Asymmetries in \sig}
\bfi[!h]
\centerline{\psfig{file=s1s2_c.ps,width=8cm,angle=270}}
\caption[\sig\, of the brightest to faintest lobe]
{\sig\, of the bright lobe against the fainter lobe.
Symbols are as in Figure \ref{zd}.}
\label{s1s2}
\efi

\index{beaming} Figure \ref{s1s2} shows there is no relationship
between \sig\, of the brighter lobe when compared to the fainter
lobe. I find that only half the sources show that the fainter lobe has
the larger \sig. This is unlike the depolarisation result where in
almost all cases the fainter lobe was more depolarised.  However, the
depolarisation is affected by changes in the path length through the
Faraday screen but \sig\, is insensitive to these changes and so would
not be expected to show any trend with the lobe asymmetries.

As with depolarisation there is little trend with lobe volume (Figure
\ref{sratio}(a)). This confirms the idea that the extent that a lobe 
can grow to (i.e the volume) has little effect on the measured
parameters such as spectral index, depolarisation and \sig. Most of
the sources show that the larger \sig\, is in the longer lobe. In only 3
out of the 23 sources is this not the case (6C 1129+37, 6C 0943+39
and 3C 280) which is similar to the trend noted with depolarisation.

\bfi[!h]
\centerline{\psfig{file=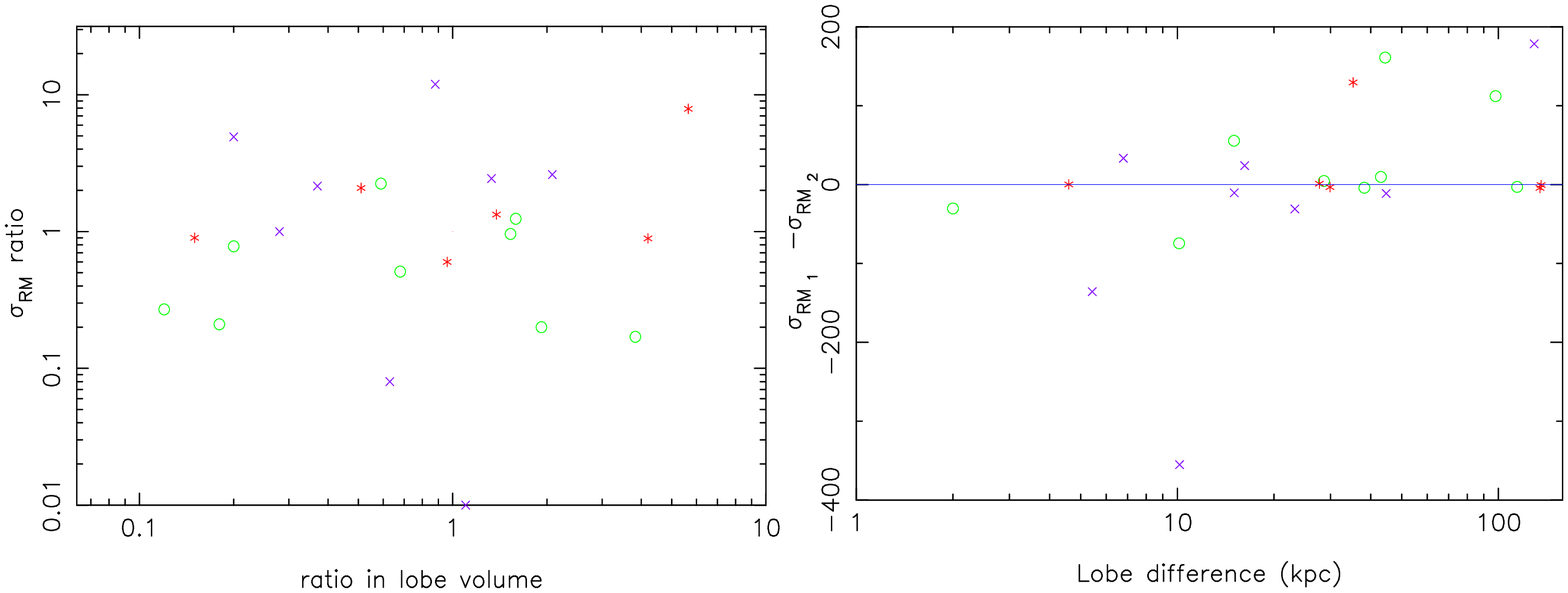,width=16cm}}
\caption[\sig\, asymmetries]{(a - left) \sig\, ratio against the lobe
volume ratio.  (b - right) Difference in \sig\, against the difference
in size of the lobes. Plotted so that \DLa-\DLb $>$ 1 in all cases.
Symbols are as in Figure \ref{zd}.}
\label{sratio}
\efi

\subsection{Summary}

I found that there is no significant LP effect, but this lack of trend
could be attributed to the larger size distribution of my sample
compared to the sample of \citet{lp91b}. This dependence on size has
also been noted by IC01. From the spectral and depolarisation
asymmetries I found evidence that many of the sources were beamed. The
\bmin\, asymmetry results could be explained by both a change in the
environment or by the presence of beaming. The \bmin\, model is
not an accurate picture of the magnetic field of a source but does
give a rough estimate of the size of the field.

The depolarisation and \sig\, asymmetries are similar,
confirming the assumption that they sample similar regions of the
Faraday screen. In general, the depolarisation and \sig\, are smaller
in the shorter lobe and almost all the fainter lobes show larger
depolarisations. The \sig\, variations are insensitive to which lobe
is brighter. 

\section[Conclusions]{Conclusions}
A number of source properties are derived directly from the radio
observations. These include the spectral index, the rotation measure
and the depolarisation measure. I find no correlation of spectral
index with any of the fundamental parameters (redshift,
radio-luminosity and radio-size). I can therefore not
confirm previous findings of a correlation of spectral index with
redshift \citep{athreya}, a trend of spectral index with radio
luminosity \citep{vv72,on89} nor a trend of spectral index with size
\citep{brw99}

In the previous chapter I argued that the overall rotation measure for
a given source is mainly Galactic in origin. This interpretation is
consistent with the absence of any correlation of rotation measure
with the fundamental parameters. Variations in the rotation measure
are quantified in order of decreasing angular scales by dRM$_z$
(difference of rotation measure between the two lobes of an individual
source), $\sigma _{RM_z}$ (rms variation of rotation measure) and
DM$_z$ (depolarisation measure). I find that the significance of a
correlation with redshift increases from dRM$_z$ to DM$_z$. Therefore
I can conclude that the variation of the rotation measure on small
angular scales is caused by a Faraday screen local to the
sources. Furthermore, these Faraday screens give rise to greater
variation of the rotation measure for sources at higher redshift
compared to their low redshift counterparts.

I have also studied the asymmetries of these derived properties with
the fundamental properties, but I have also looked at the asymmetries
in the 4.8 GHz flux data, S, and the \bmin, B$_{min}$. By definition,
S$_b$/S$_f$ $>$ 1 but several sources, of which 3 are quasars, show a
much larger flux ratio. I find evidence from spectral asymmetries that
many of the sources show indications of beaming. However, the results
from the B$_{min}$ asymmetries are inconclusive and could also explain
the asymmetries by implying a more environmental based explanation. In
general all asymmetries were found to be insensitive to changes in the
source redshift, or low frequency radio luminosity; suggesting that
the environments of the high redshift and low redshift samples are
similar and any asymmetries caused are due to localised changes in the
environment.

This interpretation is confirmed in the analysis of the depolarisation
and \sig\, asymmetries with B$_{min}$. I suggest that it is not the
environment that is causing the depolarisation and \sig\, asymmetries
but is actually the degree of disorder in the sources that is causing
a change. This fits with my findings that the shifted depolarisation,
DM$_z$, dRM$_z$ and \sig\, show trends with redshift but no trend with
the \bmin. Thus it is the degree of disorder in the magnetic field
that is changing with redshift and not the density of the source
environment.

\newpage
\thispagestyle{empty}
\chapter{Modelling of the Faraday screen} 
\index{Faraday screen modelling}
\index{variations in RM}
To sample the environment around radio sources I use the observed
polarisation and the observed RM variations. These parameters are used
to determine how the environment evolves with redshift,
radio-luminosity and radio size\footnote{The relationship between the
environment and the depolarisation measure and \sig\, are discussed in
detail in the previous chapters.}. However, the observed polarisation,
the observed RM variations and hence the observed depolarisation
measure are also effected by source based variations in the structure
of the Faraday screen. The structure of the Faraday screen combined
with the limited spatial resolution of any radio observation lead to
the observed depolarisation of the synchrotron emission.

\index{Burn's law} \index{Tribble models} \index{depolarisation} 

A uniform Faraday screen comprised of large ($>$ 50 kpc) cells will
have little effect on the underlying polarisation and RM structure.  A
cell is defined as a region in the Faraday screen where the magnetic
field has almost uniform strength and orientation. As these cells
shrink (i.e the cells size becomes smaller than the size of the
telescope beam) and the Faraday screen becomes increasingly
non-uniform the effects of the underlying structure can become more
pronounced. The effect of the Faraday screen on a source was first
investigated by \citet{burn}. I use the \citet{tribble}, hereafter
T91, models as a modern alternative to Burn's theory. The T91 models
assume a specific structure to the RM variations and calculate the
effect of the structure function on the underlying RM distribution and
hence the overall polarisation of a source at any given
wavelength. These models can be used to correct the observed
depolarisation of my sample sources to a common redshift and the
results can be compared with the Burn shifted depolarisation, DM$_z$,
in section \ref{galactic}. If the depolarisation measure is found to
be correlated with redshift and radio-luminosity using the T91, models
then this will confirm the findings in section \ref{4:dep} where
DM$_z$ is found to be a function of both redshift and
radio-luminosity.

\section{Critique of Burn's model}
The polarisation, $p$, at a given point in a radio map can be defined
as a function of position within the source, $\vec{x_{\circ}}$, and
wavelength, $\lambda$,
\begin{equation}\label{eq:tr}
p(\vec{x_{\circ}},\lambda)=\int{\rm
W}(\vec{x_{\circ}}-\vec{x})e^{2iRM(\vec{x})\lambda^2}d^2\vec{x}
\end{equation}
where $W$ is the restoring beam of the telescope and RM\footnote{See
section \ref{rm} for a full explanation of the rotation measure of a
source.} is the intrinsic rotation measure of the Faraday screen
in front of the source.

By assuming that the RM can be defined as a Gaussian distribution with
a dispersion $\sigma$ (T91), equation \ref{eq:tr} reduces to
$<p(\lambda)>=e^{-2\sigma^2\lambda^4}$ which is more commonly known as
Burn's law for a foreground depolarisation screen \citep{burn}.

Burn noted that this law only applies when the scale of the Faraday
screen fluctuations are much smaller than the physical size of the
source. Observations of Cygnus A by \citet{carilli} found that their
RM fluctuations were coherent over scales of order 10 kpc. This scale
length is comparable to the results found by \cite{prm89}, who find
scale lengths between 10-50 kpc, using a sample of high redshift, high
radio-luminosity sources. These coherence lengths are not
significantly smaller than the source size, violating one of the main
assumptions of Burns' theory.

This is not the only problem associated with Burn's theory. In his
analysis of the polarisation properties of a source, Burn used the
vector $<p(\lambda)>$ as a test for the polarisation which falls off
rapidly to zero as the direction of the polarisation field
randomises. Burn also predicts that the observed polarisation
decreases steeply as a function of wavelength, which is very rarely
observed. Table \ref{best_alpha} demonstrates than even in a small
sample of radio galaxies the decrease of the polarisation fraction
with wavelength varies widely from the Burn relationship,
$p\,\sim\lambda^{-4}$. This confirms the assumption that by using Burn's
theory to shift the depolarisation to a common redshift in section
\ref{sec:bdep} I am almost certainly over-estimating the
depolarisation.  Any observations of the polarisation of radio
emission measures the length of the polarisation vector, thus
$<|p(\lambda)|^2>$ is a better indicator of the polarisation compared
to Burn's $<p(\lambda)>$. In the next section I explain the main
points of the T91 models which use $<|p(\lambda)|^2>$ to test the
Faraday medium around a source. For full details see T91.

\section{The Tribble models}

The polarisation distribution as a function of wavelength, independent
of any source model, is given by,
\begin{equation}\label{eq:t1}
<|p\,(\lambda)\,|^2>=\int R(\vec{s})\,\xi _p(\vec{s})\,d^2\vec{s}
\end{equation}
where $R(\vec{s})$ is determined by the size of the telescope beam $W$
on the sky and $\xi _p(\vec{s})=\xi _f(\vec{s},\lambda)\,\xi
_{\varepsilon}(\vec{s})\,\xi
_{\theta}(\vec{s})$. $\xi_{\varepsilon}(\vec{s})$ determines how the
emissivity of neighbouring regions interact and becomes important when
a radio source has prominent filamentary structures \citep{feb89}.
$\xi _{\theta}(\vec{s})$ determines how the intrinsic polarisation
angle of neighbouring regions interact. For a uniform source both the
emissivity, $\xi _{\varepsilon}(\vec{s})$, and the angle, $\xi
_{\theta}(\vec{s})$, are unity, therefore $\xi _p(\vec{s})$ depends
only on the distribution of the external Faraday screen, $\xi
_f(\vec{s},\lambda)$.

$\xi _f(\vec{s},\lambda)$ is the critical component in the determination
of the polarisation as this relates directly to the structure
function of the RM fluctuations, $D(\vec{s})$.
\begin{equation}
\xi _f(\vec{s},\lambda) = e^{-2\lambda^4D(\vec{s})}
\end{equation}
where $D(\vec{s})$ determines the RM distribution in a source and how
neighbouring regions of local RM are correlated, i.e.
$D(\vec{s})=<[RM(\vec{x}+\vec{s})-RM(\vec{x})]^2>$.  The exact form
of the structure function depends on whether or not the RM field is
assumed to be dominated by Quadratic fluctuations, Gaussian
auto-correlation fluctuations or Power-law fluctuations.

The exact form of $R(\vec{s})$ is dependent on the structure of the
telescope beam,
\begin{equation}
R(\vec{s}) = \int \,W(\vec{x})\,W(\vec{x}+\vec{s})\,d^2\vec{x}
\end{equation}
assuming a Gaussian beam this reduces to,
\begin{equation}
R(\vec{s}) = \frac{1}{2\pi t^2}\;e^{-|s|^2/t^2}
\end{equation}
where $t$ is the physical size of the telescope beam. The Tribble
models are calculated in the restframe of the source. The size of the
RM fluctuations, \so, $t$ and the observing wavelength, $\lambda$ must
all be calculated in the source frame of reference.

\subsection{Quadratic model}
\index{Quadratic model}
For the quadratic model the structure function is given by,
\begin{eqnarray}
D(s) =& \hspace{1.2cm}2\sigma^2(s/s_{\circ})^2 & s<s_{\circ}\\ \nonumber
D(s) =& 2\sigma^2 & s>s_{\circ}
\end{eqnarray}
where $s_{\circ}$ is the size of the RM fluctuations within the
telescope beam, $t$. The number of independent RM cells within a
telescope beam is given by $N=(t/s_{\circ})^2$. A completely unresolved
source has $s_{\circ} \ll t$ whereas a completely
resolved source has $s_{\circ} \gg t$.

The structure function can be substituted into equation \ref{eq:t1} to
give an expression for the polarisation,
\begin{equation}\label{eq:qu}
<|p\,(\lambda\,)|^2> = \frac{1-e^{-\alpha-\beta}}{1+2\beta / \alpha}\,+\,e^{-\alpha-\beta}
\end{equation}
where $\alpha=\s_{\circ}^2/2t^2$ and $\beta=4\sigma^2\,\lambda^4$

However, to calculate the observed polarisation at any given wavelength
the intrinsic $\sigma$ of the source must also be known and this is 
affected by the choice of the structure function.

For the quadratic model,
\begin{equation}
\sigma^2_{\rm observed}=\sigma^2_{\rm RM}[1-(1-e^{-\alpha})/\alpha]
\end{equation}
where $\sigma_{{\rm observed}}$ is taken from the observations of the
RM variations but the exact value of \so\, is not known. In theory it
should be possible to estimate the value of \so\, from high resolution
RM maps as \so\, should be comparable to the scale length over which
the RM is roughly constant. This would require higher resolution maps
than those presented here, thus it is not possible to directly
estimate the value of \so\, from my observations. In section
\ref{sec:tfit} I discuss the effects of varying \so\, on the observed
polarisation and the RM variations.

\subsection{Gaussian auto-correlation model}
\index{Gaussian model}
The Gaussian auto-correlation structure function is given by,
\begin{equation}
D(s) = 2\sigma^2[1-e^{-s^2/s_{\circ}^2}]
\end{equation}
Again substituting the structure function into equation \ref{eq:t1}
gives,
\begin{equation}
<|p\,(\lambda\,)|^2> = \alpha e^{-\beta}\,\int^1_0\, t^{\alpha-1}\,e^{\beta t}dt
\end{equation}
and the dispersion in the RM is given by,
\begin{equation}
\sigma^2_{\rm observed}=\frac{\sigma^2_{\rm RM}}{1+1/\alpha}.
\end{equation}

\subsection{Power-law model}
\index{Power-law model}
The power law model is slightly more complicated and depends on the
choice of $m$, the power-law exponent. The structure function is given by,
\begin{equation}
D(s) = 2\sigma^2\,[1-(1+2s^2/ms^2_{\circ})]^{-m/2}
\end{equation}
which using equation \ref{eq:t1} gives,
\begin{equation}\label{eq:powerlaw}
<|p\,(\lambda\,)|^2> = (m/2)\,\alpha\, e^{\alpha-\beta}\,\int^{\infty}_1 e^{-m\alpha\nu/2}\,e^{\beta \nu^{-m/2}}d\nu.
\end{equation}
This form of the structure function results in the relation
$p\,\sim\lambda^{-4/m}$.

The dispersion of RM in this model is much more complex than either of
the previous models and is given by,
\begin{eqnarray}\label{eq:powerlaw2}
\sigma^2_{\rm observed}&=&\sigma^2_{\rm RM}[m\alpha/2]^{m/2}\,\Gamma(1-m/2,m\alpha/2)\,e^{m\alpha/2}\\
{\rm where} \hspace{1cm}\Gamma(a,x)&=&\int^{\infty}_x\,e^{-t}\,t^{a-1}dt. \nonumber
\end{eqnarray}
In all 3 models when the source is completely resolved $\sigma^2_{\rm
observed}=\sigma^2_{\rm RM}$.

\subsubsection{Power law exponent}
T91 uses $m=2$ in his analysis, but a sample of radio galaxies may
show a range of exponents. The only way to determine how polarisation
decreases with wavelength is to take a sample of sources with at least
3 sets of polarisation information widely separated in frequency, but
also at high enough resolution that the polarisation measurements are
not affected by beam depolarisation. The sample must also be observed
at the same resolution at all three frequencies.  Although there is a
large volume of polarisation information in the archives, it proved to
be difficult to find a large enough sample of sources with these
requirements. Table \ref{best_alpha} shows the best fit to
p$\,\sim\lambda^{-4/m}$ using sources taken from \citet{ag95} with
observations at 1.4, 4.8 and 8.4 GHz. In general I find that the
polarisation decreases with 1$\lesssim$m$\lesssim$ 4. The north lobe
of 3C 266 shows a much steeper decrease in its polarisation, but as
Figure \ref{Tgood3} demonstrates the effects of changing $m$ are
small. I use $m=2$ and $m=4$ in the following sections as limiting
cases.
\begin{table}[!h]
\centering 
\begin{tabular}{r@{ }lcc}\hline
\multicolumn{2}{c}{Source} & Component & $-4/m$\\\hline
 3C& 67 &    T &   1.03  \\
 3C& 173&    T &     1.79 \\
 3C& 213.1&  T& 1.72 \\
 3C& 266&  N&  0.26  \\
 3C& 266&  S& 1.72  \\
 3C& 299&  S &  2.34  \\
 3C& 305.1&  T&   3.17  \\
 3C& 309.1&  T&    1.29  \\
 3C& 346&  E &  3.94  \\
 3C& 380&  E   &3.16  \\
 3C& 380&  W   & 1.62  \\
 3C& 455&    T&    2.56  \\ \hline
\end{tabular}
\caption[Calculating $m$ using archival observations]
{Calculating the range of exponents ,$m$, seen in sources observed by \citet{ag95}
}\label{best_alpha}
\end{table}
\section{Fitting the T91 models to the observations}\label{sec:tfit}
The T91 models have 2 free parameters, \so\, and $\sigma_{RM}$ and two
parameters taken from observations, $t$ and the observed variations in
the RM, $\sigma_{observed}$, plus a third parameter, $m$, for the
power-law model.  There are three basic assumptions that can be made
about the values of \so\, and $\sigma_{RM}$:
\renewcommand{\labelenumi}{\roman{enumi}.)}
\begin{enumerate}
\item 
The observations of $\sigma_{{\rm observed}}$ and $z$ can be fitted
assuming a constant \so\, and a constant $\sigma_{RM}$ with respect to
redshift. If \so\, is constant at all redshifts then this assumes that
cells in the Faraday screen are always the same size independent of
the source morphology or location and will have the same effect on the
underlying RM distribution independent of redshift. By also setting
$\sigma_{RM}$ to be constant with redshift I am assuming that all
sources have the same underlying RM distribution. This is a rather
naive picture as Table \ref{rsRM} shows a strong evolution in the
observed variations in the RM with redshift.
\item 
A more realistic assumption is that the observations of $\sigma_{{\rm
observed}}$ and $z$ can be fitted assuming a constant $\sigma_{RM}$ and
varying \so\, as a function of redshift i.e. \so = $\gamma(1+z)^\mu$.
\item 
Conversely, the observations may also be fitted assuming a constant \so\, and
varying $\sigma_{RM}$ as a function of redshift, i.e. $\sigma_{RM} =
\gamma(1+z)^\mu$.
\end{enumerate}

\index{AMOEBA}
I use the Numerical Recipes subroutine {\small AMOEBA} \citep{AMEOBA}
to estimate the best fits for the three assumptions presented
above. This is done by calculating the least squares difference
between the predicted $\sigma_{{\rm observed}}$ and the true
$\sigma_{{\rm observed}}$ for each source. The best-fit is then
given by the values of \so\, and $\sigma_{RM}$ which gives the smallest least
squares value\footnote{Least squares value = (model -
observed)$^2$}. These results are presented in Table
\ref{tab:best}. Unfortunately, no best-fit solutions can be given
for either Power-law model when both free parameters are constant with
redshift. This is simply due to the fact that the Gamma function in
equation \ref{eq:powerlaw2} is very sensitive to the exact choice of
\so\, and $\sigma_{RM}$ and no reasonable fit can be achieved.
\begin{table}[!h]
\centering 
\begin{tabular}{c l@{\si\, = }r@{.}l l@{\so\, = }r@{.}l}
\multicolumn{7}{c}{\bf constant \si, constant \so}\\
Gaussian && 89&1 && 15&9\\
Quadratic && 85&2 && 3&5$\times10^3$\\
\multicolumn{7}{c}{\bf constant \si, varying \so}\\
Gaussian && 523&4 &&0&03(1+z)$^{5.5}$\\
Quadratic && 173&3&&1&20(1+z)$^{3.2}$\\
Power-law with $m=2$ && 529&6&&0&01(1+z)$^{6.1}$\\
Power-law with $m=4$&& 504&6 && 0&03(1+z)$^{5.2}$\\
\multicolumn{7}{c}{\bf constant \so, varying \si}\\
Gaussian &&6&1(1+z)$^{4.7}$&&5&5\\
Quadratic &&4&9(1+z)$^{4.6}$&&2&0\\
Power-law with $m=2$ &&6&9(1+z)$^{4.8}$&&2&4\\
Power-law with $m=4$&& 5&4(1+z)$^{4.4}$&&12&0\\
\end{tabular}
\caption[Parameters of the best fit]
{Parameters of best fit. \so\, in units of kpc and \si\, in units
of rad m$^{-2}$. All results are in the sources' frame of
reference.}\label{tab:best}
\end{table}
\subsection{Constant \so, constant $\sigma_{RM}$}
Figure \ref{Tbad} demonstrates that the models with constant \so\, and
constant $\sigma_{RM}$ do not adequately describe the data. Both
models, despite having slightly different $\sigma_{RM}$ and
drastically different \so, have similar solutions. The only
discernible difference occurs at very low redshift where the Quadratic
model quickly falls to zero. It is obvious then that there must be
some evolution with redshift in either \so\, or $\sigma_{RM}$ to
account for the observed correlation between $\sigma_{{\rm observed}}$
and redshift.
\bfi[!h]
\centerline
{\psfig{file=Plotbad_c.ps,width=10cm,angle=270}}
\caption[Tribble model for constant \so\, and \si] {Best fit solution
of the Faraday screen keeping \so\, and \si\, fixed for the Quadratic
and Gaussian models. The $\sigma_{{\rm observed}}$ values for all
lobes are plotted as `$\times$'.}
\label{Tbad}
\efi

\subsection{Varying \so\, and $\sigma_{RM}$}
Figure \ref{Tgood1}(a) and \ref{Tgood1}(b) present the best fit
solutions to the Gaussian auto-correlation model and the Quadratic
model when \so\, is varied keeping $\sigma_{RM}$ constant and also
when $\sigma_{RM}$ is varied keeping \so\, constant. Figures
\ref{Tgood1}(a) and \ref{Tgood1}(b) use the RM dispersion of the lobe
and that averaged over the entire source, respectively when calculating
the best-fit solutions. It is obvious that the solutions, although
using subtly different data, produce almost identical fits. For this
reason I use only the average source values in all further fits.

\bfi[!h]
\centerline{\psfig{file=Plot5.2a_c.ps,width=8.5cm,angle=270}
\psfig{file=Plot5.2b_c.ps,width=8.5cm,angle=270}}
\caption[Tribble models for varying \so\/ and \si\/] {Best fit
solutions of the Faraday screen for the Quadratic and Gaussian
auto-correlation models. Line styles indicate whether \so\, or \si\,
are being varied as a function of (1+z). (a - left) Solution uses all
lobe $\sigma_{{\rm observed}}$ values in the and all lobe
$\sigma_{{\rm observed}}$ values are plotted. (b - right) Solution
uses only the average $\sigma_{{\rm observed}}$ values for each source
and the average source $\sigma_{{\rm observed}}$ values are plotted.}
\label{Tgood1}
\efi

The best-fit solutions to the Power-law model with exponents $m=2$ and
$m=4$ are shown in Figure \ref{Tgood3}. Comparing all 8 solutions
presented in Figures \ref{Tgood1}(b) and \ref{Tgood3} demonstrates
that by using the $\sigma_{\rm observed}$ from observations it is impossible to
determine whether or not it is \so\, varying with redshift or
$\sigma_{RM}$ varying with redshift that is causing the observed trend
of $\sigma_{{\rm observed}}$ with redshift. All 8 solutions are
statistically identical in terms of their least squares fit value.
The only noticeable difference occurs at very low
redshifts when \so\, varies as a function of redshift and
$\sigma_{RM}$ is kept constant. Each of the different models shows a
slight upturn as $z\rightarrow$0. This is simply due to the fact that
as $z\rightarrow$0, the beam size approaches zero and hence
$\alpha\rightarrow \infty$ and so $\sigma^2_{\rm observed}=\sigma^2_{\rm
RM}$. This is similar to the case when a source is completely resolved.

\bfi[!h]
\centerline{\psfig{file=Plot3_c.ps,width=10cm,angle=270}}
\caption[Tribble models for varying $m$, the exponent of the Power law model]
{Best fit solutions of the Tribble models for the Faraday screen for a
Power-law model with exponents $m=2$ and $m=4$ plotted against the
average $\sigma_{\rm observed}$ values for each source.}
\label{Tgood3}
\efi
\index{depolarisation}
\subsection{Depolarisation}\label{section:dep}
Although there is no significant difference between the solutions with
varying \so\, as a function of redshift compared to varying
$\sigma_{RM}$, I folded the results presented in Table \ref{tab:best}
into the corresponding expressions for polarisation given in equations
\ref{eq:qu} to \ref{eq:powerlaw}. This allows the fractional
polarisation that would be observed at 4.8 GHz and also at 1.4 GHz to
be calculated. These values can then be used to calculate the
depolarisation parameter DM$^{4.8}_{1.4}$ (see section
\ref{galactic}).

\subsubsection{Constant \so, constant $\sigma_{RM}$}
Figure \ref{Dcon} shows that the calculated Quadratic depolarisation
keeping \so\, and $\sigma_{RM}$ constant with respect to redshift, is
always 1. It is obvious that this model fails to describe the data.
When the Gaussian auto-correlation model is used to calculate the
depolarisation, it predicts more depolarisation at lower redshifts,
underestimating the higher redshift depolarisation measurements. Both
models fail to describe the data confirming my earlier finding that
there must be some evolution in either \so\, or \si.
\bfi[!h]
\centerline
{\psfig{file=Plot1DM_c.ps,width=10cm,angle=270}}
\caption[Tribble model predictions for the depolarisation I] {Model
prediction for the depolarisation with constant \so\, and \si\, obtained
from the best-fit solutions using the Quadratic and Gaussian
auto-correlation models. The data points represent the average
observed source depolarisation.}
\label{Dcon}
\efi
\subsubsection{Varying \so\, and $\sigma_{RM}$}
\bfi[!h]
\centerline{\psfig{file=Plot4_c.ps,width=10cm,angle=270}}
\caption[Tribble model predictions for the depolarisation II] {Model
predictions for depolarisation using all models with a constant
\si\, and a varying \so\, with redshift, obtained from the best-fit
solutions in Table \ref{tab:best}. Data points are as in Figure
\ref{Dcon}.}
\label{Dsig}
\efi 
All solutions presented in Figures \ref{Tgood1} and
\ref{Tgood3} are equally significant independent of
whether or not \so\, or $\sigma_{RM}$ is varied. The solutions are
also independent of which model is used to calculate the observed RM
variations. However, there {\it is} considerable difference between
the calculated depolarisations as shown in Figures \ref{Dsig} and
\ref{Dso}. Figure \ref{Dsig} uses only the solutions which have a
constant $\sigma_{RM}$ and varying \so. In all cases the
depolarisation is widely over-estimated. In the most extreme cases
predicting depolarisations in excess of 30. Such high depolarisations
have never been observed, even at high resolution
\citep[e.g][]{blr97,carilli,bcg98,dtb97}. In the case of the Quadratic
model the solution even predicts a strong anti-correlation of
depolarisation with redshift which has never been observed.  
\bfi[!h]
\centerline{\psfig{file=Plot5_c.ps,width=10cm,angle=270}}
\caption[Tribble model predictions for the depolarisation III]
{As Figure \ref{Dsig} but with a constant \so\/ and varying \si\/ with
redshift.}
\label{Dso}
\efi
In striking contrast Figure \ref{Dso}, which shows the results for
constant \so\, and varying $\sigma_{RM}$, is a much better fit to the
observations. At lower redshifts ($z<0.5$) all models are equally
likely and the scatter in the depolarisation encompasses all the
solutions. It is only when the higher redshift ($z>$0.5) sources are
considered that I am able to determine a significant difference
between the models. Both the Power-law model (with an exponent $m=2$)
and the Quadratic model overestimate the observed
depolarisation. However, the $m=4$ Power-law solution and the
Gaussian auto-correlation solution give results that are consistent
with the observations. The fact that there is some degree of scatter
at all redshifts suggests that a fixed \so\, is a simplification and
\so\, may vary slightly from source to source. However, the dominant
effect is the variation of \si.

In the next section I consider the effects of varying \so\, slightly
using the Gaussian auto-correlation model and the Power-law model with
$m=4$.

\subsection{Scatter in \so}
\begin{figure}[!h]
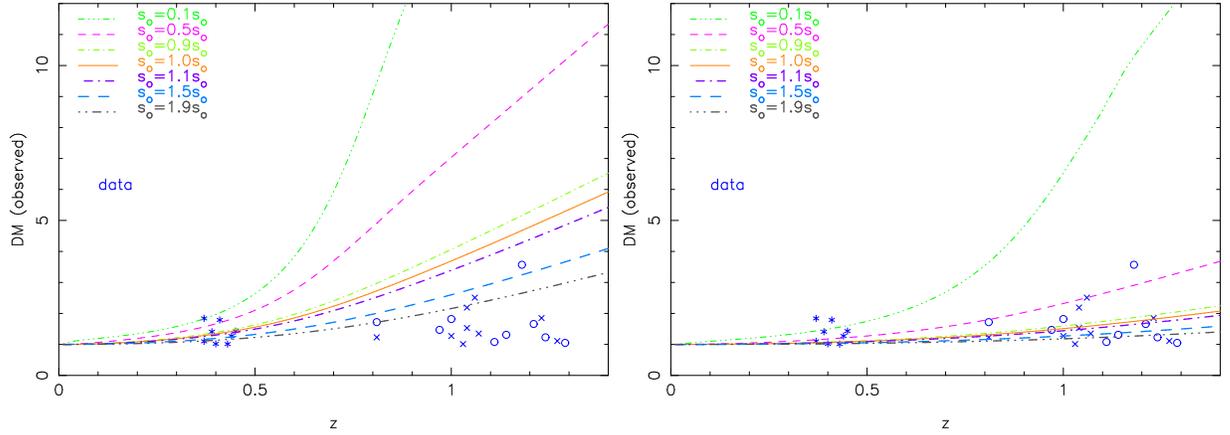

\centerline{\psfig{file=scatter2.ps,width=8cm,angle=270}
\psfig{file=scatter1.ps,width=8cm,angle=270}}
\caption[Effects of varying \so]
{(a - left) Varying the value of \so\, using the Gaussian
auto-correlation model and (b - right) the Power-law model with
$m=4$. In both plots the bold line indicates the initial \so\, taken
from Table \ref{tab:best}.}
\label{gau}
\end{figure}

Figure \ref{gau}(a) shows the Gaussian auto-correlation model with
$\sigma_{RM}=6.1(1+z)^{4.7}$ \rad\, with \so\, varying around the
best-fit value, \so=5.5 kpc. By allowing \so\, to vary by a maximum
of $\pm90\%$ from the best fit value I can determine the effects of
changes in \so\, on the data. Figure \ref{gau}(b) is the $m=4$
Power-law model with $\sigma_{RM}=5.4(1+z)^{4.4}$ \rad, again \so\,
is allowed to vary around the corresponding best-fit value of
\so=12.0 kpc.
In both cases the low redshift results ($z<$0.5) are largely
insensitive to changes in \so. This is not the case at the higher
redshifts ($z>$ 0.5). As \so$\rightarrow\pm90\%$\so\, the
depolarisation becomes increasingly unrealistic. In both models using
\so=0.1\so\, produces large over-estimates in the high redshift
depolarisations. This is also true when \so=0.5\so\, using the
Gaussian auto-correlation model. This suggests that although \so\,
doesn't vary with redshift, a small amount of variation can describe
the scatter in the observed data seen in the DM-z plots of the
previous chapters. Figure \ref{gau}(b) is the best fit to the data
suggesting that the Power-law model with $m=4$ is perhaps the most
accurate and the small variations in \so\, do aptly describe the
data. This model predicts $p\sim\lambda^{-1}$ compared to the Burn-law
$p\sim\lambda^{-4}$.

It must be noted that the fitting procedure only
uses 23 out of a possible 26 sources\footnote{As noted in the previous
chapters 3 sources have no RM information in at least one lobe.}. To
gain a better estimate of which model is the most accurate a much
larger sample, preferably with higher resolution, would have to be
used. By using a sample of radio galaxies at high resolution the
variations in \so\, would be less significant and the sources would be
well resolved and hence \si$\approx \sigma_{\rm observed}$.

\subsection{Using \so=10 kpc}
As I have already noted my RM maps are not of sufficient quality to
determine how the RM structure changes over each lobe of a source. In
several sources (e.g. 3C 299 and 6C 0943+39) I have only a few pixels
of RM data in each lobe. In section \ref{section:dep} I demonstrated
that keeping \so\, constant and varying $\sigma_{RM}$ as a function of
redshift is the most realistic of all the solutions presented. In all
cases \so\, was found to be less than 12 kpc.

The method described in the previous sections does not include any
information on how the depolarisation differs with changes in
radio-luminosity. Figures \ref{Tgood1} to \ref{Tgood3} show that the
two high redshift samples (`$\times$' = low radio-power sample A and
`$\circ$' = high radio-power sample B) are equally distributed around
the solutions. This method does not enable any determination of how
depolarisation is affected by changes in P$_{151}$ and \DL. To
overcome this problem, I set \so=10 kpc and calculate $\sigma_{RM}$
using the observed rms variation in the rotation measure, $\sigma_{\rm
observed}$, and the known beam size. This calculated $\sigma_{RM}$ is
then used to work out the depolarisation for the source if it was
located at $z=1$.  Only by doing this can I determine if any
depolarisation trends are really due to redshift effects and not
simply due to the fact that a third of my sample is located at
$z=0.4$. This is the same method as used in section \ref{sec:bdep} to
shift the depolarisation using Burn's law.

\begin{table}[!h]
\centering
\begin{tabular}{l@{ = }rl@{    }rl@{ = }r}
\multicolumn{6}{c}{\bf Quadratic model results}\\
r$_{z,DM}$ &   0.67857&&& t &  6.13\\
r$_{P,DM}$ &   0.49405&&& t &  3.77\\
r$_{D,DM}$ &  -0.21899&&& t & -1.49\\
\multicolumn{6}{c}{\bf Gaussian model results}\\
r$_{z,DM}$ &  0.67894&&& t &  6.13\\
r$_{P,DM}$ &  0.49541&&& t &  3.78\\
r$_{D,DM}$ & -0.22294&&& t & -1.52\\
\multicolumn{6}{c}{\bf Power-law model results}\\
r$_{z,DM}$ &  0.67771&&& t & 6.1\\
r$_{P,DM}$ &  0.49171&&& t & 3.75\\
r$_{D,DM}$ & -0.23059&&& t &-1.57\\
\end{tabular}
\caption{Spearman rank results for \so=10 kpc}\label{tab:tresults}
\end{table}

As Table \ref{tab:tresults} demonstrates, I always find a strong
DM$_{z=1}$--z correlation with a weaker correlation with
radio-luminosity, independent of the model used. There is never a
significant anti-correlation with size. These results are consistent
with the results found in the previous chapters, obtained from Burn's
law.

\section{Conclusions}

I have used the T91 models to determine how the size of the cells in
an external Faraday screen, \so, varies with redshift and also to
determine if there is any evolution in the intrinsic RM variations,
\si, with redshift.

I have compared the observed RM variations with the predicted RM
variations using a constant \so\, and \si\, with redshift and I find
that this model is inconsistent with the data. There
is no statistical difference between the models when \so\, is
varied compared to when \si\, is varied. Only by comparing the
predicted depolarisation with the observed depolarisation is there any
distinction between the models. In all cases when \so\, is varied,
keeping \si\, constant, the predicted depolarisation is
over-estimated.  Only by keeping \so\, constant and varying \si\, is
the predicted depolarisation similar to the observed depolarisation.

The Gaussian auto-correlation model and the Power-law model with
exponent $m=4$ are good fits to the data. When \so\, is allowed to vary
slightly in these models I find that it has little effect on the low
redshift sources, but as redshift increases the size of \so\, becomes
important. A small amount of variation (typically $<$ 50\%) can be
introduced to the models to allow for the obvious scatter in the
depolarisation-redshift plots. In all cases it is the higher redshift
sources that constrain the models.

A fixed cell size and varying \si\, with redshift gives further
evidence that the observed depolarisation is due to the increasing
disorder with redshift and not due changes in the density of the
environment with redshift (see previous chapter).

By setting \so=10 kpc and calculating \si\, using the observed
$\sigma_{\rm observed}$ and the known beam size I am able to determine
how the depolarisation is affected by changes in radio-luminosity and
radio-size. Irrespective of the model I use I always find that
depolarisation correlates strongly with redshift and with
radio-luminosity. In no cases do I find an anti-correlation with
size. These findings agree with the results presented in sections
\ref{sec:bdep} and \ref{4:dep}. Although the Burn's law over-estimates
the degree of depolarisation, I find similar results using the T91
models, which are more realistic than the Burn model. This proves that
the simplistic Burn model can still be used to find general trends
with depolarisation in a sample, irrespective of the fact that it may
overestimate the depolarisation.  Interestingly I find that
$p\sim\lambda^{-1}$ best describes the data, compared to the widely
used Burn $p\sim\lambda^{-4}$ result.

\newpage
\thispagestyle{empty}
\chapter[Modelling FR II sources]{Modelling FRII sources} 
\index{FRII modelling} 

In the previous chapters I have used depolarisation and the variations
in the rotation measure to sample the environment around powerful FRII
radio galaxies. However, In chapter 4 I found evidence that these
quantities do not sample the density of the source environment
\citep[e.g.][]{gc91}, but are indicators of the degree of disorder of
the Faraday screen.

In the following I use the model for the evolution of the radio
luminosity of FRII-type sources as a function of source age by
\citet[][hereafter KDA]{kat97} to model the source spectra. This
allows a more theoretical approach to determine source properties such
as the age of the jets and their energy transport rate or jet
power. The model also provides an estimate for the density of the
gaseous source environment. By comparing observations of the source
depolarisation and variations in the RM with model estimates of the
density, jet-power and lobe pressure allows an independent check on
the trends of the observational properties of the source environment
derived in the previous chapters.

I have obtained depolarisation and rotation measure observations for
only a small sample of 26 sources. Because the KDA model can be used
to model the radio spectra of a source I can apply it to the complete
3CRR, 6CE and 7C III samples. In practice I use flux measurements at 3
different observing frequencies, for each source, to constrain the KDA
model. I then analyse how the model parameters (jet power, density of
the environment, source age and lobe pressure) evolve as a function of
radio-luminosity, size and cosmological epoch. This gives a more
statistically significant result on the overall evolution of the
environments of powerful FRII sources.

\section{The model for FRII sources}\label{KDA:theory}
The large scale structure of FRII sources is formed from twin jets
emerging from a central AGN buried inside the nucleus of the host
galaxy. Each jet propagates outward from the core of the source and
terminates in a strong shock giving rise to the so--called radio
hotspots. The jet material inflates a lobe surrounding the jet, which
drives a bow shock into the surrounding medium. This model was first
proposed by \citet{scheuer}. \citet{ka97}, hereafter KA, showed that
in a purely dynamical model of FRII evolution, the bow shock and lobe
grow self--similarly, which has also been inferred from observations
\citep[e.g][]{lw84,lms89}. 

The dynamical model of KA assumes a simple power--law for the external
density distribution around the radio source, $\rho_x$,
\begin{equation} 
\rho_x=\rho_\circ\left(\frac{d}{a_\circ}\right)^{-\beta}
\end{equation}
where $d$ is the distance from the central AGN, $a_\circ$ is a scale height,
$\rho_\circ$ is the density at $d=a_\circ$ and $0 <
\beta < 2$.

The KA model also assumes that the rate of injection of energy into
the lobe, defined as the jet power, Q$_\circ$, is constant over the
lifetime of the source. Due to the high sound speed in the lobes, the
lobe pressure is uniform throughout except for the regions close to
the hotspots. It is the lobe pressure, \p\,, that confines the jets.

The KA model predicts that the length of a single jet, D$_{lobe}$,
grows with time as
\begin{equation}\label{eq:dlobe}
D_{lobe}=c_1\left(\frac{t^3 Q_\circ}{a_\circ^\beta\rho_\circ}\right)^{\frac{1}{5-\beta}} 
\end{equation}
where $c_1$ is a constant which depends on the geometry of the lobe,
the thermodynamic properties of the jet material and the gas in the
source environment (see KA for details). $t$ is the age of the source.

KDA added a prescription for calculating the synchrotron radio
emission of the lobes to the dynamical model of KA. For this the lobe
is split into small volume elements, $\delta V$. These are injected
into the lobe at time $t_i$ containing a population of relativistic
electrons with a power--law energy distribution with exponent $p$
ranging from $\gamma=1$ to some $\gamma_{max}$. Here, $\gamma$ is the
Lorentz factor of the relativistic electrons. This electron
distribution is then subject to energy losses due to adiabatic
expansion, synchrotron radiation and inverse Compton scattering of the
CMB. Since $t_i$ will be different for different volume elements, the
resulting energy distribution of the electrons in each $\delta$V at
time $t$ is a function of the injection time $t_i$.

The total radio emission of the lobe at a given frequency, $\nu$, is
the sum of all the contributions from the volume elements,
\begin{equation}\label{eq:p}
P_\nu=\int \frac{1}{6\pi}\sigma_T\,c\, u_B\, \frac{\gamma^3}{\nu}
n(\gamma)\,\delta V(t)
\end{equation}
where $\sigma _T$ is the Thomson cross section, $u_B$ is the energy
density of the magnetic field and $n(\gamma)$ is the density of
relativistic electrons with Lorentz factor $\gamma$.

In the model of KDA equation \ref{eq:p} becomes,
\begin{equation}\label{eq:p2}
P_\nu=\int^t_0\frac{\sigma_T\,c\,rQ_\circ\,n_\circ}{6\pi\,\nu\,(r+1)}\frac{(2R)^{\frac{2(1-\Gamma_c)}{\Gamma_c}}\gamma^{3-p}\;t_i^{a_1(p-2)/3}}{\left(t^{-a_1/3}-a_2(t,t_i)\gamma\right)^{2-p}}\left(\frac{t}{t_i}\right)^{-a_1(1/3 + \Gamma_B)}dt_i,
\end{equation}
where $n_\circ$ is the normalisation of the electron energy
distribution. $r$ is the ratio of the energy density of the magnetic
field and that of all particles at the time of injection, $t_i$. R is
the ratio of the length of the lobe and its width at half the lobe
length. The various $\Gamma$ are the adiabatic indices of the material
in the lobe, $\Gamma_c$, and the magnetic field, which is assumed to
be tangled on small scales, $\Gamma_B$.  $a_2(t,t_i) =
(t^{-a_1/3})/\gamma-(t_i^{-a_1/3})/\gamma_i$ where $a_1$
is the exponent of the power-law time dependence of the volume
elements, i.e. $\delta V \propto t^{a_1}$, which KDA find to be
$a_1=(4+\beta)/[\Gamma_c(5-\beta)]$. Finally, $\gamma_i$ is the
Lorentz factor at $t_i$ of electrons which have a Lorentz factor of
$\gamma$ at the current time, $t$.

There is no analytical solution for equation \ref{eq:p2} for general
$p$. However, for $p=2$ the integral simplifies considerably to give
\begin{equation}\label{eq:p_sol}
P_{\nu} = G
\frac{p_{lobe}^{7/4}}{s}\left[1-\left(\frac{t_{min}}{t}\right)^s\right]
\end{equation}
where
$s=a_1\left(\frac{1}{3}+\frac{3\Gamma_B}{4}\right)
+\big(\frac{8-7\beta}{20-4\beta}\big)$, and
\begin{equation}\label{eq:p_sol2}
G=\frac{\sigma_T\,r^{3/4}\,(\Gamma_x+1)\,(2R)^{2/\Gamma_c}\,D_{lobe}^3\,[ln(\gamma_{i,max})+\gamma_{i,max}^{-1}-1]^{-1}}
{108\,\pi\,\nu^{1/2}\,c_1^{(5-\beta)}\,c\,m_e^{1/2}\,(\Gamma_c-1)^{3/4}\,(k'+1)\,(r+1)^{7/4}\,(5-\beta)^{-2}}\left(\frac{2\pi^2}{e^2\mu_\circ}\right)^{1/4},
\end{equation}
where p$_{lobe}$ is the pressure in the lobe at the current time
$t$. $a_1$ determines $a_3=1-a_1(\Gamma_B+1/3)$ and
$a_4=1-a_1/3$. $\Gamma_x$ is the adiabatic index of the gas in the
source environment. The lobe may contain non-relativistic particles,
the ratio of their energy density and that of the relativistic
electrons is given by $k'$. Finally, $m_e$ is the electron mass and
$\mu_\circ$ is the magnetic permeability of the vacuum.\footnote{In
the derivation of equations
\ref{eq:p2} to \ref{eq:p_sol2} I have used the results of KA and KDA.}
\index{lobe pressure}

Older parts of the lobe will no longer radiate at the observing
frequency $\nu$ due to severe radiative losses, so it is not realistic
to integrate equation \ref{eq:p2} over the entire age of the
source. $t_{min}$ is the minimum injection time for volume elements to
still be emitting radiation at the observing frequency $\nu$. I assume
that the electrons only emit at their critical frequency
$\gamma^2(t,t_{i})\nu_L(t,t_{i})$ where $\nu_L$ is the Lamour
frequency, and use the condition that
$\nu=\gamma^2(t,t_{min})\nu_L(t,t_{min})$ to give an implicit equation
for $t_{min}$,

\begin{equation}\label{tmin}
\nu=\frac{et^{-(2a_1/3)}}{2\pi m_e}\left\{\frac{\sqrt{(2X\mu_o)}\;
t_{min}^{-(a_1\Gamma_c/2)}\;(t_{min}/t)^{\Gamma_Ba_1/2}(3m_ec)^2}
{(4\sigma_T)^{2}
\left[X\, t_{min}^{a_1(\Gamma_B-\Gamma_C)}(t^{a_3}-t_{min}^{a_3})/a_3
+u_c(t^{a_4}-t_{min}^{a_4})/a_4\right]^2}\right\}
\end{equation}
where $X=(r \,p_{lobe}\, t^{\Gamma_c\,a_1})/[(r+1)\,(\Gamma_c\,-1)]$,
u$_c$ is the energy density of the cosmic microwave background, $u_c
\propto (1+z)^4$, (see also KDA). There is no analytical solution for 
$t_{min}$, but it is possible to numerically find a $t_{min}$, for any
given source age and frequency, that solves equation \ref{tmin} to a
high degree of accuracy (i.e. within $1\times 10^{-4}$ or less.)

The model calculates all the properties in the sources' frame of
reference. This means that the observing frequency and $P_\nu$ have to
be transformed using the chosen cosmology, see appendix \ref{Cosm} for
details.

\subsection{Application of the model}
\label{model2}
The model for an individual radio lobe depends on a number of
parameters, some of which are determined from the radio
observations. These are the observing frequency, $\nu$, the length of
the lobe, D$_{lobe}$, the aspect ratio of the lobe, R, and the
monochromatic radio luminosity, $P_\nu$. The source redshift is taken
from archival observations, see section \ref{setup} and appendices
\ref{ap:3CRR} to \ref{ap:7C}.
Another set of parameters are not directly accessible by observations,
but their values are either well-constrained in general or are simply
set to reasonable values. The adiabatic indices are set to
$\Gamma_x=\Gamma_c=5/3$, assuming that the material in the lobe is
mainly non-relativistic. The magnetic field in the lobe is assumed to
be tangled on scales much smaller than D$_{lobe}$, therefore
$\Gamma_B=4/3$. Any particles in the lobe other than the electrons
responsible for the emission of synchrotron radiation are neglected,
so $k'= 0$. I further assume that the relativistic electrons and the
magnetic field are initially, at time $t_i$, given by their values
appropriate for minimum energy conditions
\citep[e.g.][]{longair}. Thus $r=3/4$ for the chosen slope of the
electron energy distribution with $p=2$. I set the high--energy
cut--off of this distribution to $\gamma_{i,max}=10^{18}$ and
$\gamma_{i,min}=1$. It is clear from equation \ref{eq:p_sol2} that the
exact value of $\gamma_{i,max}$ does not significantly influence the
results. Finally, the gaseous atmospheres around powerful radio
galaxies imply $1 \leq \beta \leq 2$ and so I set $\beta=1.5$. This
leaves three free model parameters: The current pressure in the lobe,
p$_{lobe}$, the age of the source, $t$, and the minimum injection
time, $t_{min}$.

D$_{lobe}$, the lobe ratio and the rest frame frequency of the
observations are used to determine the constant G (equation
\ref{eq:p_sol2}). Although p$_{lobe}$ and $t$ are the principal free
parameters of the model, in the fitting process
$a_\circ^\beta\rho_\circ$ is used instead of the lobe pressure. This
is due to the fact that p$_{lobe}$ and ty (the age of a source in
years) are highly anti-correlated, see Table \ref{rspc}, and applying
the model using these parameters could create a bias in the data. As
Table \ref{rsty} shows there is no significant
$a_\circ^\beta\rho_\circ$--ty correlation, making these two parameters
a good choice for the input parameters for the fitting
process. Initial `guesses' for $a_\circ^\beta\rho_\circ$ and ty are
used to calculate the lobe pressure, which depends on the jet
length and the lobe ratio.
\begin{eqnarray}
p_{lobe}&=&\frac{18\,c_1^{2-\beta}\,t^{-a_1\Gamma_c}}{4R^2(\Gamma_x+1)(5-\beta)^2}\left(\frac{a_\circ^\beta\rho_\circ}{Q_\circ^{(\beta-2)/3}}\right)^{3/(5-\beta)}\label{eq:ao}\\
&=&\frac{9a_\circ^\beta\rho_\circ
D_{lobe}^{2-\beta}}{2(\Gamma_c+1)(5-\beta)^2R^2ty^2}\label{eq:ao2}
\end{eqnarray}

\index{AMOEBA}
The age of a source and its density, $a_\circ^\beta\rho_\circ$, are
used to calculate the minimum injection time, $t_{min}$, (equation
\ref{tmin}) and the monochromatic luminosity, $P_\nu$, using equation 
(\ref{eq:p_sol}). This monochromatic luminosity is converted to a
specific flux, S$_\nu$ using equation (\ref{transform}). Comparing
S$_\nu$ with the observed flux it is possible to obtain a $\chi^2_\nu$
deviation of the model at frequency $\nu$. This gives three
measurements of $\chi^2_\nu$ at the three different frequencies. The
Numerical Recipes subroutine {\small AMOEBA}
\citep{AMEOBA}, is used to find the best estimate for ty and
$a_\circ^\beta\rho_\circ$ for a lobe by
minimising $\chi^2 = \chi_{\nu_1}^2+ \chi_{\nu_2}^2 +
\chi_{\nu_3}^2$. 

\index{jet power}
The jet power, Q$_\circ$, is then calculated from the final values of
ty, $a_\circ^\beta\rho_\circ$ and hence p$_{lobe}$,

\begin{equation}\label{eq:Qo}
Q_\circ = \frac{2p_{lobe}(\Gamma_c+1)(5-\beta)^2D_{lobe}^3 R}{9c_1^{5-\beta}t}
\end{equation}

For full details of the above model see KA and KDA.

\section[Modelling the small sample]
{Modelling the small sample of 26 sources}\label{model} 

The model uses the three flux measurements around 1.4 GHz, 1.6 GHz and
4.8 GHz and their corresponding errors to constrain the free
parameters of the model. D$_{lobe}$, is calculated from the observed
angle of the source on the sky, and the measured redshift using the chosen
cosmology. The lobe ratio is taken from the total intensity maps.

\subsection{Angle to the line of sight}
\index{orientation angle}
It is not possible to determine from the radio data whether or not the
jets in a given source lie in the plane of the sky or whether the
measured length of the lobe is projected. The length of the jet and
the lobe ratio are both affected by this angle dependence, however it
is the jet--length that is most affected.

To test how sensitive the model is to a change of the viewing angle,
each source is modelled with a number of viewing angles ranging from
10$^\circ$ to 90$^\circ$ to the line--of--sight. I found that in most
cases the source is most probably angled around 57.3$^\circ$ to the
line--of--sight, which is the average angle expected for a sample of
randomly oriented sources. In these cases there is no significant
difference in the results for any orientation unless the source is
angled to less than 23$^\circ$ to the line--of--sight. These very
small angles produced highly unlikely results, i.e. extremely large
sizes, improbable jet--powers and/or extremely high densities. Sources
which are already large, e.g. 3C 46, are unlikely to be orientated at
angles much less than 57$^\circ$ as this makes the source larger than
a few Mpc across. In several cases when the sources became large
(small orientation angle) the source age was not consistent with the
source size, i.e. less than 10$^6$ years for sources of order of a Mpc
across.

There are a few sources that all angles seemed equally probable
(e.g. 6C 0943+39 and 3C 280), but these are in the minority. To assume a
different angle for each source is impractical, especially in section
\ref{large} where 211 sources taken from the combined 3CRR,
6CE and 7C III samples are modelled. The model does not directly depend on the
line--of--sight so it is not possible to determine an orientation
angle from the fitting procedure.

For a sample of randomly orientated sources the most likely
orientation is 57.3$^\circ$ to the line--of--sight and only a few of
the sources will be orientated at angles smaller than 23$^\circ$. Thus
these sources will not have a large effect on the statistics. In all
the following sections the results have been calculated using an
orientation angle of 57.3$^\circ$ for all sources.

\section{Results}

The KDA model provides an estimate for the jet-power, density, lobe
pressure and source age for each lobe in the small sample. However, in
the next section I only consider how the density parameter, \ao\,,
relates to changes in depolarisation and variations in the rotation
measure. How the model parameters are affected by changes in the
`fundamental' parameters (i.e. redshift, radio-luminosity and source
size) is presented in detail in section \ref{large} using the combined
3CRR, 6CE and 7C III samples. By using a large sample any results found
will be more significant than with only 26 sources. However, the large
samples do not have depolarisation and rotation measure information
for each source. Therefore I use my smaller sample to determine if
depolarisation and density are related.

\subsection{Comparing \ao\, with DM$_z$ and \sig}\label{small}
\index{depolarisation}
\index{variations in RM}
\index{density of source environment}
In section \ref{4:dep} I found that the Burn corrected depolarisation
measure, DM$_z$, correlated strongly with redshift. By analysing the
\bmin\, asymmetries in section \ref{bmin:theory} I found evidence that
the observed depolarisation of a source was unaffected by changes in
the density of the environment. The KDA model gives an estimate of the
gas density around each lobe which can be compared with the source
depolarisation.
\begin{figure}[!h]
\centerline{\psfig{file=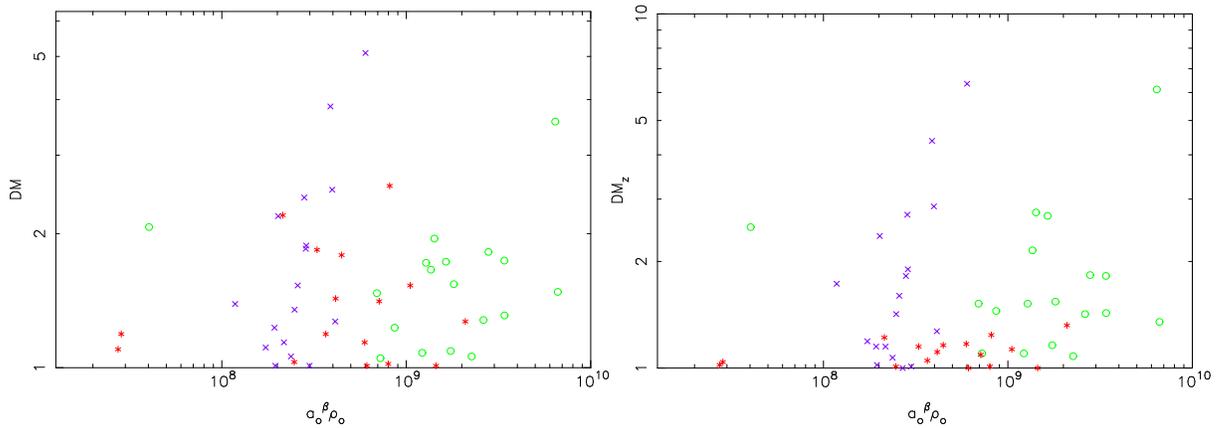,width=16cm}}
\caption[Relationship between depolarisation and \ao] 
{(a - left) Unshifted depolarisation against \ao\, and (b - right)
depolarisation shifted using Burn's law. Symbols are as in Figure
\ref{zd}.}
\label{aoDM}
\end{figure}
\begin{figure}[!h]
\centerline{\psfig{file=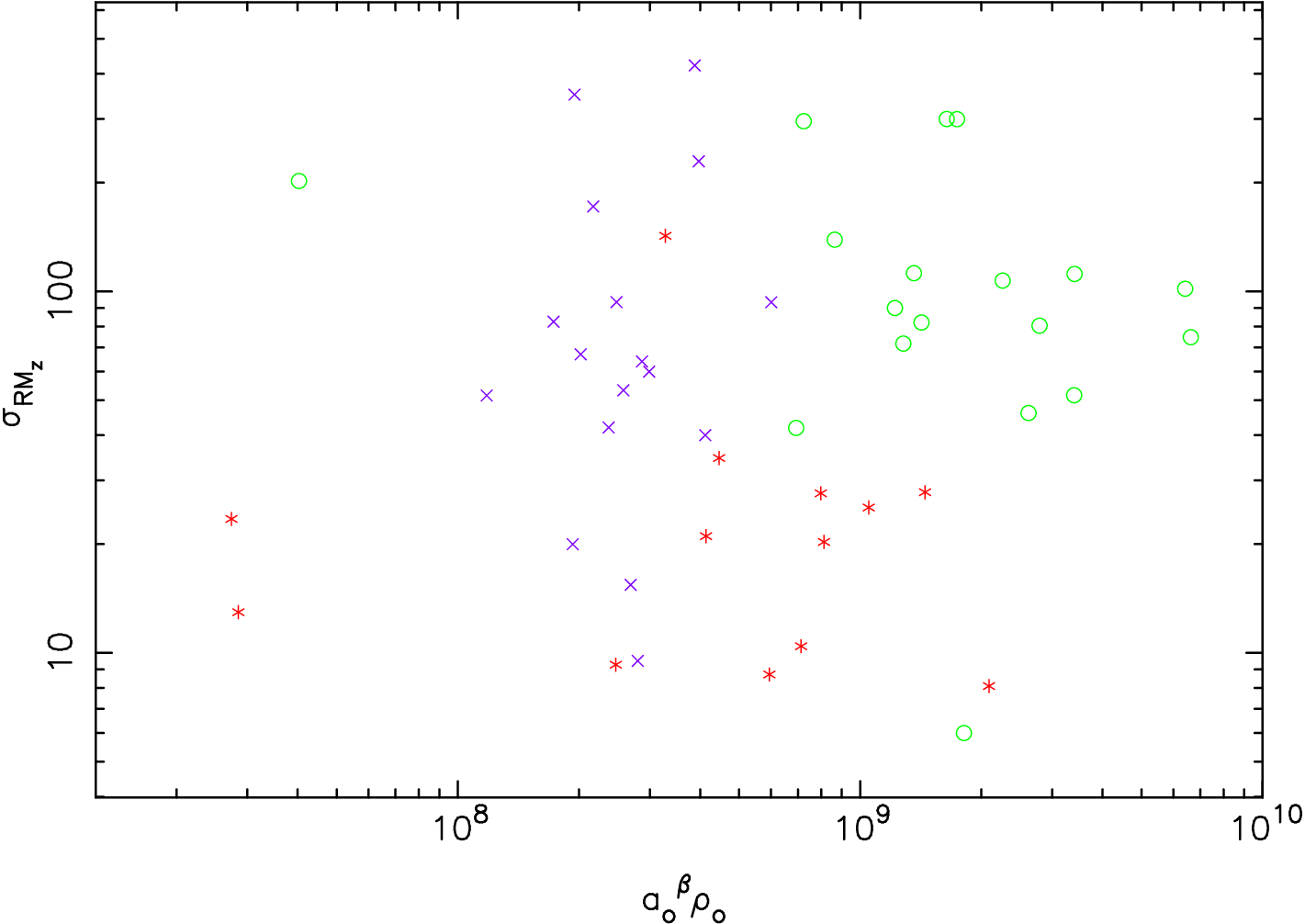,width=10cm,angle=270}}
\caption[Relationship between \sig\, and \ao\,] 
{\sig\, against \ao. Symbols are as in Figure \ref{zd}.}
\label{aosig}
\end{figure}

Figure \ref{aoDM} shows that there is no relationship between
depolarisation and \ao, independent of whether I use the observed
depolarisation (Figure \ref{aoDM}(a)) or I use the shifted
depolarisation from section \ref{galactic} (Figure
\ref{aoDM}(b)). Figure \ref{aosig} shows that there is also no trend
between the variations in the RM in the sources' frame of reference,
\sig, and \ao.

\begin{figure}[!h]
\centerline{\psfig{file=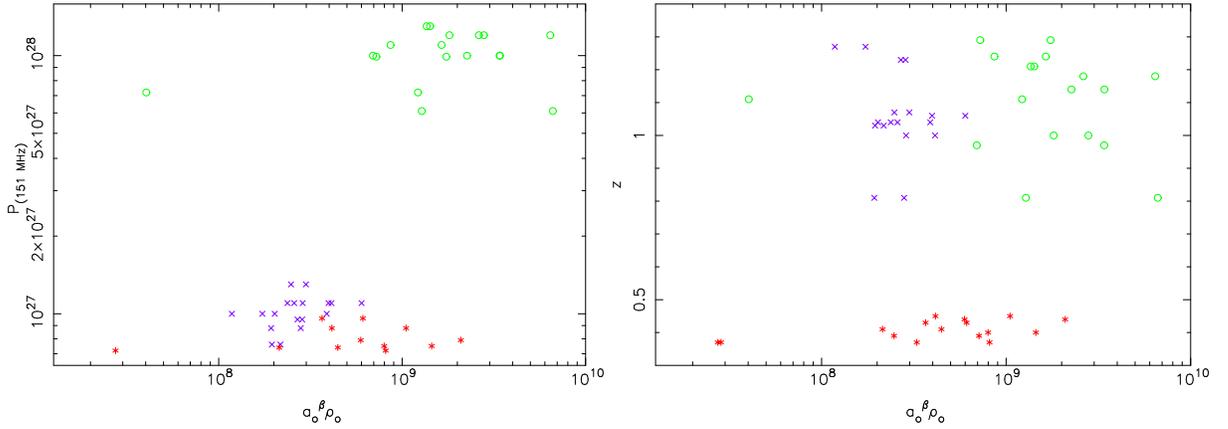,width=16cm}}
\caption[Relationship between redshift and radio luminosity with \ao] 
{(a - left) Radio-luminosity against \ao\, and (b - right) redshift
against \ao. Symbols are as in Figure \ref{zd}.}
\label{aozP}
\end{figure}

Figure \ref{aozP} shows a weak correlation between radio luminosity
and \ao\,, but no corresponding correlation with redshift. The
relationships between, z, \Po\, and \ao\, are discussed in detail in
section \ref{aoB} using the large sample. However, it is worth noting
that in the small sample there is no indication of any evolution of
the environment density with redshift. Thus the evolution of
depolarisation with redshift cannot be attributed to changes in
density.
In section \ref{4:dep} I also found a strong \Po-\DMz\, correlation
that was comparable to the z-\DMz\, correlation. As the Partial
Spearman results show, at any given radio-luminosity there is still no
significant \DMz-\ao\, correlation.
\begin{eqnarray}
r_{ a_\circ^\beta\rho_\circ\hspace{1mm}DM_z,\hspace{1mm}P_{151}}=&
   -0.119\hspace{2cm}& D = 0.83\nonumber\\
r_{a_\circ^\beta\rho_\circ \hspace{1mm}P_{151},\hspace{1mm}DM_z} =&
 \hspace{3mm} 0.526\hspace{2cm}& D = 4.05\nonumber\\
r_{P_{151} \hspace{1mm}DM_z,\hspace{1mm}a_\circ^\beta\rho_\circ} =&
  \hspace{3mm}  0.504\hspace{2cm}& D = 3.84\nonumber
\end{eqnarray}
At any given density the \Po-\DMz\, correlation still exists, thus any
changes in \DMz\, with either z or \Po\, are independent of changes in
the density of the source environment. The lack of correlation between
\ao\, and the observed polarisation parameters gives further
evidence that depolarisation and \sig\, should {\it not} be used as
tracers of the density of the environment
\citep[e.g][]{burn,gc91a,brw99} and should only be used as a measure for
the disorder in the environment.

\subsection{Minimum energy, \ao\, and p$_{lobe}$}\label{6:bmin}
\index{minimum energy} \index{lobe pressure} 

\begin{figure}[!h]
\centerline{\psfig{file=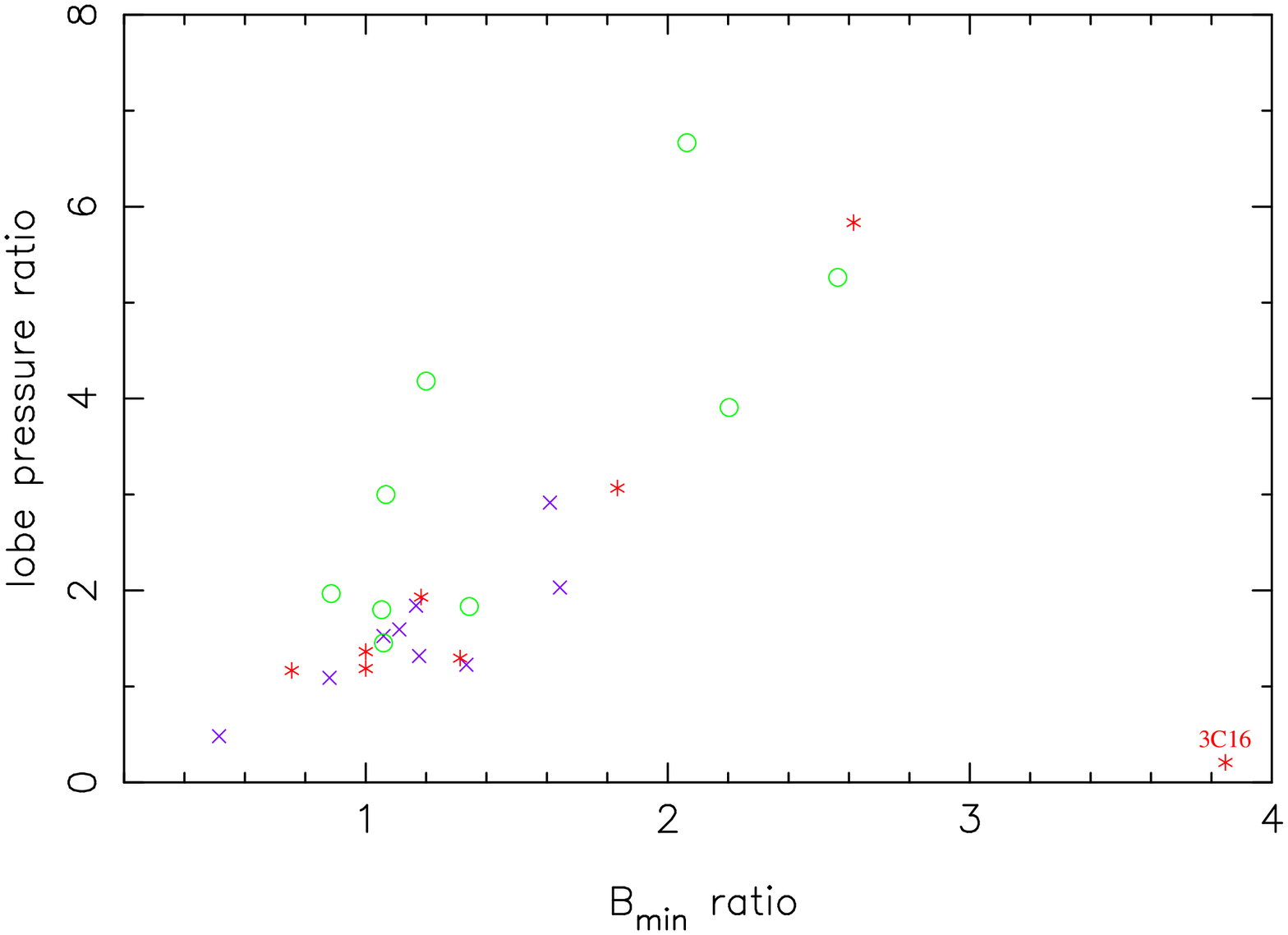,width=8cm}
\psfig{file=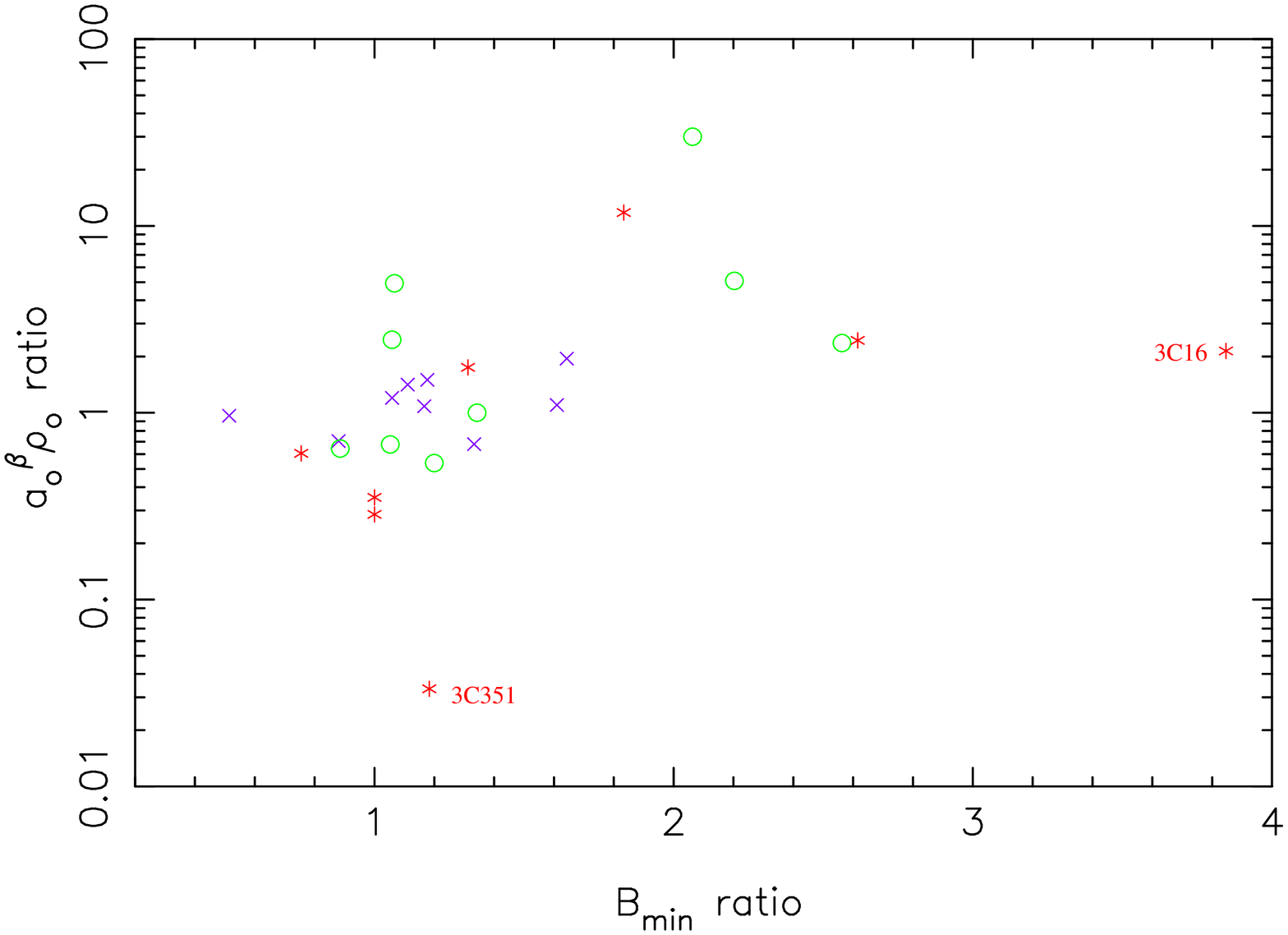,width=8cm}}
\caption[Relationship between \sig\, and \ao] 
{(a - left) Ratio of \bmin\, against the ratio of lobe pressures and (b
- right) against the ratio in \ao. Symbols are as in Figure \ref{zd}.}
\label{Bminao}
\end{figure}

Figure \ref{Bminao}(a) shows that there is a tight correlation between
the ratio of pressures in the lobes of a given source and the \bmin\,
ratio. The only notable exception being 3C 16, which has a high \bmin\,
ratio, but a relatively low ratio of lobe pressures. This agrees with
the suggestion that 3C 16 is in the process of switching back on
\citep[][also see section \ref{3C16}]{hh98}. Figure \ref{Bminao}(a)
demonstrates that the simple \bmin\, model can be used to give an
estimate of the lobe pressure. For the sources in the large sample,
discussed in section \ref{large}, there is no information on the
individual lobes of each source so it is impossible to estimate the
\bmin\, for individual lobes. However, as the \bmin\,and p$_{lobe}$ are
tightly correlated, I can be confident that the results using the KDA
model give a good estimation of the magnetic field in a source.

Figure \ref{Bminao}(b) shows that there is a weaker correlation
between the density ratio and the \bmin\, ratio which corresponds to a
significance of $\sim$99.5\%. By removing 3C 351 and 3C 16 the
correlation becomes $>$99.9\% significant. In section \ref{fluxasym} I
noted that 3C 351 has a very distorted northern lobe, significantly
extending to the west of the double hotspots. This could be attributed
to a very low density on the western side of the north lobe, allowing
the source to propagate westward. The southern lobe is much more
constrained, implying a larger density, which could be the cause of
the large ratio of densities between the two lobes. The fact that the
\bmin\, ratio and \ao\, ratio are correlated suggests that it is the
difference in the environment of the two lobes of a source that is
causing the difference in the \bmin\, and {\it not} beaming as
suggested in section \ref{bmin:theory}.

\section[Modelling the complete 3CRR, 6CE and 7C III FRII samples]{Modelling the complete 3CRR, 6CE and \\7C III FRII samples}\label{large}
\subsection{Observations}
\label{obser}
\begin{figure}[!h]
\centerline{\psfig{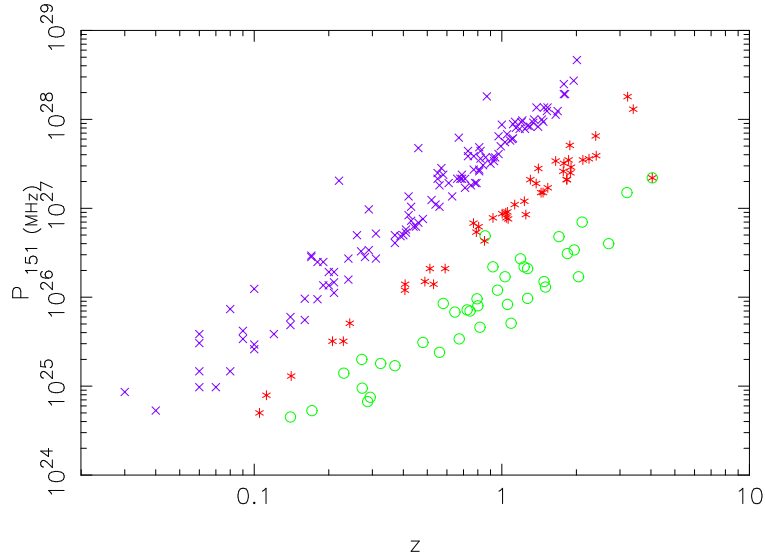}}
\caption
[Redshift against the radio luminosity at 151 MHz for the combined
sample] {Redshift against the radio luminosity at 151 MHz. The 3CRR
sample is represented by `$\times$', the 6CE sample by `$\circ$' and
the 7C III sample by `*'. A spectral index of -0.75 is used to shift
the 3CRR luminosity data at 178 MHz to 151 MHz.}\label{zpB}
\end{figure}

The large sample contains a sample of 211 FRII sources taken from the
complete 3CRR, 6CE and 7C III samples, see Figure \ref{zpB}. Only
sources larger than 10 kpc are included in the analysis as below this
limit the model may not be applicable \citep{paul}.  For consistency
all sources smaller than 10 kpc are excluded using the stated
cosmology.

Appendices \ref{ap:3CRR} to \ref{ap:7C} give the archival redshift,
size, flux measurements, lobe ratio and their associated references
for each source used in the analysis. In section \ref{model} I stated
that the model uses three flux measurements and their corresponding
errors to constrain the free parameters. I apply the model to the flux
measurements at 178 MHz\footnote{The different lower frequency of the
3CRR sample is simply due to the fact that it is defined at 178 MHz
instead of 151 MHz.}, 365 MHz and 1400 MHz for the 3CRR sample; 151
MHz, 365 MHz and 1400 MHz for the 6CE sample and 151 MHz, 327 MHz and
1490 MHz for the 7C III sample.  $\alpha_{low}$ is defined as the
spectral index between the lower frequency and 365 MHz and
$\alpha_{high}$ is defined as the spectral index between 365 MHz and
1400 MHz\footnote{For the 7C III sample I use 327 MHz and 1490 MHz
instead of 365 MHz and 1490 MHz. This is due to the fact that is only
at 327 MHz and 1490 MHz that there is flux information for each source
in the sample.}.

In section \ref{KDA:theory} I use $p=2$ in the determination of the
monochromatic radio luminosity, $P_\nu$. This means that only sources
with a spectral index, $\alpha_{boundary} < -0.5$ (including errors)
can be accurately modelled where $\alpha_{boundary}$ is determined by
the choice of $p$, i.e. $p=2\alpha+1$ \citep{longair}. For this reason
3C 382 and 6C 0922+36 are excluded, since they have
$\alpha_{high}=-0.22$ and $\alpha_{high}=-0.30$ respectively, which
even when the errors are accounted for have
$\alpha_{high}>\alpha_{boundary}$.

\subsection{Application of the model to the combined sample}
The source size, D$_{source}$, is calculated from the observed angle
of the source on the sky, and the source redshift. The lobe ratio for
each lobe ($R_{\rm{lobe 1}}$ and $R_{\rm{lobe 2}}$) is taken from high
resolution maps, where they exist, in the literature (see appendices
\ref{ap:3CRR} to \ref{ap:7C}). In the few cases where there are no
high resolution maps, particularly the faint 7C III sources, the lobe
ratio is set to be the average sample lobe ratio, using the sample
that the source is taken from.

For every source the flux of the lobe, at each frequency, is taken to
be half the total flux. This a simplification, but there are no flux
measurements for individual lobes, at all 3 frequencies, for each of
the 211 sources in the literature. For this reason the lobe length is
also taken to be D$_{lobe}$= D$_{source}$/2 and the lobe ratio is
averaged over the source i.e. $R_{\rm{average}}=(R_{\rm{lobe
1}}+R_{\rm{lobe 2}})/2$. This ensured that the results are not biased
towards sources that are well observed and have high resolution maps
(e.g. the majority of the 3CRR sample). This is particularly important
when dealing with the sources of the 7C III sample for which very
little archival information exists.

\subsection{Fundamental Parameters}
\label{fundamentals2}
\index{fundamental parameters} In the following sections I investigate
the relationship of the model properties of my sources with the
`fundamental' properties redshift, z, radio luminosity at 151 MHz,
P$_{151}$, and physical size, D$_{lobe}$. Before I can understand how
the model parameters evolve with the fundamental properties I need to
understand the relationship between the fundamental properties. This
is similar to the analysis in section \ref{fundamentals}. Table
\ref{zpd2} contains the associated Spearman Rank results of the
fundamental parameters. As expected, I find redshift and radio
luminosity to be highly correlated (see Figure \ref{zpB}), to a
significance of $> 99.99\%$, but this is simply due to the flux limits
of the samples and cannot be avoided. Using the statistical techniques
described in chapter \ref{statistics} this correlation can be isolated
from any other correlations.
\begin{table}[!h]
\centering 
\begin{tabular}{ r@{ }l r@{.}l r@{.}l }\hline
\multicolumn{2}{c}{Parameters}&
\multicolumn{2}{c}{rs value}&\multicolumn{2}{c}{ t value}\\ \hline
 z &P$_{151}$&0&69302 &  13&93\\
 z &D$_{lobe}$  & -0&40682 &  6&45\\
 P$_{151}$ &D$_{lobe}$ &-0&29855 &  4&53\\\hline
\end{tabular}
\caption{Spearman Rank values for  z, P$_{151}$ and D$_{lobe}$ in the combined samples.}\label{zpd2}
\end{table}

\begin{figure}[!h]
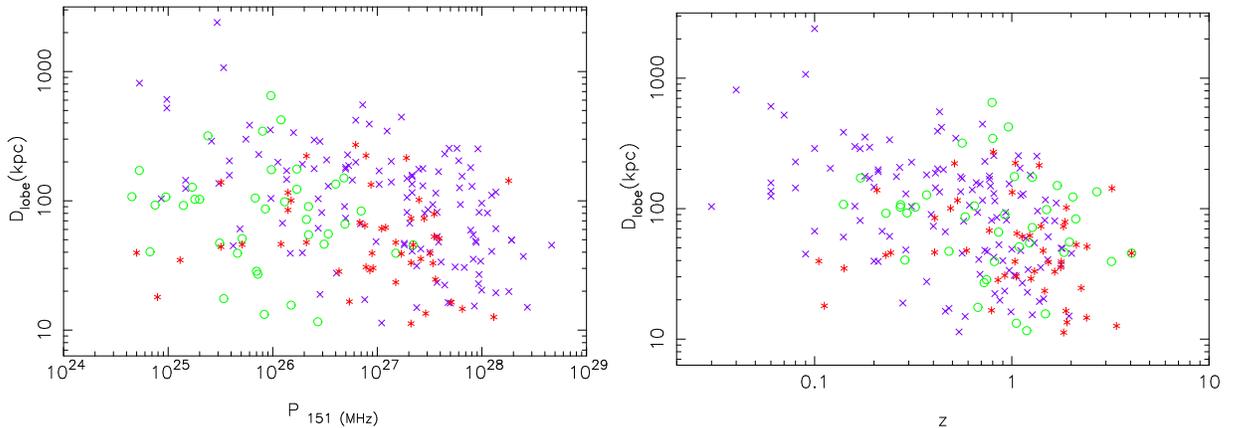

\centerline{\psfig{file=3PD_c.ps,width=8cm,angle=270}
\psfig{file=3zD_c.ps,width=8cm,angle=270}}
\caption[Relationship between z, P\li\, \& \DL.] 
{(a - left) Source size against radio-luminosity and (b - right)
against redshift. Symbols are as in Figure \ref{zpB}.}
\label{zpDB}
\end{figure}

By using the Partial Spearman Rank test I find that even with 3 large
complete samples there is no significant \Po-\DL\, correlation and
that the size-redshift anti--correlation is weak when it is compared
to the strength of the redshift - radio luminosity correlation.
\begin{eqnarray}
r_{z \hspace{1mm}D_{lobe},\hspace{1mm}P_{151}} =&
   -0.288\hspace{2cm}& D = 4.27\nonumber\\
r_{z \hspace{1mm}P_{151},\hspace{1mm}D_{lobe}} =&
  \hspace{3mm} 0.655\hspace{2cm}& D = 11.27\nonumber\\
r_{P_{151} \hspace{1mm}D_{lobe},\hspace{1mm}z} =&
  -0.019\hspace{2cm}& D = 0.27\nonumber
\end{eqnarray}
This is also evident in Figure \ref{zpDB}(a), the weak \Po-\DL\,
anti-correlation is almost indistinguishable from the scatter.  This
confirms that even with such a large sample of sources there is only a
weak relationship between the low frequency radio-luminosity and lobe
size and agrees with the findings of \citet{bam88}.\footnote{The
\Po-\DL\, anti-correlation is discussed in detail in section
\ref{fundamentals}.}

As already noted, redshift and lobe size are only weakly related, see
Figure \ref{zpDB}(b). The weak correlation can be see in Table
\ref{zpd2} and also in the Partial Spearman Rank results
above. However the z-\DL\, correlation is much weaker than the z-\Po\,
correlation and will have little effect on the model results.

\subsection{Source age, ty}
\label{sty}
\begin{figure}[!h]
\centerline{\psfig{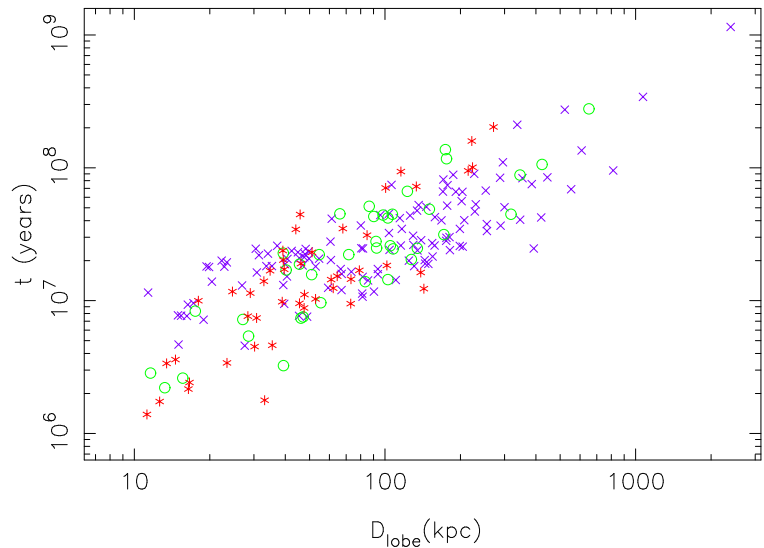}}
\caption[Age of a source in relation to radio size] 
{Source age against source size. Symbols are as in Figure \ref{zpB}.}
\label{Dty}
\end{figure}

\index{source age}
An important test of the model is to determine if the age of a lobe,
ty, correlates with its size, D$_{lobe}$. Clearly larger sources
should be older. As Table \ref{rsty} and Figure \ref{Dty} demonstrate,
D$_{lobe}$ and ty correlate to a high degree, $>99.99\%$
significant. The PCA results shown in Table \ref{pcaty} are dominated
by this strong correlation and do not show any other significant
trends. A strong D$_{lobe}$-ty correlation is expected from the model,
but it is the scatter caused by redshift and radio-luminosity that is
important in determining how the source environment affects its age.

\begin{table}[!h]
\centering 
\begin{tabular}{ r@{ }l r@{.}l r@{.}l }\hline
\multicolumn{2}{c}{Parameters}&
\multicolumn{2}{c}{rs value}&\multicolumn{2}{c}{ t value}\\ \hline
ty& D$_{lobe}$ & 0&80452 &  19&63\\
ty& z & -0&40583 &  6&43\\
ty& P$_{151}$ & -0&29238 &  4&43\\
ty&\ao&0&01446&0&21\\
ty& $\alpha_{low}$ & -0&24527 &  3&67\\\hline
\end{tabular}
\caption{Spearman Rank values for the age of a lobe}\label{rsty}
\end{table}

\begin{figure}[!h]
\centerline{\psfig{file=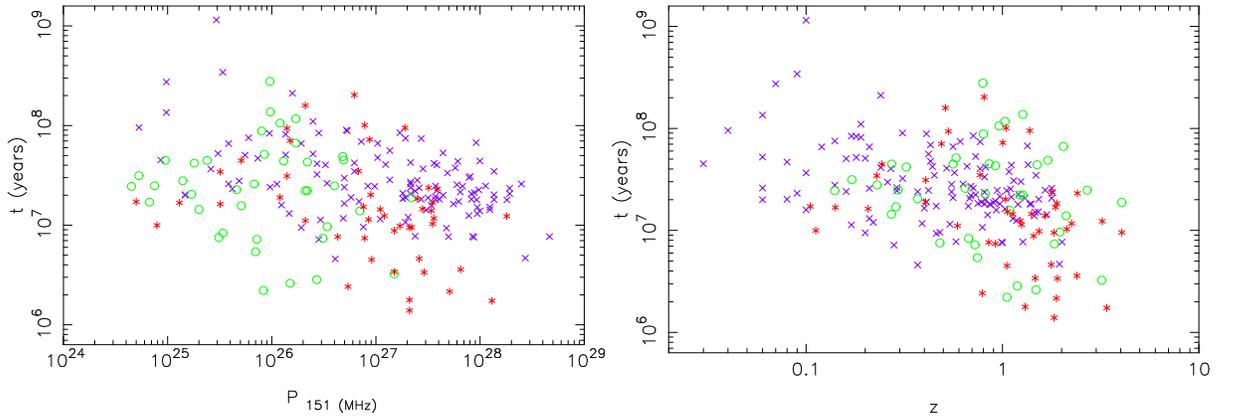,width=16cm}}
\caption[Age of a source in relation to redshift and radio-luminosity] 
{(a - left) Source age against radio-luminosity and (b - right)
against redshift. Symbols are as in Figure \ref{zpB}.}
\label{zptyB}
\end{figure}

Table \ref{rsty} shows an anti-correlation of the source age with
redshift and also a slightly weaker anti-correlation with
radio-luminosity. Figure \ref{zptyB}(a) shows that the \Po-ty
correlation is weak. The trend with redshift could be explained by the
inclusion of low-redshift giants.
\begin{table}[!h]
\centering
\begin{tabular}{l r@{.}l r@{.}l r@{.}l  r@{.}l }\hline
Parameter  & \multicolumn{2}{c}{1} &  \multicolumn{2}{c}{2}    &  
 \multicolumn{2}{c}{3}  &   \multicolumn{2}{c}{4}     \\\hline
z & 0&378&0&575&0&724&0&054\\
P$_{151}$&0&312&0&658&-0&685&0&001\\
D$_{lobe}$ &-0&625 &0&318&0&021&0&713\\
 ty & -0&608&0&368&0&078&-0&699\\ \hline
Eigenvalue& 53&09\%& 30&96 \%&13&70\% & 2&25 \% \\ \hline 
\end{tabular}
\caption{Eigenvectors and Eigenvalues for the age of the lobe}\label{pcaty}
\end{table}
In section \ref{fundamentals2} I have shown that there is a weak
z-\DL\, anti-correlation which combined with the strong evolution of
source size with age could cause the apparent z-ty anti-correlation
seen in Figure \ref{zptyB}(b).

The Partial Spearman Rank analysis shown below demonstrates, that for
any given radio-size the z-ty correlation is barely
significant. Interestingly at any given source age, the z-\DL\, also
becomes barely significant. This suggests that the observed z-\DL\,
correlation (and hence also the z-ty correlation) is simply a by
product of the Malmquist bias arising from the flux-limited nature of
the samples. Only by observing at fainter flux-limits, e.g TOOT
\citep{hr03}, can this observational bias be removed.
\begin{eqnarray}
r_{z \hspace{1mm}D_{lobe},\hspace{1mm}ty} =& -0.148  
\hspace{2cm}& D = 2.15\nonumber\\
r_{z \hspace{1mm}ty,\hspace{1mm}D_{lobe}} =&-0.145   
\hspace{2cm}& D = 2.10\nonumber\\
r_{D_{lobe} \hspace{1mm}ty,\hspace{1mm}z} =& \hspace{3mm}  0.766
\hspace{2cm}& D = 14.57\nonumber
\end{eqnarray}
It is also worth noting that when I use Partial Spearman rank to
determine the relationship between redshift, radio-luminosity and age
the \Po-ty anti-correlation vanishes at a fixed source age.
\begin{eqnarray}
r_{z \hspace{1mm}ty,\hspace{1mm}P_{151}} =&   -0.295
\hspace{2cm}& D = 4.38\nonumber\\
r_{z \hspace{1mm}P_{151},\hspace{1mm}ty} =& \hspace{3mm}  0.657
\hspace{2cm}& D = 11.36\nonumber\\
r_{P_{151} \hspace{1mm}ty,\hspace{1mm}z} =&  -0.017
\hspace{2cm}& D = 0.24\nonumber
\end{eqnarray}
This suggests that both the z-ty and \Po-ty anti-correlations are due
to the strong \DL-ty and z-\Po\, correlations. Sources at higher
redshifts and/or higher radio-luminosities are not younger than their
lower redshifts/lower radio-luminosity cousins.

\begin{figure}[!h]
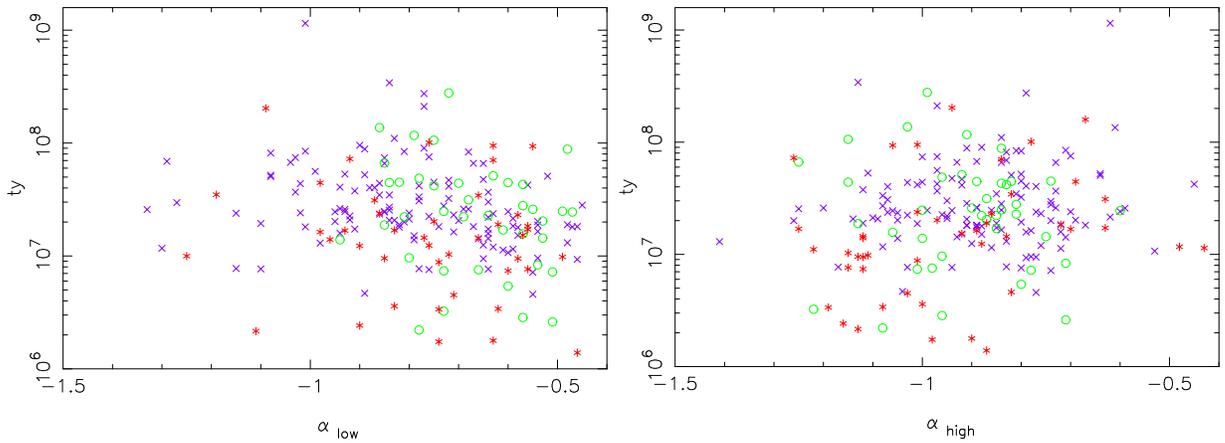

\centerline{\psfig{file=alow_c.ps,width=8cm,angle=270}
\psfig{file=ahigh_c.ps,width=8cm,angle=270}}
\caption[Spectral ageing]
{(a - left) Source age against $\alpha_{low}$ and (b - right)
against \ah. Symbols are as in Figure \ref{zpB}.}
\label{spec}
\end{figure}

Table \ref{rsty} also shows that there is a weak anti-correlation of the
low-frequency spectral index with the age of the source. However, I
would expect \ah\, to show the stronger correlation over
$\alpha_{low}$ due to spectral ageing. Although Table \ref{rsty} shows
that there is a weak anti-correlation, Figure \ref{spec}(a) indicates
that the $\alpha_{low}$-ty anti-correlation is insignificant. Figure
\ref{spec}(b) also shows that there is no \ah-ty
anti-correlation. Thus, between 151 MHz and 1400 MHz spectral ageing
appears to be insignificant.
\index{spectral ageing}

\subsection{Lobe pressure, p$_{lobe}$}
\label{section:pc}
\index{lobe pressure}
Table \ref{rspc} shows that there is a strong \DL-\p\,
anti-correlation, at a significance exceeding 99.99\%. This is not
surprising as a small source will have a high initial lobe pressure,
driving the source to expand which in turn will cause the lobe
pressure to fall. Table \ref{rspc} also shows a somewhat weaker
ty-\p\, anti-correlation, but this is simply due to the strong \DL-ty
correlation from section \ref{sty} and the strong \DL-\p\,
anti-correlation already discussed.
\begin{table}[!h]
\begin{tabular}{ r@{ }l r@{.}l r@{.}l }\hline
\multicolumn{2}{c}{Parameters}&
\multicolumn{2}{c}{rs value}&\multicolumn{2}{c}{ t value}\\ \hline
p$_{lobe}$& z  & 0&70061 &  14&23\\
p$_{lobe}$& P$_{151}$  & 0&75192 &  16&53\\
p$_{lobe}$& D$_{lobe}$  & -0&81074 & -20&07\\
p$_{lobe}$& ty & -0&64674 & -12&29\\
p$_{lobe}$& \ao  & 0&67819 &  13&37\\
p$_{lobe}$& $\alpha_{high}$  & -0&33934 &  -5&23\\\hline
\end{tabular}
\centering\caption{Spearman Rank values for the lobe pressure.}\label{rspc}
\end{table}

\begin{figure}[!h]
\centerline{\psfig{file=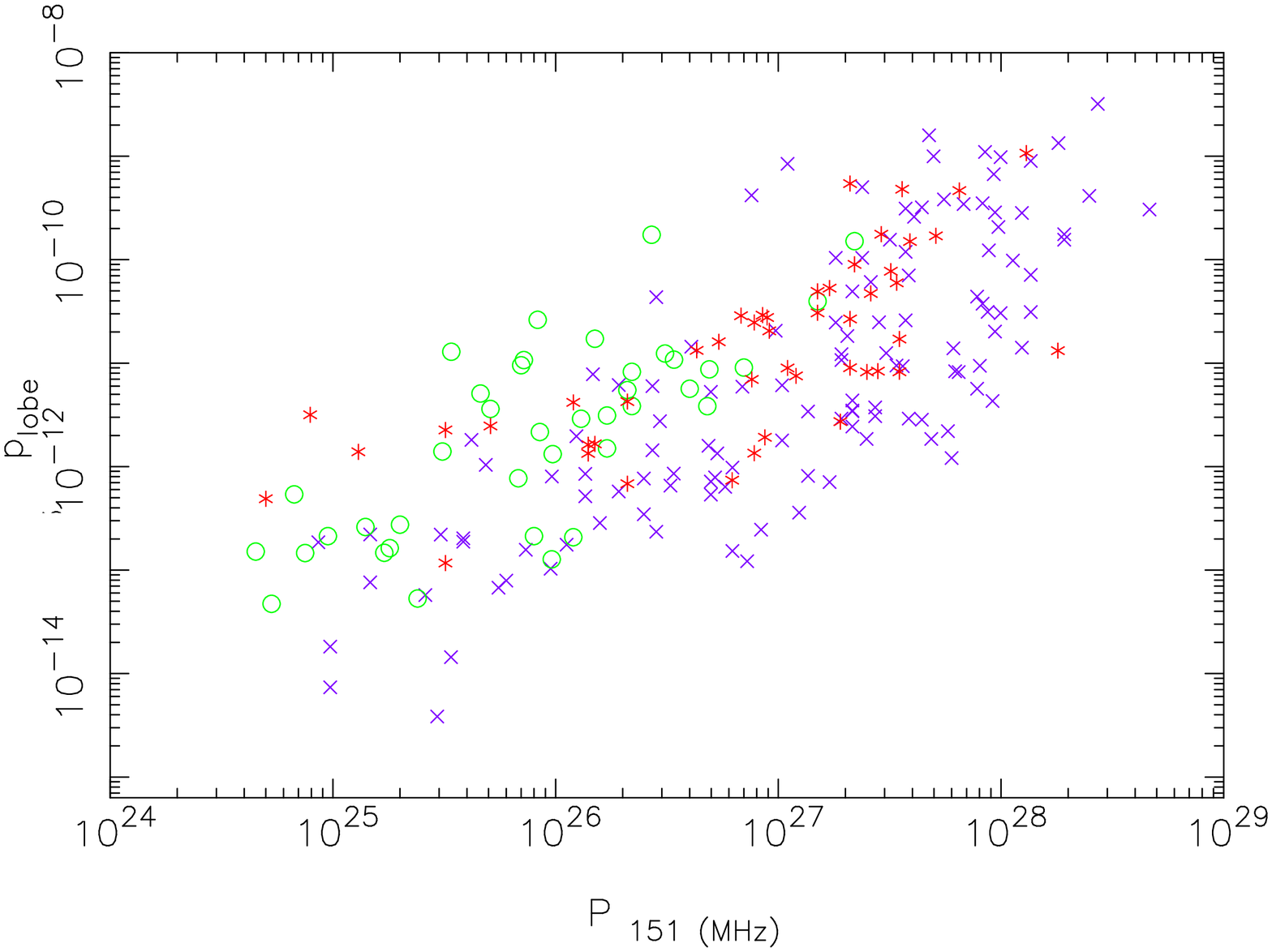,width=8cm}
\psfig{file=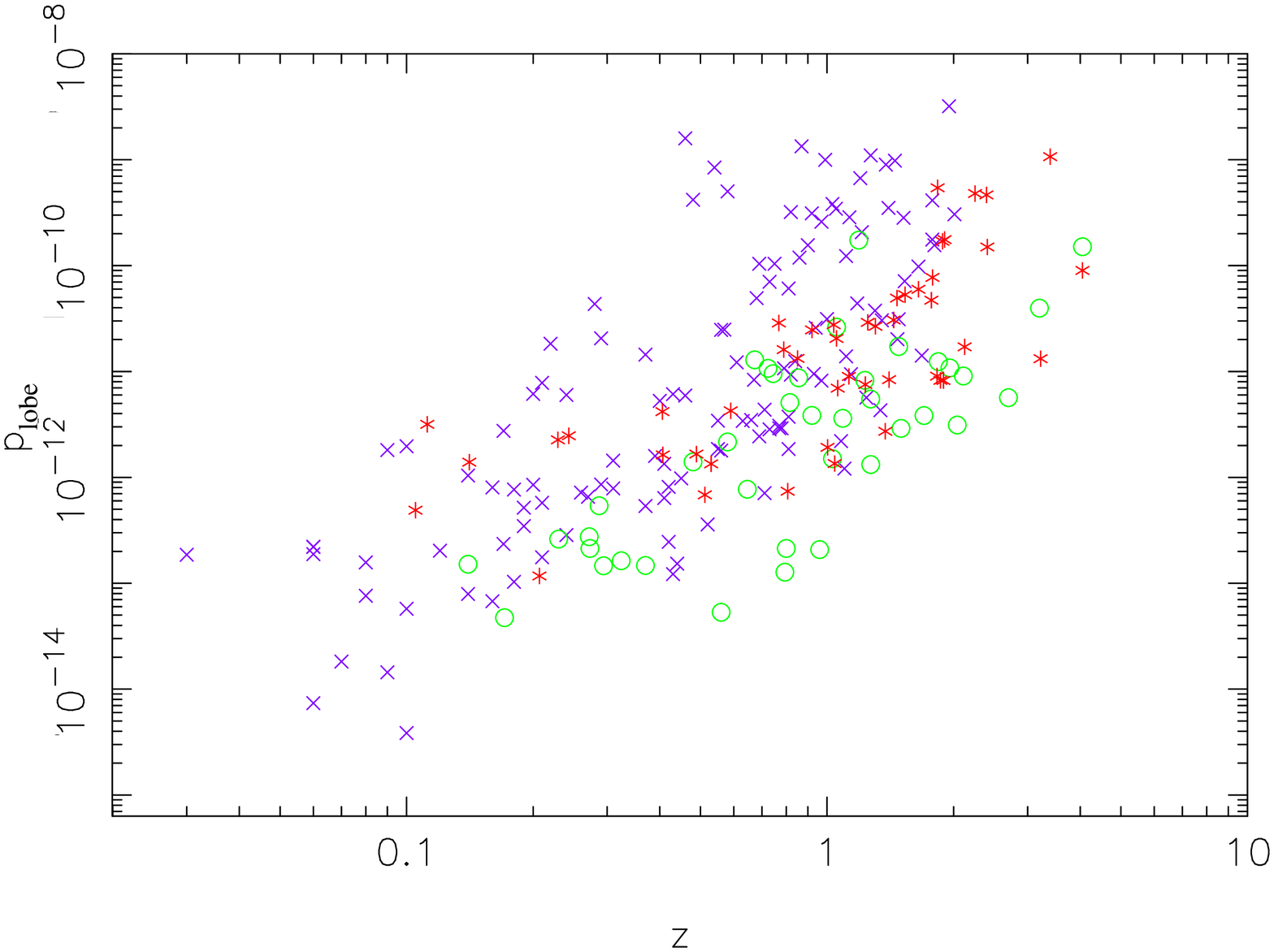,width=8cm}}
\caption[Pressure of a lobe in relation to its redshift and radio-luminosity] 
{(a - left) Lobe pressure against radio-luminosity and (b - right)
against redshift. Symbols are as in Figure \ref{zpB}.}
\label{pc}
\end{figure}

Table \ref{rspc} also reveals a strong \Po-\p\, correlation and a
slightly weaker \p-z correlation. Both of these strong correlations
can be seen in Figures \ref{pc}(a) and \ref{pc}(b) respectively.
This suggests that the lobe pressure is dependent on both the redshift
and radio-luminosity of a source. By using Partial Spearman Rank on
\p\,, \Po\, and z it is possible to disentangle the relationship with
lobe pressure.
\begin{eqnarray}
r_{z \hspace{1mm}p_{lobe},\hspace{1mm}P_{151} }= &  0.378\hspace{2cm}
&D = 5.72\nonumber\\
r_{z \hspace{1mm}P_{151},\hspace{1mm}p_{lobe}} = &   0.353\hspace{2cm}
&D = 5.31\nonumber\\
r_{P_{151} \hspace{1mm}p_{lobe},\hspace{1mm}z} = &  0.518\hspace{2cm}
&D = 8.27\nonumber
\end{eqnarray}
I find that neither the \p-z nor the \Po-\p\, correlation vanish at
any given radio-luminosity or redshift, respectively. However, it is
worth noting that in both the Spearman Rank results (Table
\ref{rspc}) and the Partial Spearman Rank results the \Po-\p\,
correlation is always the stronger correlation.

The Partial Spearman Rank results below show that at any given
radio-size, the redshift - lobe pressure correlation still exists and
is not due to any other independent correlation with the fundamental
parameters. This suggests that although the \Po-\p\, correlation is
stronger than the z-\p\, correlation, the z-\p\, correlation is
significant and indicates that there is some evolution in the lobe
pressure of a source with redshift.
\begin{eqnarray}
r_{z \hspace{1mm}D_{lobe},\hspace{1mm}p_{lobe}}= &\hspace{3mm}
0.386\hspace{2cm}& D = 5.86\nonumber\\ r_{z
\hspace{1mm}p_{lobe},\hspace{1mm}D_{lobe}}
=&\hspace{3mm}0.693\hspace{2cm} &D = 12.28\nonumber\\ r_{D_{lobe}
\hspace{1mm}p_{lobe},\hspace{1mm}z} =& -0.807\hspace{2cm} &D =
16.09\nonumber
\end{eqnarray}
The result is consistent with \citet{bcf92} who find that the pressure
around $z \sim 1$ radio-loud quasars is an order of magnitude higher
than the pressures around a lower luminosity set of quasars at $z \sim
0.3$. Unfortunately the study by \citet{bcf92} cannot determine
whether it is the redshift or the radio-luminosity of the quasars that
is connected to the change in the lobe pressure.

The PCA results in Table \ref{pcapc} show that the first eigenvector
holds 48.2\% of the variation in the data. Suprisingly I find that
this eigenvector is dominated by the \Po-\p\, and z-\p\, correlations,
with \DL-\p\, the weakest. In the second eigenvector, only the
\Po-\p\, correlation does not reverse. However, it is the \DL-\p\, {\it
correlation} that dominates this eigen-vector. The PCA results
indicate that the \Po-\p\, correlation is the dominant correlation and
that the z-\p\, is not highly significant.

\begin{table}[!h]
\centering
\begin{tabular}{l r@{.}l r@{.}l r@{.}l  r@{.}l }\hline
Parameter  & \multicolumn{2}{c}{1} &  \multicolumn{2}{c}{2}    &  
 \multicolumn{2}{c}{3}  &   \multicolumn{2}{c}{4}     \\\hline
   z   & 0&518&-0&182&0&716&-0&431\\
    P$_{151}$&0&587&0&322&0&100&0&736\\
   D$_{lobe}$ &-0&347&0&845&0&380&-0&144\\
    p$_{lobe}$&  0&516&0&386&-0&577&-0&501\\ \hline
Eigenvalue& 48&2\%&22&9\% & 18&1\% & 10&8 \%  \\ \hline
\end{tabular}
\caption{Eigenvectors and Eigenvalues for the lobe pressure}\label{pcapc}
\end{table}

The complicated relationship between \Po, \p\, and \DL\, is revealed
by using the Partial Spearman Rank statistic. I find that the
\Po-\p\/ correlation is almost as significant as the \DL-\p\/
anti-correlation. The fact that both relationships are almost equally
significant means that PCA is unable to distinguish between the two
separate relationships.
\begin{eqnarray}
r_{P_{151} \hspace{1mm}D_{lobe},\hspace{1mm}p_{lobe}} =&
\hspace{3mm}0.805\hspace{2cm}& D = 16.06\nonumber\\ r_{P_{151}
\hspace{1mm}p_{lobe},\hspace{1mm}D_{lobe}}=
&\hspace{3mm}0.912\hspace{2cm} &D = 22.22\nonumber\\ r_{D_{lobe}
\hspace{1mm}p_{lobe},\hspace{1mm}P_{151}} =& -0.931\hspace{2cm} &D =
24.08\nonumber
\end{eqnarray}

Using the \bmin\, argument or the KDA model it is possible to show
that the density of the lobe environment directly influences the lobe
pressure of a source \citep[e.g][]{longair}. It is not surprising then
that I find \p\, and \ao\, to be correlated, see Table \ref{rspc}. The
KDA model results demonstrate that by deriving the pressure in a lobe,
we can obtain an accurate estimate of the gas density in the
surroundings of a lobe.

Interestingly Table \ref{rspc} shows a weak anti-correlation between
$\alpha_{high}$ and \p\,, but no corresponding relationship with
$\alpha_{low}$. In section \ref{6:bmin} I stated that a stronger
magnetic field should cause stronger radiative losses in a source and
would lead to a steeper spectral index. This explains the weak
$\alpha_{high}$-\p\, anti-correlation seen in Table \ref{rspc}. The
fact that I do not find any relationship of $\alpha_{low}$ with \p\,
indicates a lower level of radiative losses between 151 MHz and 365
MHz.

\subsection{Density of the source environment, $a_\circ^\beta\rho_\circ$}
\label{aoB}
\index{density of source environment}
In section \ref{small} I found that in the small sample of 26 sources
there was a weak \Po-\ao\, correlation and no significant trend with
redshift.  Table \ref{rsao} demonstrates that with the large sample containing 211 sources I find a much stronger \Po-\ao\, correlation
(see Figure \ref{zpB}(a)) and a weak z-\ao\, correlation (see Figure
\ref{zpB}(b)).

\begin{table}[!h]
\centering 
\begin{tabular}{ r@{ }l r@{.}l r@{.}l }\hline
\multicolumn{2}{c}{Parameters}&
\multicolumn{2}{c}{rs value}&\multicolumn{2}{c}{ t value}\\ \hline
  \ao &z & 0&47835 & 7&89\\ \ao &P$_{151}$ & 0&66294 & 12&83\\
  $a_\circ^\beta\rho_\circ$ & D$_{lobe}$&-0&33283 & -5&11\\
 \ao &$\alpha_{high}$ & -0&31872 &  -4&87\\\hline
\end{tabular}
\caption{Spearman Rank values for \ao}\label{rsao}
\end{table}

\begin{figure}[!h]
\centerline{\psfig{file=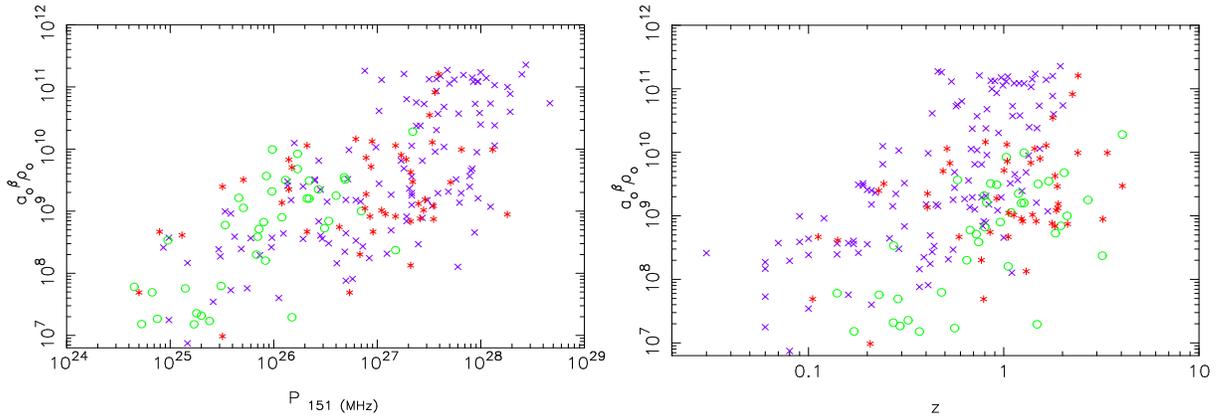,width=16cm}}
\caption[Density of the source environment in relation to its redshift 
and radio-luminosity] 
{(a - left) Source density against radio-luminosity and (b- right) against
redshift. Symbols are as in Figure \ref{zpB}.}
\label{zpaoB}
\end{figure}

However, as the Partial Spearman Rank results below show, the z-\ao\,
correlation vanishes at any given radio-luminosity. This demonstrates
that the apparent correlation of density with redshift is simply due
to the strong z-\Po\, correlation and the independent \ao-\Po\,
correlation present in the samples.
\begin{eqnarray}
r_{z \hspace{1mm}a_\circ^\beta\rho_\circ,\hspace{1mm}P_{151}} =&  0.035
\hspace{2cm}
&D = 0.51\nonumber\\
r_{z \hspace{1mm}P_{151},\hspace{1mm}a_\circ^\beta\rho_\circ} =&  0.572
\hspace{2cm}
&D = 9.38\nonumber\\
r_{P_{151} \hspace{1mm}a_\circ^\beta\rho_\circ,\hspace{1mm}z} =&  0.524
\hspace{2cm}
&D = 8.39\nonumber
\end{eqnarray}
The PCA results in Table \ref{pcaao} shows only a weak z-\ao\,
correlation which reverses in the second eigenvector. This adds
further evidence to the lack of any significant evolution of the
source density with redshift. The \Po-\ao\, correlation means that a
more powerful source resides in a denser atmosphere. By comparing
sources with a range of radio-luminosities, any bias with respect to
density of the source environment in a sample is effectively
removed. However, studies which only select radio sources with a
narrow range of radio luminosities
\citep[e.g.][]{prm89} may be affected by the \Po-\ao\, correlation.
\begin{table}[!h]
\centering
\begin{tabular}{l r@{.}l r@{.}l r@{.}l  r@{.}l }\hline
Parameter  & \multicolumn{2}{c}{1} &  \multicolumn{2}{c}{2}    &  
 \multicolumn{2}{c}{3}  &   \multicolumn{2}{c}{4}     \\\hline
   z   & 0&515&-0&213&0&677&-0&481\\
    P$_{151}$& 0&587&0&349&0&168&0&711\\
   D$_{lobe}$ &-0&372&0&813&0&406&-0&187\\
    $a_\circ^\beta\rho_\circ$& 0&501&0&415&-0&590&-0&478\\\hline
Eigenvalue& 47&2\%& 22&5\%&19&6\%  & 10&7 \%  \\ \hline
\end{tabular}
\caption{Eigenvectors and Eigenvalues for \ao}\label{pcaao}
\end{table}

\citet{hw00}, \citet{pp88} and \citet{ls79} found that low-z (hence low 
radio-luminosity) FRII sources do not live in rich
environments. \citet*{ymp89,cf89} and also \citet{yg87} find that the
most powerful radio-galaxies inhabit rich environments, generally rich
clusters. This is consistent with my results presented here. However,
it is worth noting that all of the observational studies suffer from
the redshift-radio luminosity degeneracy. Thus it is hard to determine
whether the density of the source environment changes with redshift or
radio-luminosity. A recent study by \citet{brw99} found that when they
modelled the radio-luminosity, size, redshift and spectral index plane
of the 3CRR, 6C and 7C sample their models did not require any
evolution of the source density with redshift to match the
observations. This may indicate that any evolution of density with
redshift is relatively unimportant.

Table \ref{rsao} also shows that there is a weak anti-correlation of
density with $\alpha_{high}$. The Partial Spearman Rank results
between $\alpha_{high}$, \ao\, and \p\, below, demonstrate that the
apparent \ao-$\alpha_{high}$ correlation vanishes at any
given lobe pressure and is simply due to the \ao-\p\,
correlation coupled with the \p-$\alpha_{high}$ anti-correlation. This
proves that the spectral index of a source is insensitive to direct
changes in the environment. However, since density and lobe pressure
are correlated, any change in the source density is reflected in a
change in the lobe pressure and hence $\alpha_{high}$.
\begin{eqnarray}
r_{\alpha_{high} \hspace{1mm}a_\circ^\beta\rho_\circ,\hspace{1mm}p_{lobe}} 
=&  -0.128\hspace{2cm}&D = 1.86\nonumber\\
r_{\alpha_{high} \hspace{1mm}p_{lobe},\hspace{1mm}a_\circ^\beta\rho_\circ}
=&  -0.177\hspace{2cm}&D = 2.58\nonumber\\
r_{p_{lobe} \hspace{1mm}a_\circ^\beta\rho_\circ,\hspace{1mm}\alpha_{high}}
=& \hspace{3mm} 0.638\hspace{2cm}&D = 10.92\nonumber
\end{eqnarray}

\subsection{Jet power, Q$_\circ$}
\begin{figure}[!h]
\centerline{\psfig{file=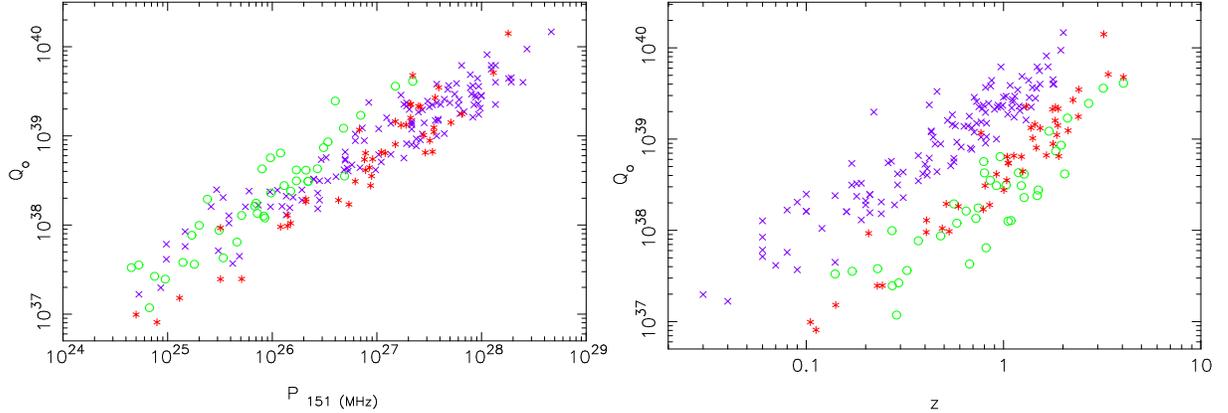,width=16cm}}
\caption[Jet power in relation to redshift and radio-luminosity] 
{(a - left) Jet power against radio-luminosity and (b - right) against
redshift. Symbols are as in Figure \ref{zpB}.}
\label{zpQoB}
\end{figure}

\index{jet power}
Figure \ref{zpQoB}(a) shows that jet power, Q$_\circ$, and
radio-luminosity are highly correlated, with a significance exceeding
99.99\% (see Table \ref{Qors}). This is expected from the model as it predicts that the more luminous sources have a stronger
jet driving them. However, the KDA model does not predict any
relationship between jet-power and redshift.
\begin{table}[!h]
\centering 
\begin{tabular}{ r@{ }l r@{.}l r@{.}l }\hline
\multicolumn{2}{c}{Parameters}&
\multicolumn{2}{c}{rs value}&\multicolumn{2}{c}{ t value}\\ \hline
Q$_\circ$ &p$_{lobe}$ & 0&64330 &  12&18\\
Q$_\circ$& D$_{lobe}$ &-0&14354 &  -2&10\\
Q$_\circ$ &z & 0&70195 &  14&28\\
Q$_\circ$ &P$_{151}$ & 0&92486 &  35&24\\
Q$_\circ$ &$\alpha_{high}$ & -0&30324 &  -4&61\\
 \hline
\end{tabular}
 \caption{Spearman Rank values for the jet power}\label{Qors}
\end{table}
Table \ref{Qors} shows that there is a strong
correlation of jet-power with redshift. This strong correlation is
evident in Figure \ref{zpQoB}(b). The PCA results in Table \ref{pcaQo}
are dominated by the \qo-\Po\, correlation which only
reverses in the final eigenvector. Conversely the
correlation with redshift reverses in the second eigenvector and shows
a much weaker correlation in comparison to the \qo-\Po\,
correlation. There is no significant trend with the size of the lobe.
\begin{table}[!h]
\centering
\begin{tabular}{l r@{.}l r@{.}l r@{.}l  r@{.}l }\hline
Parameter  & \multicolumn{2}{c}{1} &  \multicolumn{2}{c}{2}    &  
 \multicolumn{2}{c}{3}  &   \multicolumn{2}{c}{4}     \\\hline
   z   &  0&501&0&156&0&829&0&912\\
    P$_{151}$&0&578&-0&227&-0&455&0&638\\
   D$_{lobe}$ &-0&222&-0&927&0&291&0&079\\
    $Q_\circ$& 0&604&-0&254&-0&146&-0&741\\\hline
Eigenvalue& 58&3\%& 24&3 \%&14&0\% &  3&4\%   \\ \hline
\end{tabular}
\caption{Eigenvectors and Eigenvalues for jet power}\label{pcaQo}
\end{table}

By using the Partial Spearman Rank test, it is possible to disentangle
the relationship between redshift and radio-luminosity and jet-power.
\begin{eqnarray}
r_{z \hspace{1mm}Q_\circ,\hspace{1mm}P_{151}} =&  0.223\hspace{2cm}
&D = 3.26\nonumber\\
r_{z \hspace{1mm}P_{151},\hspace{1mm}Q_\circ} =&  0.162\hspace{2cm}
&D = 2.35\nonumber\\
r_{P_{151} \hspace{1mm}Q_\circ,\hspace{1mm}z} =& 0.854\hspace{2cm}
&D = 18.32\nonumber
\end{eqnarray}
It is obvious that although the z-\qo\, correlation is significantly
weaker than the \Po-\qo\, correlation it does not completely vanish at
any given radio-luminosity. FRII radio sources are powered by jets
created in the vicinity of the most massive black holes known in the
universe. It is commonly assumed that the mass of the black hole
determines the jet power. \citet{lacy01} find that for their sample of
sources taken in the redshift range $0.3< z < 0.5$ the black hole mass
- 5 GHz radio luminosity relation can be described by $L_{5\, {\rm
GHz}} \propto M^{1.9\pm0.2}_{\rm bh}$ whereas \citet{dunlop03} find
$L_{5\, {\rm GHz}} \propto M^{2.5}_{\rm bh}$ between $0.1<z<0.25$.  A
recent study by \citet{mclure04}, at a narrow range of redshift
($z\approx 0.5$), find that to 3$\sigma$ there is a trend between the
low frequency luminosity, $L_{151\/ {\rm MHz}}$, and black-hole
mass. The fact that I also find a correlation between \qo\, and z
suggests that the mass of the black-hole powering the radio sources
changes with redshift as well as with the low-frequency
radio-luminosity. This is evidence that the mass of a black-hole
powering FRII sources is not constant with redshift. The difference in
the findings of \citet{lacy01} and \citet{dunlop03} may be attributed
to the slight difference in the cosmological epoch of their samples.

Table \ref{Qors} shows that there is a weak \ah-\qo\, anti-correlation
but, as the Partial Spearman Rank results show, this is simply due to
the independent \qo-\p\, correlation and \p-\ah\, anti-correlation.
 \begin{eqnarray}
r_{p_{lobe} \hspace{1mm}Q_\circ,\hspace{1mm}\ahm} =&\hspace{3mm}  0.603\hspace{2cm}
&D = 10.04\nonumber\\
r_{p_{lobe} \hspace{1mm}\ahm,\hspace{1mm}Q_\circ} =& -0.198\hspace{2cm}
&D = 2.88\nonumber\\
r_{\ahm \hspace{1mm}Q_\circ,\hspace{1mm}p_{lobe}} =&-0.118\hspace{2cm}
&D =  1.71\nonumber
\end{eqnarray}
\section{Conclusions}
I find no trend of the polarisation parameters \DMz\, and \sig\, with
the density of the lobe environment, \ao. This provides further
evidence that the polarisation trends do not map changes in the
density of the source environment, but are indicators of the degree of
disorder in the structure of the magnetic field in these environments evolving
with redshift.

It was suggested in chapter 4 that many of the sources from the small
sample show evidence of beaming. However, the density estimate from
the KDA model disagreed with this finding, providing evidence instead
that it is changes in the lobe environment that causes the observed
changes in the \bmin. The lobe pressure and the \bmin\, are tightly
correlated, since the pressure from the KDA models is directly related
to estimates of the \bmin.


I find little evidence of any spectral ageing between 151 MHz and 1400
MHz. There is no trend of the lobe pressure with the low frequency
spectral index, but a weak trend with the higher frequency spectral
index which indicates, as expected, that radiative losses become more
important at higher observing frequencies.

Unsurprisingly I find a strong correlation with lobe pressure and
size. The lobe pressure of a source also shows a strong anti-correlation 
with radio-luminosity and a weaker correlation with redshift. 
The lobe pressure of a source
was also found to correlate with the density of the source
environment. However, density is found to show a strong trend with
radio-luminosity but there is no corresponding trend with redshift. 
This indicates that
although density and pressure are related, it is only the lobe
pressure that is significantly affected by changes in redshift. There
is no indication of any evolution of the source environment with
cosmological epoch.

Finally, I find a very strong correlation between jet power and
radio-luminosity with a much weaker correlation with redshift. This
provides evidence of the evolution of black hole masses with
low-frequency luminosity, but more importantly with redshift.

\newpage
\thispagestyle{empty}
\chapter{Conclusions} 
\index{conclusions}

I have presented a complete data set of three samples of radio
galaxies and radio-loud quasars. The three samples were defined such
that two of them overlap in redshift and two have similar radio
luminosities, allowing a study into the trends of various properties
with redshift and radio luminosity. The spectral index, the rotation
measure and the depolarisation measure were derived directly from the
radio observations.

The spectral index was found to be insensitive to the fundamental
parameters (redshift, radio-luminosity and radio-size) and thus I can
not confirm previous findings of a correlation of spectral index with
redshift \citep{athreya}, a trend of spectral index with radio
luminosity \citep{vv72,on89} nor a trend of spectral index with size
\citep{brw99}. This is most probably due to the comparatively small size 
of my sample.

All sources were found to have an external Faraday medium that was
local to each source and is responsible for variations of RM on small
angular scales. The observed rotation measure itself was insensitive
to changes in any of the fundamental parameters which is consistent
with a Galactic origin. This Galactic contribution dominates the RM
properties of the sources on large scales. At small angular scales,
e.g. $\sigma _{RM_z}$ (rms variation of rotation measure) and DM$_z$
(depolarisation measure), the Galactic contribution is effectively
removed. The higher redshift sources display a greater variation in
their rotation measures properties. Only the observed depolarisation
shows any significant trend with radio-luminosity. My results are
consistent with both \citet{mt73} and \citet{kcg72}, as I find both
redshift and radio-luminosity to correlate with depolarisation.

The lack of any asymmetries in \sig\, and DM with the \bmin\, suggests
that the environment has no direct impact on the depolarisation or
\sig\, in a source, but it is the degree of disorder in the structure of the
magnetic field that is changing with redshift.  This is consistent
with my findings that there is no strong correlation between the
projected sizes of the sources and their depolarisation measure which
agrees with the findings by \citet{jane}, but disagrees with findings by
\citet{strom,sj88,prm89} and \citet{ics98}. The lack of any trend between 
\sig\, and \DMz\, and the density of the source environment (from the KDA
models) adds further evidence against the assumption that
depolarisation and density are related.

Asymmetries in spectral index show evidence of beaming in the majority
of my sources. The results from the \bmin\, asymmetries are
inconclusive and could also be explained by an environmental
effect. The density estimates from the KDA model suggest that it is
more likely that environmental differences and not beaming that causes the
observed asymmetries.

In general all asymmetries were found to be insensitive to changes in
the source redshift, or low frequency radio luminosity; suggesting
that the environments of the high redshift and low redshift samples
are similar and any asymmetries caused are due to localised changes in
the environment.

I did not find any significant spectral ageing between 151 MHz and
1400 MHz for the sources in the 3CRR, 6CE and 7C III samples. The lobe
pressure shows only a weak trend with the spectral index between 365
MHz and 1400 MHz. This indicates, as expected, that radiative losses
are more important at higher observing frequencies.

The 3CRR, 6CE and 7C III sources also show that the density of the
source environment correlates with radio-luminosity but seemed to be
insensitive to changes in redshift. This is consistent with the
findings of \citet{ls79,pp88,ymp89} and \citet{yg87} that there is no
indication of any evolution of the source environment with
cosmological epoch.

Finally, I find that jet power and radio-luminosity are tightly
correlated. This is expected as the more luminous sources have larger
black-holes and thus have stronger jets emerging from their accretion
disk \citep{up95,mclure04,whs03,lacy01}. However, I also find evidence
of jet-power and hence black-hole mass evolution with cosmic epoch.

\section{Evolution of the Faraday screen}
The observational trends noted above are further complicated by the
presence of an external Faraday screen. Using my observations of
variations in the rotation measure and the \citet{tribble} models I
was able to determine the effects and evolution of the Faraday screen
with redshift, radio-luminosity and source size.

The observed depolarisation-redshift trend was found to be best
explained in terms of the intrinsic rotation measure, \si\,, varying
as a function of redshift and a constant cell size, \so. This is
consistent with my earlier findings that it is the increasing disorder
in the magnetic field and not density, that changes with redshift. A
slight variation in the cell size, typically no more than $\pm 50\%$
around the best fit value, describes the scatter in the depolarisation
results at any given redshift. In fact, it is generally the high
redshift results that constrain the model. This suggests that to map
the evolution of the Faraday screen, for all epochs, we must use
samples containing sources from a large range of redshifts.

Although the \citet{burn} law over-estimates the degree of
depolarisation, I find similar results using the \citet{tribble}
models, which are more realistic. The simplistic Burn model however,
can still be used to find the basic trends with depolarisation in a
sample, with fewer assumptions about the underlying distribution.  I
find that linear polarisation is proportional to $\lambda^{-1}$ best
describes the data, compared to the widely used Burn result
$p\sim\lambda^{-4}$.

This work indicates that there is no significant evolution of the
density of the source environment with cosmic epoch. Almost all of the
observed trends with redshift are directly related to the changes in
the degree of disorder in the magnetic field with redshift. The weak
trend noted between jet-power and redshift could be attributed to a
difference in the formation of black-holes at higher redshifts, i.e. a
higher rate of mergers at higher redshifts.

The significant parameter in determining source characteristics seems
to be the radio-luminosity of the source. This is evident in both the
small and large samples. This agrees with the findings of many other
authors in different wavebands
\citep[e.g][]{al87,ymp89,pp88,ibr03}.

Thus in conclusion, I find that there is evidence for a relationship
between radio-luminosity and the environment in which a given radio
source lives, but there is no significant evolution of the source
environment with redshift.

\section{Further Work}
As with any significant body of work there are many avenues in which
to extend the research. Ideally I would like to make the same
observations between 1.4 GHz and 4.8 GHz for all sources in the 3CRR,
6CE and 7C III. This is highly unrealistic and would also be too time
consuming. A compromise would be to extend my small sample to
lower-luminosities and higher redshifts. A sample of around 50-60
sources would be feasible and would represent a large cross section of
the FRII population.

It would also be interesting to do a multi-waveband study of the
sources already in my small sample. A X-ray study of each source could
determine if the gas halo around each source was contributing to the
depolarisation and if there was any interaction between the radio
lobes and the gas detected due to its X-ray emission.

Several of the high redshift 7C sources show rather twisted
morphologies and the signal-to-noise level was too low to get detailed
depolarisation and rotation measure maps. Higher resolution and longer
observations would show if there is any underlying physical reason for
their distorted structure.

\newpage

\appendix
\chapter{Sample A maps}\label{Amap}
\thispagestyle{empty}\index{Sample A maps}

Please see http://www.astro.soton.ac.uk/\~jag/Thesis.zip for full images.

\clearpage

\newpage
\chapter{Sample B maps}\label{Bmap}
\thispagestyle{empty}
\index{Sample B maps}
Please see http://www.astro.soton.ac.uk/\~jag/Thesis.zip for full images.

\newpage
\chapter{Sample C maps}\label{Cmap}
\thispagestyle{empty}
\index{Sample C maps}

Please see http://www.astro.soton.ac.uk/\~jag/Thesis.zip for full images.

\chapter{Cosmology}\label{Cosm}
\thispagestyle{empty}
\index{cosmology}
\section{Basic assumptions used}
The frequency at which I observe my sources is redshift dependent,
\begin{equation}\label{D.1}
\nu_{\rm emit} = (1+z)\nu_{\rm observed}
\end{equation}
where $\nu_{\rm emit}$ is the emitted frequency, $\nu_{\rm observed}$ is the
observed frequency and $z$ is the redshift of the source.
Although I observe at 1.4 GHz and 4.8
GHz the emitted frequency will be slightly different for each
source. This is the reason why the polarisation observations are
shifted to a common redshift, thus removing any residual redshift effects.

To convert from the angular size of a source on the sky, $\theta$, to
a physical size, D$_{source}$, I use the angular diameter distance formula,
\index{angular diameter distance}

\begin{eqnarray}
D_{source}&=& \frac{R_\circ s_r\,\theta}{1+z}\\
{\rm where,} \hspace{3mm}
R_\circ s_r &=& \bigg(\frac{2c}{H_\circ}\bigg)
\frac{\Omega z+(\Omega-2)(\sqrt{1+\Omega z}-1)}{\Omega^2(1+z)}.
\end{eqnarray}
where $\Omega$ is the density parameter of the Universe, $c$ is the
speed of light and $H_\circ$ is the Hubble constant.

The flux density of a source is also dependent on the redshift of a
source,
\begin{equation}
\label{transform}
S_\nu =\frac{P_\nu}{(1+z)(R_os_r)^2}
\end{equation}
where $P_\nu$ is the radio-luminosity at a specific frequency $\nu$.

\newpage
\thispagestyle{empty}
\chapter{3CRR sample}\label{ap:3CRR}
\index{3CRR}
\renewcommand{\thefootnote}{\alph{footnote}}
\begin{longtable}{r@{ }l r@{.}l c r@{.}l r@{.}l r@{.}l r@{.}l  r}\hline

\multicolumn{2}{c}{Source}
&\multicolumn{2}{l}{z}
&$\theta$
&\multicolumn{2}{c}{S$_{178{\rm MHz}}$}
&\multicolumn{2}{c}{S$_{365 {\rm MHz}}$}
&\multicolumn{2}{c}{S$_{1400 {\rm MHz}}$}
&\multicolumn{3}{c}{R}\\
\multicolumn{4}{c}{ }&(arc sec.)&\multicolumn{2}{c}{Jy}
&\multicolumn{2}{c}{Jy}
&\multicolumn{2}{c}{Jy}&
\multicolumn{2}{c}{ }&REF\\
\hline\endfirsthead
4C&12.03 &   0&16   &     240 &   10&9 &	4&45$^B$&	2&01$^A$& 1&77	& LP91    \\
3C&6.1   &   0&84   &     26  &   14&9 &	9&98$^D$&   3&56$^C$    &1&73	& NRH95    \\
3C&9     &   2&01   &     14  &   19&4 &	9&50& 	     2&18$^E$   &3&34	& F02    \\
3C&13    &   1&35   &     28  &   13&1 &	7&72& 	     1&87$^E$   &1&50	& LLA95    \\
3C&14    &   1&47   &     24  &   11&3 &	6&83& 	    1&98$^E$    &1&31	& ALB94    \\
3C&16    &   0&41   &     78  &   12&2 &	6&15&	    1&88$^C$    &2&62	& G04    \\
3C&19    &   0&48   &     7   &   13&2 &	8&98& 	     3&22$^E$   &3&27	& DRAGN    \\
3C&20    &   0&17   &     54  &   46&8 &	30&30&      11&53$^E$   &2&24     & HPPR97    \\
3C&22    &   0&94   &     24  &   13&2 &	8&58&      2&25$^C$	& 2&51	& BLR97    \\
3C&33    &   0&06   &     255 &   59&3 &	20&42&     13&65$^C$    &1&70	& LP91    \\
3C&33.1  &   0&18   &     227 &   14&2 &	5&34$^C$&      3&23$^C$	&  1&93	& LCF01    \\
3C&34    &   0&69   &     49  &   13&0 &	7&09&         1&65$^E$	&  2&72	& NRH99    \\
3C&35    &   0&07   &     750 &   11&4 &	3&78$^C$&      2&31$^C$	&  1&84	& DRAGN    \\	
3C&41    &   0&79   &     25  &   11&6 &	7&53$^G$& 	3&71$^E$& 1&81	& NRH99    \\
3C&42    &   0&40   &     29  &   13&1 &	8&92&          2&89$^E$	&  2&22	& G04    \\
3C&46    &   0&44   &     158 &   11&1 &	2&16$^C$&      1&17$^C$	&  3&70	& G04    \\
3C&47    &   0&43   &     77  &   28&8 &	13&02& 	      3&85 $^G$ & 1&22	& HLL94    \\
3C&49    &   0&62   &     1   &   11&2 &	7&74&	      2&74$^C$  & 0&78	& PPR85    \\
3C&55    &   0&74   &     72  &   23&4 &	11&98&      2&69$^C$    & 3&34	& FBB93    \\
3C&61.1  &   0&19   &     186 &   34&0 &	20&54$^D$&6&00 $^G$     & 2&69	& LP91    \\
3C&65    &   1&18   &     17  &   16&6 &	10&43&          3&11$^E$& 1&50	& G04    \\	
3C&67    &   0&31   &     3   &   10&9 &	8&01&          3&02$^E$	& 1&72	& AG95    \\
3C&68.1  &   1&24   &     53  &   14&0 &	8&87&          2&49$^C$	& 1&88	& HLL94    \\
3C&79    &   0&26   &     89  &   33&2 &	13&30&         4&943$^G$& 2&25	& HPPR97    \\
3C&98    &   0&03   &     310 &   51&4 &	26&67 $^G$&   10&20 $^G$& 1&95    & DRAGN    \\
3C&109   &   0&31   &     103 &   23&5 &	13&47&         4&103$^G$& 2&05	& G04    \\
4C&14.11 &   0&21   &     116 &   12&1 &	4&67&           2&10$^A$& 1&62	& HPPR97    \\
\\\hline

\multicolumn{2}{c}{Source}
&\multicolumn{2}{l}{z}
&$\theta$
&\multicolumn{2}{c}{S$_{178{\rm MHz}}$}
&\multicolumn{2}{c}{S$_{365 {\rm MHz}}$}
&\multicolumn{2}{c}{S$_{1400 {\rm MHz}}$}
&\multicolumn{3}{c}{R}\\
\multicolumn{4}{c}{ }&(arc sec.)&\multicolumn{2}{c}{Jy}
&\multicolumn{2}{c}{Jy}
&\multicolumn{2}{c}{Jy}&
\multicolumn{2}{c}{ }&REF\\\hline
3C&123   &   0&22    &     38 &   206&0&	122&53$^G$&    47&96$^G$& 1&98	& HPPR97   \\
3C&132   &   0&21    &     22 &   14&9 &	10&54	&      3&43$^E$	& 1&97	& HPPR97   \\
3C&153   &   0&28    &     9  &   16&7 &	11&25	&      4&13$^G$	& 1&53	& HPPR97   \\
3C&171   &   0&24    &     33 &   21&3 &	12&69	&      3&68$^E$	& 1&64	& HPPR97   \\
3C&172   &   0&52    &     121&   16&5 &	8&40	&      2&89$^H$	& 1&88	& SC85   \\
3C&173.1 &   0&29    &     61 &   16&8 &	9&52$^D$&      2&60$^C$	& 1&78	& HPPR97   \\
3C&175   &   0&77    &     52 &   19&2 &	10&22	&      2&44$^C$	& 1&71	& HLL94   \\
3C&175.1 &   0&92    &     7  &   12&4 &	5&64	&      1&93$^E$	& 1&39	& NRH99   \\
3C&181   &   1&38    &     6  &   15&8 &	7&62	&      2&33$^C$	& 0&93	& PH74   \\
3C&184   &   0&99    &     5  &   14&4 &	9&08	&      2&58$^E$	& 2&06	& LP91   \\
3C&184.1 &   0&12    &     182&   14&2 &	9&57$^D$&      3&30$^G$	& 2&84	& LP91   \\
3C&186   &   1&06    &     3  &   15&4 &	6&58	&      1&24$^E$	& 2&00	& AG95   \\
DA&240   &   0&04    &    2109&   23&2 &	130&00$^I$&    0&19$^E$	& 1&48	& T82   \\ 
3C&190   &   1&20    &     7  &   16&4 &	9&09	&      2&73$^E$	& 1&21	& AG95   \\
3C&191   &   1&95    &     5  &   14&2 &	7&49	&      1&85$^E$	& 2&62	& AG95 \\
3C&192   &   0&06    &     200&   23&0 &	14&39$^G$&     4&80$^G$	& 3&18	&  BHB88  \\
3C&196   &   0&87    &     6  &   74&3 &	49&02	&      15&01$^E$& 0&75	&  ASZ91  \\
3C&200   &   0&46    &     25 &   12&3 &	6&41	&      2&04$^H$	& 1&90	& DRAGN \\
4C&14.27 &   0&39    &     36 &   11&2 &	4&41	&      1&03$^A$	& 2&30	&  G04    \\
3C&204   &   1&11    &     37 &   11&4 &	5&49	&      1&38$^A$	& 2&92	&  HLL94  \\
3C&205   &   1&53    &     18 &  13&7  &	9&27	&      2&26$^E$	& 1&29	&  BMS88 \\
3C&207   &   0&69    &     13 &  14&8  & 	8&92	&      2&61$^E$	& 1&37	&  PH74   \\
3C&208   &   1&11    &     14 &  18&3  & 	10&33	&      2&36$^C$	& 1&91	&  HLL94  \\
3C&212   &   1&05    &     9  &  16&5  & 	8&34	&      2&37$^E$	& 1&50	&  ASZ91  \\
3C&215   &   0&41    &     56 &  12&4  & 	6&30$^C$&      1&59$^C$	& 1&52	&  HLL94  \\
3C&216   &   0&67    &   30   &  22&0  & 	13&93	&      4&23$^E$	& 1&70	&  AG95   \\
3C&217   &   0&90    &     12 &  12&3  & 	8&37	&      2&16$^C$	& 2&88	&  BLR95  \\
3C&219   &   0&17    &     190&  44&9  & 	14&00$^H$&      8&46$^C$& 1&50	&  NRH99  \\
3C&220.1 &   0&61    &     35 &  17&2  & 	9&23$^D$&      2&24$^E$	& 2&29	&  JPR77  \\
3C&220.3 &   0&69    &     10 &  17&1  & 	11&59	&      2&89$^C$	& 1&19	&  JPR77  \\
3C&223   &   0&14    &     306&  16&0  & 	8&51$^G$&      3&58$^C$	& 2&95	&  LP91   \\
3C&225B  &   0&58    &     5  &  23&2  & 	10&17	&      3&34$^E$	& 1&98	&  -   \\
3C&226   &   0&82    &     35 &  16&4  & 	9&34	&      2&39$^C$	& 1&80	&  BLR97 \\
4C&73.08 &   0&06    &    1004&  15&6  & 	24&20$^J$&      2&65$^A$& 1&33	&  DRAGN  \\
3C&228   &   0&55    &     46 &  23&8  & 	10&44	&      3&69$^A$	& 2&40	&  JLG95  \\
3C&234   &   0&18    &     112&  34&2  & 	15&78	&      5&38$^G$	& 3&29	&  HPPR97 \\
3C&236   &   0&10    &    2478&  15&7  & 	7&60	&      3&28$^K$ & 1&13	&  ALB94   \\
3C&239   &   1&78    &     11 &  14&4  & 	7&37	&      1&46$^C$	& 1&38	&  BLR94  \\
4C&74.16 &   0&81    &     40 &  12&8  & 	25&90$^J$&      2&30$^A$& 1&98	&   -   \\
3C&241   &   1&62    &     1  &  12&6  & 	7&18	&      1&69$^C$	& 0&80	&  PPR85\\
3C&244.1 &   0&43    &     51 &  22&1  & 	13&89	&      3&94$^C$	& 4&32	&  F02   \\
	\\\hline
\multicolumn{2}{c}{Source}
&\multicolumn{2}{l}{z}
&$\theta$
&\multicolumn{2}{c}{S$_{178{\rm MHz}}$}
&\multicolumn{2}{c}{S$_{365 {\rm MHz}}$}
&\multicolumn{2}{c}{S$_{1400 {\rm MHz}}$}
&\multicolumn{3}{c}{R}\\
\multicolumn{4}{c}{ }&(arc sec.)&\multicolumn{2}{c}{Jy}
&\multicolumn{2}{c}{Jy}
&\multicolumn{2}{c}{Jy}&
\multicolumn{2}{c}{ }&REF\\\hline
3C&245   &   1&03    &     9   	 &  15&7  &	9&45  	&      3&31$^E$   & 1&98	& -    \\
3C&247   &   0&74    &     15  	 &  11&6  &	7&61 	&      2&88$^E$   & 4&15	& ALB94   \\
3C&249.1 &   0&31    &     47    &  11&7  &	7&55$^D$ &      2&34$^A$  & 1&59	& HLL94   \\
3C&252   &   1&11    &     60 	 &  12&0  &	6&56 	&      1&21$^C$   & 1&12	& G04    \\
3C&254   &   0&73    &     15  	 &  21&7  &	12&57 	&      3&13$^E$   & 1&98	& -    \\
3C&263   &   0&66    &     51    &  16&6  &	11&12 	&      3&00$^G$	  & 1&98	& -    \\
3C&263.1 &   0&82    &     7   	 &  19&8  &	16&96 	&      3&13$^E$   & 0&98	& NRH99    \\
3C&265   &   0&81    &     78 	 &  21&3  &	12&18 	&      2&84$^C$   & 2&95	& G04    \\
3C&266   &   1&28    &     5  	 &  12&1  &	6&91 	&      1&43$^E$   & 1&98	& AG95    \\
3C&267   &   1&14    &     38 	 &  15&9  &	9&07 	&      2&16$^C$   & 1&77	& G04    \\
3C&268.1 &   0&97   &      46 	 &  23&3  &	10&02$^C$&      6&80$^G$  & 1&87	& G04    \\
3C&268.3 &   0&37   &      1  	 &  11&7  &	11&27 	&      3&72$^E$   & 1&56	& AG95    \\
3C&268.4 &   1&40   &      10  	 &  11&2  &	6&59 	&      1&98$^E$   & 1&98	& -    \\
3C&270.1 &   1&52   &      12    &  14&8  &	9&74 	&      2&85 $^E$  &1&69	& ALB94    \\
3C&274.1 &   0&42   &      150   &  18&0  &	5&12$^C$&      2&78$^C$	  & 3&56	& AL87    \\
3C&275.1 &   0&56   &      16  	 &  19&9  &	9&31$^B$&      2&90$^E$   & 1&29	& ALB94    \\
3C&277.2 &   0&77   &      55 	 &  13&1  &	7&09 	&      1&85$^C$   & 1&67	& PRL97    \\
3C&280   &   1&00   &      15  	 &  25&8  &	14&94 	&      5&10$^E$   & 0&98	& G04    \\
3C&284   &   0&24   &      178   &  12&3  &	7&07	&      1&93$^A$    &3&14	& HPPR97   \\
3C&285   &   0&08   &      180	 &  12&3  &	3&14$^C$&      2&04$^C$   & 1&25	& AL87    \\
3C&289   &   0&97   &      10  	 &  13&1  &	8&10 	&      2&40$^E$   & 2&34	& BLR97    \\
3C&292   &   0&71   &      140	 &  11&0  &	5&34 	&      2&07$^H$   & 3&82	& AL87    \\
3C&294   &   1&78   &      15  	 &  11&2  &	6&12 	&      1&32 $^E$  & 1&98	& -    \\
3C&295   &   0&46   &      6     &  91&0  &	65&44$^G$&      22&82$^C$ &1&73	& AL87    \\
3C&299   &   0&37   &      11 	 &  12&9  &	8&17	&      3&15$^E$   & 1&17	& G04    \\
3C&300   &   0&27   &      101   &  19&5  &	10&47 	&      3&70$^G$	  & 3&23	& HPPR98   \\
3C&303   &   0&14   &      47 	 &  12&2  &	7&24 	&      2&67$^C$   & 1&14	& HPPR98   \\
3C&318   &   1&57   &      1   	 &  13&4  &	9&21 	&      2&69$^E$   & 1&25	& AG95    \\
3C&319   &   0&19   &      105   &  16&7  &	7&66 	&      2&50$^C$	  & 2&49	& HPPR97   \\
3C&321   &   0&10   &      307   &  14&7  &	9&31$^G$&      3&50$^G$	  & 2&04	& BHB88    \\
3C&322   &   1&68   &      33 	 &  11&0  &	6&94 	&      1&98$^C$   & 1&53	& LLA95    \\
3C&324   &   1&21   &      10  	 &  17&2  &	8&91 	&      2&44 $^C$  & 1&07	& G04    \\
3C&326   &   0&09   &     1206   &  22&2  &	6&60$^F$&      3&10$^F$	  & 3&23	& DRAGN    \\
3C&325   &   0&86   &      16  	 &  17&0  &	12&12 	&      3&56$^E$   & 3&70	& FBP97    \\
3C&330   &   0&55   &      60    &  30&3  &	21&88 	&      7&10$^G$	  & 1&20	& F02    \\
3C&334   &   0&56   &      58 	 &  11&9  &	6&80	&      2&15$^C$   & 1&58	& HLL94    \\
3C&336   &   0&93   &      28 	 &  12&5  &	7&88 	&      2&74$^C$   & 1&65	& HLL94    \\
3C&341   &   0&45   &      74 	 &  11&8  &	6&35	&      2&15$^C$   & 3&14	& HLL94    \\
3C&340   &   0&78   &      46    &  11&0  &	7&79 	&      2&49$^C$	  & 2&10	& JPR97    \\
3C&337   &   0&63   &      46 	 &  12&9  &	9&98 	&      2&94$^C$   & 2&07	& BLR97    \\
3C&343.1 &   0&75   &      0.3	 &  12&5  &	8&00$^C$&      4&83$^C$   & 0&89	& AG95    \\
\\ \hline
\multicolumn{2}{c}{Source}
&\multicolumn{2}{l}{z}
&$\theta$
&\multicolumn{2}{c}{S$_{178{\rm MHz}}$}
&\multicolumn{2}{c}{S$_{365 {\rm MHz}}$}
&\multicolumn{2}{c}{S$_{1400 {\rm MHz}}$}
&\multicolumn{3}{c}{R}\\
\multicolumn{4}{c}{ }&(arc sec.)
&\multicolumn{2}{c}{Jy}
&\multicolumn{2}{c}{Jy}
&\multicolumn{2}{c}{Jy}
&\multicolumn{2}{c}{ }&REF\\
\hline
3C&349   &   0&21   &      86   &  14&5  &	9&95 	&      3&30$^G$	 & 2&77	& HPPR97     \\
3C&351   &   0&37   &      74   &  14&9  &	8&37 	&      3&27$^G$	 & 1&34	& G04     \\
3C&352   &   0&81   &      13   &  12&3  &	7&85 	&      1&87$^E$ & 1&57	& ALB94     \\
3C&356   &   1&08   &      75   &  12&3  &	6&39 	&      1&41$^C$ & 3&00	& FBB93     \\
4C&16.49 &   1&30   &      16   &  11&4  &	6&27 	&      1&46 $^E$ & 1&10	& G04     \\
4C&13.66 &   1&45   &      6    &  12&3  &	6&95 	&      1&70$^E$  & 1&49	& LGS98     \\
3C&368   &   1&13   &      8    &  15&0  &	7&40 	&      1&11$^C$ & 2&67	& BLR95     \\
3C&381   &   0&16   &      73   &  18&1  &	8&71	&      3&910$^G$& 2&11	& HPPR97     \\
3C&382   &   0&06   &      185  &  21&7  &	6&55 	&      5&83$^C$ & 0&86	& F02     \\
3C&388   &   0&09   &      50   &  26&8  &	16&04 	&      5&74$^C$	  & 1&21& PH74     \\
3C&390.3 &   0&06   &      229  &  51&8  &	30&19$^D$ &      11&90$^G$& 1&40& F02     \\
3C&401   &   0&20   &      24   &  22&8  &	14&34 	&      5&07$^E$  & 1&49	& HPPR97     \\
3C&427.1 &   0&57   &      27   &  29&0  &	15&52$^D$ &      3&80$^G$& 2&78	& NRH95     \\
3C&432   &   1&81   &      15   &  12&0  &	6&48 	&      1&58$^E$  & 1&33	& BHL94    \\
3C&433   &   0&10   &      68   &  61&3  &	31&27 	&      11&80$^G$& 1&98	& -    \\
3C&436   &   0&21   &      109  &  19&4  &	5&74$^C$&      3&40$^C$ & 2&75	& HPPR97     \\
3C&437   &   1&48   &      34   &  15&9  &	8&68 	&      2&76$^C$ & 1&94	& BLR97     \\
3C&438   &   0&29   &      23   &  48&7  &	26&40 	&      6&37$^E$  & 1&81	& HPPR97     \\
3C&441   &   0&71   &      37   &  13&7  &	8&34 	&      2&52$^C$ & 1&58	& FBP97     \\
3C&452   &   0&08   &      280  &  59&3  &	32&67$^G$&      10&54$^C$ & 2&16& DRAGN     \\
3C&455   &   0&54   &      4    &  14&0  &	9&18 	&      2&89$^C$ & 1&46	& AG95     \\
3C&457   &   0&43   &      210  &  14&3  &	5&68 	&      1&75$^A$	& 2&45	& G04     \\
3C&469.1 &   1&34   &      74   &  12&1  &	6&46$^D$ &      1&76$^C$& 1&98	& -     \\	
3C&470   &   1&65   &      24   &  11&0  &	5&98 	&      1&94$^E$ & 3&35	& BLR97 \\\hline
\\

\caption[3CRR data]{Parameters used in the fitting on the KDA model in chapter 6. Column
1: 3CRR source name. Column 2: redshifts. Column 3: Angular size in arcseconds. Column 4: Flux
density at 178 MHz. Column 5: Flux density at 365 MHz taken from
\citet{texas96} unless otherwise noted. Column 6: Flux
density at 1400 MHz and Column 7: Average source lobe ratio. Columns 2, 3
and 4 use data taken from \citet{3CRR}. Sources with no high quality maps
have lobe ratios which are the average for the 3CRR sample. See Table
\ref{3.2} for an explanation of the reference codes used.}
\end{longtable}		

\begin{table}
\centering
\begin{tabular}{lr}\\ \hline
\\
Listing & Full Reference\\ \hline
A  & \citet{wb92}\\
B & \citet{lml81}\\
C & \citet{lp80}\\
D & \citet{wensp}\\
E & \citet{ccg98}\\
F & \citet{kpt69}\\
G & \citet{kwp81}\\
H & \citet{ptw66}\\
I & \citet{ve75}\\
J&\citet{hwr95}\\
K & \citet{bwh95}\\
LP91 & \citet{lp91a}\\
G04 &  \citet{me}\\ 
 LLA95& \citet{lla95}\\
NRH95 & \citet{nrh95}\\
F02 &\citet{f02}\\
ALB94 &\citet{alb94}\\
DRAGN &\citet{dragon}\\
HPPR97&\citet{hppr97}\\
HPR98&\citet{hppr98}\\
HLL94&\citet{bhl94}\\
ASZ91 &\citet{asz91}\\
BLR97&\citet{blr97}\\
LCF01 &\citet{lcf01}\\
PPR85&\citet{ppr85}\\
FBB93&\citet{fbb93}\\
AG95 &\citet{ag95}\\
SC85&\citet{sc85}\\
PH74&\citet{ph74}\\
T82&\citet{t82}\\
BHB88&\citet{bhb88}\\
JPR77&\citet{jpr77}\\
BMS88&\citet{bms88}\\
JLG95&\citet{jlg95}\\
AL87&\citet{al87}\\
FBP97&\citet{fbp97}\\
LGS98&\citet{lgs98}\\
\hline
\end{tabular}
\caption{Reference table}\label{3.2}
\end{table}

\newpage

\newpage
\thispagestyle{empty}

\chapter{6CE sample}
\index{6CE}
\begin{longtable}{r@{ }l r@{.}l c r@{.}l r@{.}l l  l@{.}r r}\hline
\multicolumn{2}{c}{Source}&\multicolumn{2}{c}{z}&$\theta$
&\multicolumn{2}{c}{S$_{151 {\rm MHz}}$}
&\multicolumn{2}{c}{S$_{365 {\rm MHz}}$}&
\multicolumn{1}{c}{S$_{1400 {\rm MHz}}$}&\multicolumn{3}{c}{R}\\
\multicolumn{4}{c}{ }&(arc sec.)
&\multicolumn{2}{c}{Jy}&\multicolumn{2}{c}{Jy}
&\multicolumn{1}{c}{mJy}&\multicolumn{2}{c}{ }&REF\\
\hline\endfirsthead
6C&0820+364 &1&86  &  24 &  2&39 &  1&15     & 213$^A$ &    1&58 & \\
6C&0822+341 &0&41  &  18 &  3&06 &  1&77     & 552$^A$ &    1&83 &-\\
6C&0822+343 &0&77  &  21 &  2&93 &  1&03     & 111$^A$ &    1&64 & \\
6C&0823+375 &0&21  &  81 &  3&35 &  1&42     & 425$^A$ &    1&41 & \\
6C&0824+353 &2&25  &  8  &  2&42 &  1&83     & 958 $^E$&    1&80 & NAR92 \\
6C&0825+345 &1&47  &  7  &  2&10 &  1&22     & 284$^A$ &    1&95&   \\
6C&0847+375 &0&41  &  33 &  3&07 &  1&43     & 614$^A$ &    2&67&MMC97 \\
6C&0857+390 &0&23  &  24 &  2&71 &  1&51     & 503$^A$ &    1&83& -  \\
6C&0901+355 &1&90  &  4  &  2&07 &  1&08     & 219$^K$ &    1&24&NAR92 \\
6C&0902+341 &3&40  &  5  &  2&14 &  1&12     & 298$^K$ &    2&42&COM94 \\
6C&0905+395 &1&88  &  5  &  2&82 &  0&94$^L$ & 233$^A$ &    2&85&LEL95 \\
6C&0908+373 &0&11  &  39 &  2&33 &  1&33     & 614$^A$ &    1&08&PRF86 \\
6C&0913+390 &1&25  &  9  &  2&27 &  1&80     & 1005$^E$&    0&80 &HBW95  \\
6C&0919+380 &1&65  &  10 &  2&72 &  1&05$^L$ &  65$^N$ &    1&84&NAR92\\
6C&0922+364 &0&11  &  17 &  3&27 &  1&09     & 725$^K$ &    1&85&MPC97\\
6C&0930+385 &2&40  &  5  &  2&21 &  1&06     & 278$^A$ &    1&84&FPD01\\
6C&0943+395 &1&04  &  12 &  2&31 &  1&20     & 323$^A$ &    2&01&G04\\
6C&0955+384 &1&41  &  22 &  3&45 &  1&75     & 388$^A$ &    1&66& \\
6C&1011+363 &1&04  &  66 &  2&10 &  1&07     & 374$^A$ &    2&07& \\
6C&1016+363 &1&89  &  31 &  2&28 &  1&39     & 528 $^E$&    1&61& \\
6C&1017+371 &1&05  &  9  &  2&68 &  1&44     & 359$^A$ &    1&83& \\
6C&1018+372 &0&81  &  83 &  2&52 &  0&97     & 271$^A$ &    2&07& \\
6C&1019+392 &0&92  &  9  &  2&99 &  1&76     & 392$^A$ &    2&05 &BEL99\\
6C&1025+390 &0&36  &  1  &  2&97 &  1&98     & 659$^E$ &    2&6 & NAR92\\
6C&1031+340 &1&83  &  3  &  2&33 &  1&55     & 478$^A$ &    2&72&  \\
6C&1042+391 &1&77  &  11 &  2&68 &  1&96     & 655$^E$ &    1&57&NAR92 \\
6C&1043+371 &0&79  &  5  &  2&62 &  1&18     & 249$^K$ &    1&07& NAR92\\
\\\hline
\multicolumn{2}{c}{Source}&\multicolumn{2}{c}{z}&$\theta$
&\multicolumn{2}{c}{S$_{151 {\rm MHz}}$}
&\multicolumn{2}{c}{S$_{365 {\rm MHz}}$}&
\multicolumn{1}{c}{S$_{1400 {\rm MHz}}$}&\multicolumn{3}{c}{R}\\
\multicolumn{4}{c}{ }&(arc sec.)
&\multicolumn{2}{c}{Jy}&\multicolumn{2}{c}{Jy}
&\multicolumn{1}{c}{mJy}&\multicolumn{2}{c}{ }&REF\\
\hline
6C&1045+340 &1&83  &  22 &  2&00 &  1&20     & 266$^A$ &    1&98& \\
6C&1045+355 &0&85  &  9  &  2&07 &  1&26     & 267 $^K$&    1&44&NAR92 \\
6C&1045+351 &1&60  &  0.1&  3&03 &  2&44     & 1035$^E$&    1&83&- \\
6C&1100+350 &1&44  &  14 &  2&26 &  1&18     & 304$^A$ &    4&25&  \\
6C&1108+395 &0&59  &  16 &  2&10 &  2&01     & 390$^A$&    1&83&- \\
6C&1113+345 &2&41  &  17 &  2&33 &  1&39     & 440$^A$ &    2&80&   \\
6C&1123+340 &1&25  &  0.2&  3&40 &  3&80     &1378$^E$ &    0&98&NAR92 \\
6C&1125+374 &1&23  &  18 &  2&07 &  1&06     & 325$^K$ &    1&83&- \\
6C&1129+371 &1&06  &  19 &  2&36 &  1&42     & 416$^K$ &    1&72&  \\
6C&1130+345 &0&51  &  78 &  3&20 &  2&10$^M$ & 524$^K$ &    2&31&  \\
6C&1134+365 &2&13  &  17 &  2&07 &  1&10     & 235$^K$ &    1&27&  \\
6C&1141+352 &1&78  &  12 &  2&40 &  1&13     & 290$^K$ &    1&65&  \\
6C&1143+370 &1&96  &  0.1&  2&06 &  1&51     & 448$^K$ &    1&83&- \\
6C&1148+363 &0&14  &  27 &  3&21 &  1&41     & 550$^A$&    1&83&- \\
6C&1148+384 &1&30  &  10 &  3&83 &  2&20     & 652$^E$ &    2&22&NAR92 \\
6C&1158+343 &0&53  &  40 &  2&12 &  1&30     & 314$^K$ &    1&83&- \\
6C&1159+365 &1&40  &  2  &  2&20 &  1&40     & 352$^K$ &    1&28&  \\
6C&1204+351 &1&38  &  63 &  3&43 &  1&96     & 505$^A$ &    1&49&  \\
6C&1205+391 &0&24  &  24 &  3&83 &  1&61     & 634$^A$ &    1&55&MMC82 \\
6C&1212+380 &0&95  &  0.6&  2&14 &  1&27     & 287$^K$ &    1&20&NAR92  \\
6C&1213+350 &0&86  &  0.1&  2&39 &  2&32     & 1507$^E$&    1&25&XRP95\\ 
6C&1217+364 &1&09  &  0.5&  2&40 &  1&32     & 396$^K$ &    0&93&  \\
6C&1220+372 &0&49  &  36 &  2&52 &  1&44     & 463$^A$ &    1&83 &-\\
6C&1230+345 &1&53  &  12 &  2&90 &  1&89     & 424$^A$ &    2&30 &  \\
6C&1232+394 &3&22  &  51 &  3&27 &  1&47     & 1130$^L$&    1&70 &NAR92 \\
6C&1255+370 &0&71  &  0.6&  3&66 &  2&33     & 759$^E$ &    1&83&- \\
6C&1256+364 &1&13  &  18 &  2&88 &  1&61     & 527$^E$ &    1&53&  \\
6C&1257+363 &1&00  &  40 &  2&40 &  1&07     & 196$^A$&    1&80&   \\
6C&1301+381 &0&47  &  28 &  3&46 &  2&01     & 536$^A$ &    1&83&- \\\hline
\caption[Archival 6CE data] {Parameters used in the fitting on the KDA model in
chapter 6. Column 1: 6CE source name. Column 2: redshifts. Column 3:
Angular size in arcseconds. Column 4: Flux density at 151 MHz. Column
5: Flux density at 365 MHz taken from \citet{texas96} unless otherwise
stated. Column 6: Flux density at 1400 MHz and Column 7: Average
source lobe ratio, high resolution maps taken from \citet{lla95}
unless otherwise stated. Columns 2, 3 and 4 use data taken from
\citet{rel01}. See Table \ref{6.2} for explanation of reference
codes.}
\end{longtable}
\begin{table}
\centering
\begin{tabular}{lr}\\ \hline
\\
Listing & Full Reference\\ \hline
A  & \citet{wb92}\\
E & \citet{ccg98}\\
I & \citet{ve75}\\
J&\citet{hwr95}\\
K & \citet{bwh95}\\
L & \citet{fgt85}\\
M&\citet{ps65}\\
N&\citet{bwe91}\\
G04 &  \citet{me}\\ 
HBW95&\citet{hbw95}\\
PRF86&\citet{prf86}\\
COH94&\citet{coh94}\\
LEL95&\citet{lel95}\\
MPC97&\citet{mpc97}\\
FPD01&\citet{fpd01}\\
BEL99&\citet{bel99}\\
NAR92&\citet{nar92}\\
MMC82&\citet{mmc82}\\
XRP95&\citet{xrp95}\\
\hline
\end{tabular}
\caption{6CE Reference table}\label{6.2}
\end{table}

\newpage
\thispagestyle{empty}
\chapter{7C III sample}\label{ap:7C}
\index{7C III}
\begin{longtable}{cccccrrc}\hline
\multicolumn{2}{c}{Source}&z&$\theta$&S$_{151 {\rm MHz}}$&S$_{327 {\rm MHz}}$&S$_{1490 {\rm MHz}}$&R\\
&&&(arc sec.)& Jy&mJy&mJy&\\

\hline\endfirsthead
7C&1732+6535&	0.86	& 20	& 6.17	  &  2979	&  1058		&1.80	\\
7C&1733+6719&	1.84	& 3	& 1.84	  &  966	&  247 	        &1.80	\\
7C&1741+6704& 	1.05	& 4	& 0.73	  &  366	&  80	 		&1.80	\\
7C&1742+6346&	1.27	& 51	& 0.62	  &  290	&  68			&1.80	\\    
7C&1743+6344&	0.32	& 14	& 1.33	  &  685	&  214			&3.16	\\
7C&1743+6431&	1.70	& 45	& 1.89	  &  948	&  247		&1.25	\\
7C&1743+6639&	0.27	& 50	& 1.97    &  1238	&  433		&1.51	\\
7C&1745+6415&	0.67	& 6	& 0.64	  &  399	&  148		&1.50	\\
7C&1745+6422&	1.23	& 16	& 1.51	  &  823	&  237		&1.07	\\
7C&1747+6533&	1.52	& 0.7	& 2.92	  &  1605	&  375		&1.80	\\
7C&1748+6703&	3.20	& 14	& 2.29	  &  1199	&  217		&1.94	\\	
7C&1748+6657&	1.05	& 0.3	& 1.21	  &  1523	&  530		&1.80	\\
7C&1748+6731&	0.56	& 108	& 0.64	  &  376	&  118		&2.75	\\
7C&1751+6809&	1.54	& 2	& 0.74	  &  480	&  134		&1.80	\\    
7C&1751+6455&	0.29	& 43	& 0.65	  &  421	&  130		&1.19	\\
7C&1753+6311&	1.96	& 17	& 1.09	  &  536	&  138	   	        &2.35	\\	
7C&1753+6543&	0.14	& 84	& 1.62	  &  1072	&  461		&1.64	\\	
7C&1754+6420&	1.09	& 15	& 0.42	  &  255	&  57		&2.11	\\
7C&1755+6830&	0.74	& 9	& 1.11	  &  654	&  211		&1.80	\\
7C&1756+6520&	1.48	& 5	& 0.73	  &  465	&  171		&1.15	\\    
7C&1758+6535&	0.80	& 106	& 1.13	  &  739	&  226		&2.50	\\
7C&1758+6553&	0.17	& 115	& 1.30	  &  715	&  210		&1.80	\\
7C&1758+6307&	1.19	& 4	& 1.94	  &  1174	&  304		&1.50	\\
7C&1758+6719&	2.70	& 45	& 0.76	  &  400	&  98			&2.19	\\    
7C&1801+6902&	1.27	& 21	& 1.37	  &  673	&  203		&1.40	\\
7C&1802+6456&	2.11	& 26	& 1.97	  &  857	&  211		&1.71	\\
7C&1804+6313&	1.50	& 29	& 0.62	  &  333	&  66			&1.80	\\    
\\ \hline
\multicolumn{2}{c}{Source}&z&$\theta$&S$_{151 {\rm MHz}}$&S$_{327 {\rm MHz}}$&S$_{1490 {\rm MHz}}$&R\\
&&&(arc sec.)& Jy&mJy&mJy&\\ \hline

7C&1805+6332&	1.84	& 14	& 1.08	  &  569	&  137			&2.08	\\	
7C&1807+6831&	0.58	& 29	& 2.12	  &  1216	&  334		&1.44	\\
7C&1807+6841&	0.81	& 12	& 0.63	  &  359	&  115		&1.38	\\
7C&1811+6321&	0.27	& 52	& 0.95	  &  452	&  129		&2.56	\\
7C&1813+6439&	2.04	& 38	& 0.5	  &  237	&  41			&2.00	\\
7C&1813+6846&	1.03	& 52	& 1.51	  &  754	&  209			&1.93	\\
7C&1814+6529&	0.96	& 126	& 1.22	  &  629	&  125		&1.80	\\
7C&1815+6805&	0.23	& 50	& 1.96	  &  1188	&  381		&1.70	\\
7C&1815+6815&	0.79	& 200	& 1.37	  &  725	&  181		&2.00	\\
7C&1816+6710&	0.92	& 27	& 2.44	  &  1476	&  456			&1.67	\\
7C&1816+6605&	0.92	& 2	& 1.29	  &  723	&  162	        &1.92	\\
7C&1819+6550&	0.72	& 9	& 1.21	  &  772	&  257			&1.19	\\
7C&1822+6601&	0.37	& 52	& 0.97	  &  607	&  178		&1.45	\\
7C&1826+6602&	2.38	& 3	& 1.63	  &  605	&  144		&1.10	\\    
7C&1826+6510&	0.65	& 34	& 1.39	  &  852	&  240			&1.83	\\
7C&1826+6704&	0.29	& 19	& 0.61	  &  355	&  108		&1.19	\\
7C&1827+6709&	0.48	& 17	& 1.10	  &  614	&  155		&2.15	\\
7C&1814+6702&	4.05	& 18	& 2.32	  &  1097	&  223		&1.34	\\\hline
\\	
\caption[Archival 7C III data] {Parameters used in the fitting on the
KDA model in chapter 6. Column 1: 7C III source name. Column 2:
redshifts taken from \citet{7CI}. Column 3: Angular size in arcseconds
taken from \citet{lrw92}. Column 4: Flux density of 151 MHz. Column 5:
Flux density of 327 MHz. Column 6: Flux density of 1490 MHz and Column
7: Average source lobe ratio. Sources with a lobe ratio of 1.8 do not
have high enough quality maps. All other lobe ratios use maps taken
from \citet{lrw92}. Columns 4-6 are taken from \citet{7CI}.}
\end{longtable}		

\newpage
\thispagestyle{empty}

\newpage
\thispagestyle{empty}
\input{Thesis.ind}

\end{document}